\documentclass[conference,keeplastbox]{IEEEtran}
\usepackage{cite}
\usepackage{amsmath,amssymb,amsfonts}
\usepackage{algorithmic}
\usepackage{graphicx}
\usepackage{textcomp}
\usepackage{xcolor}
\usepackage{float}

\newcommand{\rotatecurvearrowleft}{%
            \mathrel{\raisebox{0em}{%
            \reflectbox{\rotatebox[origin=c]{90}{$\curvearrowleft$}}}}}

\newcommand{\rotatecurvearrowright}{%
            \mathrel{\raisebox{0em}{%
            \reflectbox{\rotatebox[origin=c]{270}{$\curvearrowright$}}}}}

\usepackage{subfigure}
\PassOptionsToPackage{hyphens}{url}\usepackage{hyperref}

\usepackage{graphicx}  %Required
\usepackage{caption}
 
\DeclareCaptionLabelFormat{lc}{{#1}~#2}
\def\tablename{Table}

\makeatletter
\def\endthebibliography{%
  \def\@noitemerr{\@latex@warning{Empty `thebibliography' environment}}%
  \endlist
}
\makeatother

 \usepackage{epsfig,endnotes}
\let\labelindent\relax
\usepackage{floatpag,enumitem}
\setlist{leftmargin=10pt}
\usepackage{relsize}
 \usepackage{tikz}

 \usepackage{mathtools}
\usepackage{multirow}
\usepackage{setspace}
\usepackage{mathrsfs}
\usepackage{bbm}
\usepackage{pifont}
\usepackage{amsfonts}

\usepackage{amsmath, amsthm}
\newcommand{\qeda}{~\hfill\ensuremath{\blacksquare}}

\usepackage{tabularx}
\usepackage{listings}
\usepackage{nicefrac}       % compact symbols for 1/2,
\usepackage{graphicx}
\usepackage{amssymb,booktabs}
\usepackage{array}
 \usepackage{fancyhdr}
\usepackage{cases}

 \usepackage{supertabular}
\usepackage{mathrsfs}

\usepackage{psfrag}
\usepackage{bbding}

\usepackage{float}

\usepackage{amsbsy}

\makeatletter
\providecommand{\leftsquigarrow}{%
  \mathrel{\mathpalette\reflect@squig\relax}%
}
\newcommand{\reflect@squig}[2]{%
  \reflectbox{$\m@th#1\rightsquigarrow$}%
}
\makeatother

\usepackage{algorithm}
 \usepackage{algorithmic}

\makeatletter
\newcommand{\newalgname}[1]{%
  \renewcommand{\ALG@name}{#1}%
}

\newcommand {\C} {{\rm I\kern-5.5pt C}}

\renewcommand{\arraystretch}{1}

\newcommand{\bp}[1]{{\mathbb{P}}\left[{#1}\right]}

\newcommand{\bE}[1]{{\mathbb{E}}\left[{#1}\right]}

\newenvironment{myitemize}
{ \begin{itemize}[leftmargin=10pt]
    \setlength{\itemsep}{0pt}
    \setlength{\parskip}{0pt}
    \setlength{\parsep}{0pt}      }
{ \end{itemize}                  }

\def\centerhack#1{\hbox to 0pt{\hss\footnotesize #1\hss}}
\def\centerhackn#1{\hbox to 0pt{\hss #1\hss}}
\def\dchack#1{\vbox to 0pt{\vss{\hbox to 0pt{\hss#1\hss}}\vss}}
\newcommand{\pr}[1]{{\mathbb{P}}\left[{#1}\right]} % probability measure

\usepackage{etoolbox}
\AtBeginEnvironment{align}{\setcounter{subeqn}{0}}% Reset subequation number at start of align
\newcounter{subeqn} %
\usetikzlibrary{positioning}

\newcounter{mysub}

\newtheorem{defn}{Definition}

\newtheorem{lem}{Lemma}

\newtheorem{thm}{Theorem}

\newtheorem{rem}{Remark}

\newtheorem{Claima}{Claim}
\newtheorem{prop}{Proposition}
\newtheorem*{proposition1.1}{Proposition 1.1}
\newtheorem*{proposition1.2}{Proposition 1.2}
\newtheorem*{proposition1.3}{Proposition 1.3}
\newtheorem*{proposition2.1}{Proposition 2.1}
\newtheorem*{proposition2.2}{Proposition 2.2}
\usepackage{subfigure}

\begin{document}
% \title{A Markov Chain based Approach to Analyze Blockchain Consistency in Asynchronous Networks: Deriving a Neat Bound}
\title{An Analysis of Blockchain Consistency in Asynchronous Networks: Deriving a Neat Bound\\[-39pt]}

% \author{\IEEEauthorblockN{Authors (To be decided and anonymized): Jun Zhao, Anupam, Yang Liu, Lam, Jeff Yan, Zengxiang Li, Delta} %Shengli Zhang, 
% \IEEEauthorblockA{Nanyang Technological University
% \\
% junzhao@ntu.edu.sg
% }}

%\author{\IEEEauthorblockN{Anonymized} }

% \author{\IEEEauthorblockN{Paper ID: 182} }

\author{Jun Zhao$^{1}$, Jing Tang$^2$, Zengxiang Li$^3$, Huaxiong Wang$^{4}$, Kwok-Yan Lam$^{1}$, Kaiping Xue$^{5}$
\vspace{1.3mm}
\\
\fontsize{10}{10}
%\selectfont\itshape
$^1$School of Computer Science and Engineering, Nanyang Technological University, Singapore\\
$^2$Department of Industrial Systems Engineering and Management, National University of Singapore, Singapore\\
$^3$Institute of High Performance Computing (IHPC), Agency for Science, Technology and Research (A*STAR), Singapore\\
$^4$School of Physical and Mathematical Sciences, Nanyang Technological University, Singapore\\
$^5$Department of Information Security, University of Science and Technology of China, China\vspace{1.3mm}
\\
\fontsize{9}{9}
%\selectfont\ttfamily\upshape 
$^1$\{junzhao,~kwokyan.lam\}@ntu.edu.sg, $^2$isejtang@nus.edu.sg, 
$^3$liz@ihpc.a-star.edu.sg, 
$^4$hxwang@ntu.edu.sg, 
$^5$kpxue@ustc.edu.cn}
\maketitle
 \thispagestyle{fancy}
\pagestyle{fancy}
\lhead{This paper appears in the Proceedings of IEEE International Conference on Distributed Computing Systems (\textbf{ICDCS}) 2020.\\ Please feel free to contact us for questions or remarks.}
\cfoot{\thepage}
\renewcommand{\headrulewidth}{0.4pt}
\renewcommand{\footrulewidth}{0pt}


% \section{Proving Lemmas ?}

% \begin{align}
%   & \bp{C(t_0, t_0+T-1) > A(t_0, t_0+T-1)} \nonumber  \\ & \geq \bp{\begin{array}{l}
% C(t_0, t_0+T-1) > (1-\delta_2) \cdot T \hspace{1pt} {\overline{\alpha}}^{2\Delta} \alpha_1 \\
% \geq (1+\delta_4) \cdot T  p \nu n > A(t_0, t_0+T-1)  \end{array}} \nonumber  \\ & \geq \bp{\begin{array}{l} \left(C(t_0, t_0+T-1) > (1-\delta_2) \cdot T \hspace{1pt} {\overline{\alpha}}^{2\Delta} \alpha_1\right) \\ \cap\, \left(A(t_0, t_0+T-1) <    (1+\delta_4) \cdot T  p \nu n\right)  \end{array}}
%   \nonumber  \\ &    \nonumber  \\ & \geq \bp{A(t_0, t_0+T-1) \geq (1+\delta_4) \cdot T  p \nu n}  \nonumber  \\ & \leq \exp\left(- T \nu n \cdot D\left( (1+\delta_4) p ||p\right)  \right) . \label{a}
% \end{align}

\begin{abstract}
Formal analyses of blockchain protocols have received much attention recently. Consistency results of Nakamoto's blockchain protocol are often expressed in a quantity $c$, which denotes the expected number of network delays before some block is mined. With $\mu$ (resp., $\nu$) denoting the fraction of computational power controlled by benign miners (resp., the adversary), where $\mu + \nu = 1$, we prove for the first time that to ensure the consistency property of Nakamoto's blockchain protocol in an asynchronous network, it suffices to have $c$ to be just slightly greater than $\frac{2\mu}{\ln (\mu/\nu)}$. Such a result is both neater and stronger than existing ones. In the proof, we formulate novel Markov chains which characterize the numbers of mined blocks in different rounds.
\end{abstract}
 %(e.g., no, one, or over one mined block)
% https://www.wolframalpha.com/input/?i=plot+2x%2F(ln(x%2F(1-x))),+for+0.5%3Cx%3C1 
% plot (1-x)/(10^(13))/(1-(1+1/(10^(15)))*(x/(1-x))^(1/(2*10^(13)))), 0<x<1/2

%computation offloading for mobile-edge server computing in emerging  wireless networks with Reconfigurable intelligent surfaces
\begin{IEEEkeywords}
Blockchain, consistency, asynchronous networks, Markov chains. \vspace{-5pt}
\end{IEEEkeywords}

\section{Introduction}

% With $\mu$ (resp., $\nu$) denoting the fraction of computational power controlled by benign miners (resp., the adversary), where $\mu + \nu= 1$,
% we prove for the first time that to ensure the consistency
% property of Nakamoto's blockchain protocol in an asynchronous network, (slightly loosely speaking) it suffices to have $c$ denoting $\frac{1}{p n \Delta}$ to be just slightly greater than $\frac{2\mu}{\ln (\mu / \nu)}$ for most $\nu \in (0, \frac{1}{2})$. In the proof, we formulate novel Markov chains which result from the situation of mined blocks (e.g., no, one, or over one mined block) in different rounds. 

Nakamoto's blockchain protocol~\cite{nakamoto2008bitcoin} supports the Bitcoin application and relies on the proof of work (POW). POW means that to create a block, a player needs to  provide a solution of a cryptographic puzzle based on hash functions. Formal analyses of the protocol have received considerable interest recently~\cite{garay2015bitcoin,pass2017analysis,pass2017fruitchains,shi2018analysis}.

Garay, Kiayas and Leonardos~\cite{garay2015bitcoin} propose the first formal modeling for Nakamoto's blockchain protocol. They also identify conditions which enable   Nakamoto's protocol to achieve a common prefix-property, where honest players' blockchain views have a large
common prefix.

The model of~\cite{garay2015bitcoin} assumes a \textit{synchronous} network. Removing such a strong assumption, Pass, Seeman, and Shelat~\cite{pass2017analysis} consider an \textit{asynchronous} network by allowing the adversary to adaptively and individually delay messages up
to a delay limit $\Delta$. We refer to this as the \mbox{$\Delta$-delay} model.

One of the desired properties in a blockchain protocol is consistency. In this paper, we follow~\cite{pass2017analysis,kiffer2018better} to define consistency as the property that for any positive integer $T$,  with overwhelming probability in $T$, for any two rounds $r$ and $s$ with $r < s$, all but the last $T$ blocks in the chain of any honest player $i$ at round $r$ is a prefix of the chain of any honest player $j$ at round $s$. For an event to have an overwhelming probability in $T$, the probability of its complementary event should decay at least exponentially with respect to $T$. 

Consistency results of Nakamoto's blockchain protocol are typically expressed in a quantity $c$ defined as $\frac{1}{p n \Delta}$, where $p$ denotes the hardness of the proof of work, $n$ is the number of players, and $\Delta$ is the maximum delay of a message by the adversary (the notation will be summarized in Table~\ref{table-Notation} on Page~\pageref{table-Notation}). Roughly speaking, $c$ means the expected number of network delays before some block is mined. 
% More specifically, $c$ is defined as $\frac{1}{p n \Delta}$, where as we have explained:

In this paper, we present a result for the consistency
property of Nakamoto's blockchain protocol. Our consistency result is stronger than   existing ones in the literature (e.g., the result of~\cite{pass2017analysis}). Under the \mbox{$\Delta$-delay} model, with $\mu$ (resp., $\nu$) denoting the fraction of computational power controlled by benign miners (resp., the adversary), where $\mu+\nu=1$ and $0< \nu < \mu$, we show that it suffices to achieve consistency for $c$ denoting $\frac{1}{p n \Delta}$ to be just slightly greater than $\frac{2\mu}{\ln (\mu / \nu)}$. Our work is the first one in the literature to derive such a neat expression $\frac{2\mu}{\ln (\mu / \nu)}$. In Section~\ref{sec-related1},, we will explain the superiority of our consistency result over existing results.

\textbf{Contributions.} Our contributions are as follows:
\begin{itemize}
\item \textbf{\textit{(Major) Contribution~1 of proving Theorem~\ref{thm-bound-c2}:}} Our Theorem~\ref{thm-bound-c2} to be presented on Page~\pageref{thm-bound-c2} gives the following neat condition to ensure the consistency property of Nakamoto's blockchain protocol: $c$ denoting the expected number of network delays before some block is mined just needs to be slightly greater than $\frac{2\mu}{\ln (\mu/\nu)}$, where $\mu$ (resp., $\nu$) denotes the fraction of computational power controlled by benign miners (resp., the adversary).
\item \textbf{\textit{(Secondary) Contribution~2 of fixing~\cite{kiffer2018better} and  proving Theorem~\ref{thm-alpha}:}} We show an issue in the analysis of~\cite{kiffer2018better}: the probability that \textit{only one} honest miner succeeds in solving a puzzle is computed as the probability that \textit{at least one} honest miner succeeds in solving a puzzle in one round. Although~\cite{kiffer2018better} mentions ``a single honest mined block'', but its calculation actually uses ``at least one honest mined block''. After we fix the above issue and correct some minor notation typos of~\cite{kiffer2018better}, the result of~\cite{kiffer2018better} will become the same as our Theorem~\ref{thm-alpha} on Page~\pageref{thm-alpha} (We emphasize that Theorem~\ref{thm-alpha} is our secondary contribution while Theorem~\ref{thm-bound-c2} is our major contribution). Yet, in an effort to present an clearer explanation than that of~\cite{kiffer2018better}, we formulate two novel Markov chains to prove Theorem~\ref{thm-alpha}. With a state of a round characterizing the number of mined blocks (e.g., no, one, or over one mined block), our first  Markov chain models the transition of a variable denoting the suffix of the concatenation of the previous states and the current state. Our second Markov chain models the transition of a variable which denotes the concatenation of i) the suffix of previous states before the $\Delta$ to last state, ii) the previous $\Delta$ states, and iii) the current state. 
\end{itemize}

% We show an issue in the analysis of~\cite{kiffer2018better}: ?. In the presence of the above issue, their condition to ensure the consistency  of Nakamoto's blockchain protocol is that ?. After we fix the above issue and correct \mbox{$\mu \cdot p$} to $\alpha$ in many places of~\cite{kiffer2018better}, the result of~\cite{kiffer2018better} will become the same as our Theorem~\ref{thm-alpha} on Page~\pageref{thm-alpha} giving the following condition to ensure the consistency  of Nakamoto's blockchain protocol:   (We emphasize that Theorem~\ref{thm-alpha} is our secondary contribution while Theorem~\ref{thm-bound-c2} is our major contribution). which leads to the

\textbf{Organization of this paper.} In Section~\ref{sec-related}, we survey related studies. Section~\ref{sec-model} explains the model for Nakamoto's blockchain protocol. Section~\ref{sec-results-consistency} presents our results for the consistency
property of Nakamoto's blockchain protocol. In Sections~\ref{sec-Proof-Theorem-thm-bound-c2} and~\ref{sec-Proof-Theorem-thm-alpha}, we discuss the proofs of Theorems~\ref{thm-bound-c2} and~\ref{thm-alpha}, respectively. We conclude the paper in Section~\ref{sec-Conclusion}. Additional proof details are given in the Appendices of the online full version~\cite{fullpaper}.

\textbf{Notation.} Table~\ref{table-Notation} lists the notation and their meanings.

\begin{table}[!t]
\caption{Notation and their meanings.}
\label{table-Notation}
\setlength{\tabcolsep}{1pt} \renewcommand{\arraystretch}{1.15}
\begin{tabular}{l|l}
\hline
\hspace{-5pt}\textbf{Notation} & \textbf{Meanings} \\ \hline
$p$ & the hardness of the proof of work \\ \hline
$n$ & \begin{tabular}[c]{@{}l@{}}the number of miners (either honest or corrupted),\\[-1pt] each with identical computing power \end{tabular}  \\ \hline
$\Delta$ & the maximum delay of a message by the adversary  \\ \hline
$c$ & \begin{tabular}[c]{@{}l@{}} $c:=\frac{1}{p n \Delta}$. Roughly speaking, $c$ means the expected\\[-1pt] number of \mbox{$\Delta$-delays} before some block is mined.  \end{tabular}  \\ \hline
$\mu$ & \begin{tabular}[c]{@{}l@{}}the fraction of computational power controlled by\\[-1pt] benign miners (i.e., the fraction of benign miners)\end{tabular}    \\ \hline
$\nu$ & \begin{tabular}[c]{@{}l@{}}the fraction of computational power controlled by\\[-1pt] the adversary (i.e., the fraction of corrupted miners)\end{tabular}    \\ \hline
$\alpha$ & \begin{tabular}[c]{@{}l@{}} $\alpha$ denotes the probability that \textit{at least one} honest miner \\[-1pt] succeeds in solving a puzzle in one round.\\[-1pt] $\alpha = 1 - (1- p)^{\mu n}$. \end{tabular} \\ \hline
$\overline{\alpha}$ & \begin{tabular}[c]{@{}l@{}} $\overline{\alpha}$ denotes the probability that \textit{no} honest miner \\[-1pt] succeeds in solving a puzzle in one round.\\[-1pt] $\overline{\alpha} = (1- p)^{\mu n}$. \end{tabular} \\ \hline
$\alpha_1$ & \begin{tabular}[c]{@{}l@{}} $\alpha_1$ denotes the probability that \textit{only one} honest  \\[-1pt] miner succeeds in solving a puzzle in one round.\\[-1pt] $\alpha_1 = p \mu n  \times (1-p)^{\mu n - 1}$.\end{tabular} \\ \hline
$\beta$  & \begin{tabular}[c]{@{}l@{}}$\beta$ denotes the expected number of blocks mined  \\[-1pt] in each round by the adversary controlling  \\[-1pt] $\nu $ fraction of computational power. \\[-1pt] $\beta:=p \nu n$.  \end{tabular} \\  \hline 
\end{tabular}
\end{table}

\begin{figure}[!t]
 \centering
\includegraphics[scale=0.5]{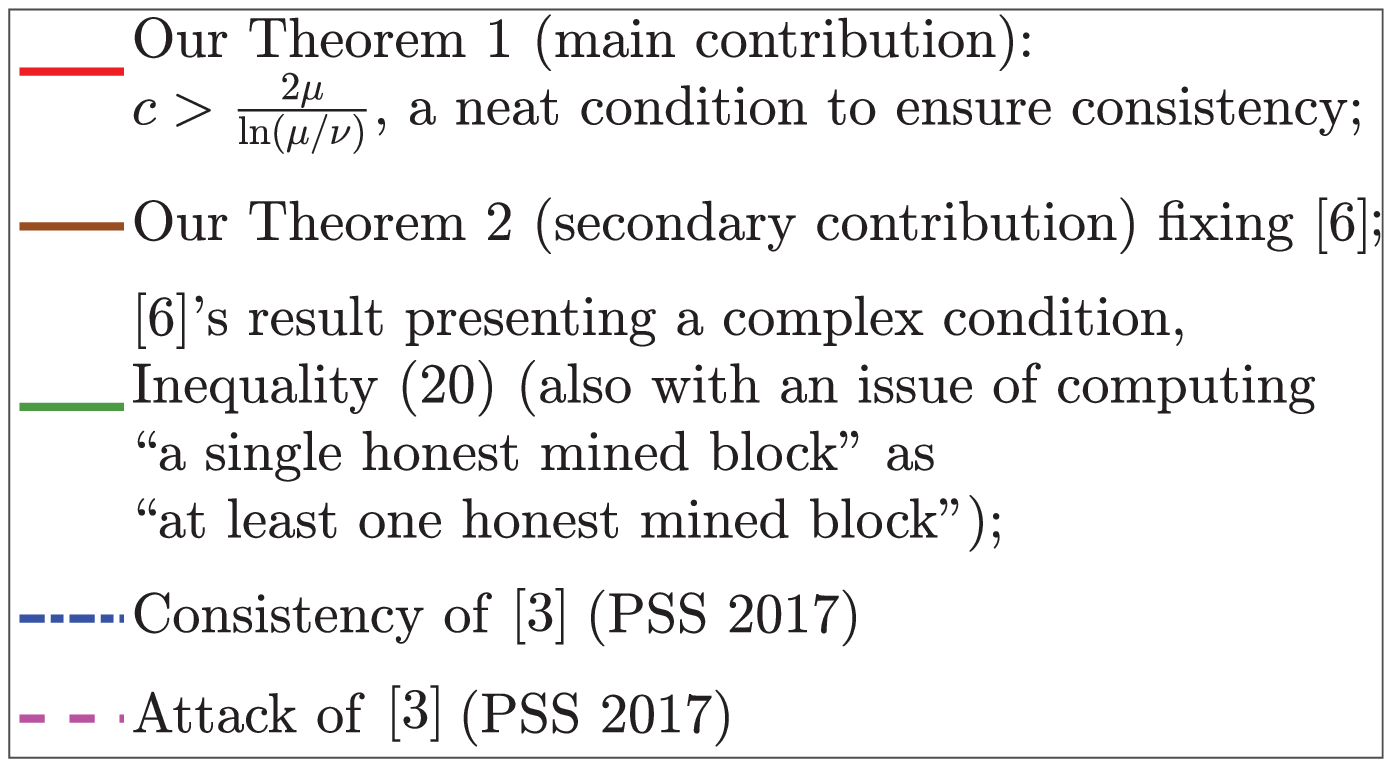} 
\includegraphics[scale=0.5]{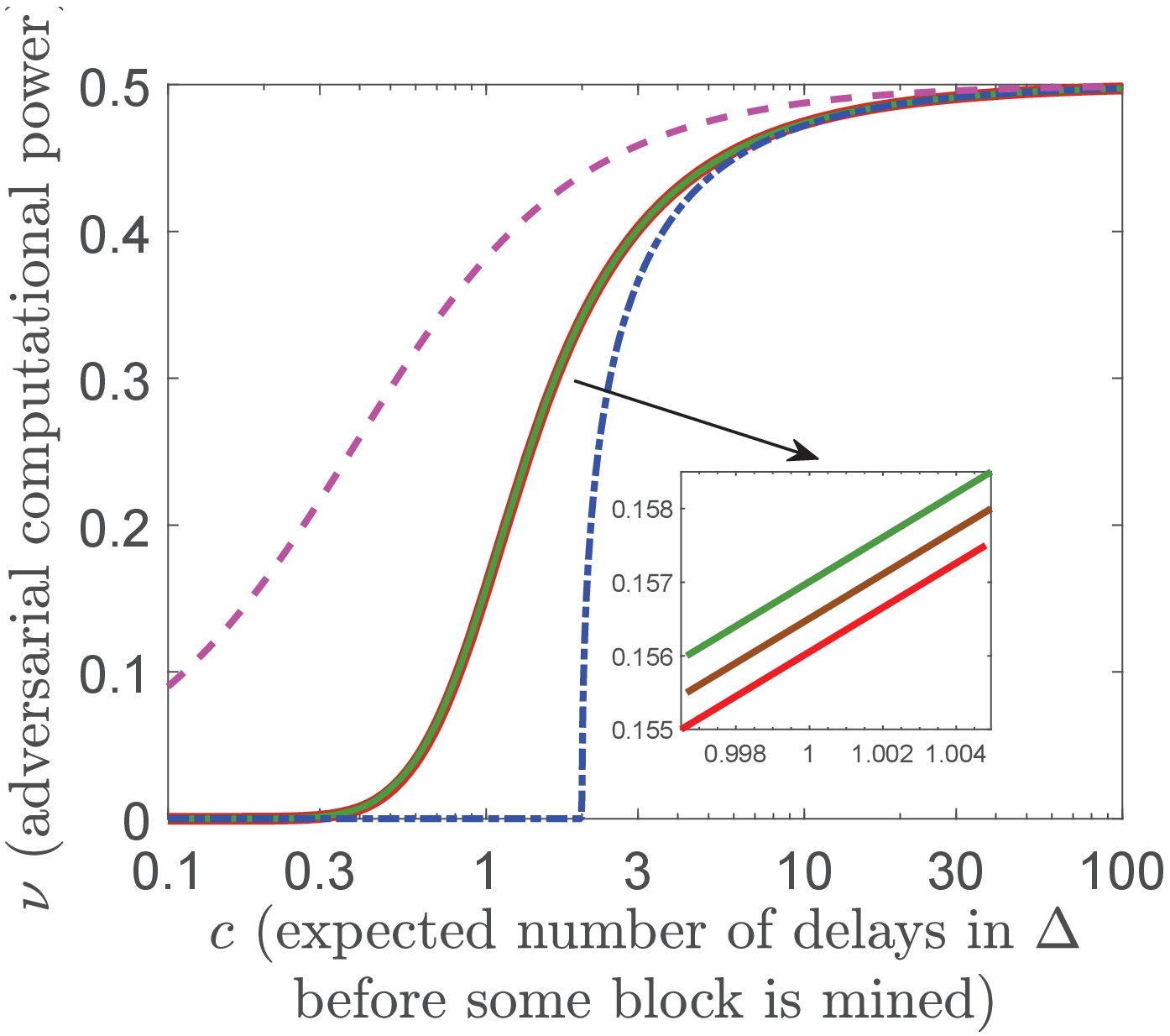} 
\caption{A comparison of our consistency result with consistency  of~\cite{kiffer2018better} by Kiffer, Rajaraman, and Shelat in ACM CCS 2018 as well as consistency and attack of~\cite{pass2017analysis} by Pass, Seeman, and Shelat (PSS) in Eurocrypt 2017. We adopt $n=10^5$ and $\Delta=10^{13}$ from Figure~1 of~\cite{pass2017analysis}. $c$ denoting $\frac{1}{p n \Delta}$ roughly means the expected number of network delays before some block is mined. See Table~\ref{table-Notation} on the left-hand column for the meanings of the notation. \vspace{-20pt}} \label{figcompre}
\end{figure}

\section{Related Work} \label{sec-related}

This section is organized as follows. In Section~\ref{sec-related1}, we elaborate the comparison between our results and related ones, where Figure~\ref{figcompre} is plotted. Section~\ref{sec-relatedother} presents additional related studies.

\subsection{Comparing our results and related ones} \label{sec-related1}

We compare our consistency results with~\cite{pass2017analysis,kiffer2018better} and use Figure~\ref{figcompre}  to illustrate the comparison. Our   Figure~\ref{figcompre} adopts $n=10^5$ and $\Delta=10^{13}$ from Figure 1 of~\cite{pass2017analysis}. In Figure~\ref{figcompre}, all lines except the magenta line illustrate conditions used in different results to ensure the consistency
property of Nakamoto's blockchain protocol. In particular, these red, brown, green, and blue lines   plot the allowed maximum (or the limit superior) value for the fraction $\nu$ of computational power controlled by the adversary  with respect to $c$, the expected number of network delays before some block is mined, in order to not break consistency according to the respective results. More details are as follows. % to show the most extreme cases that preserve consistency according to the respective results, 

The red, brown, and green lines almost overlap in Figure~\ref{figcompre}. Hence, in the lower right corner of Figure~\ref{figcompre}, we also zoom some parts to show the (negligible) separation between the lines. The red line shows a neat condition on $c$ to ensure consistency: $c > \frac{2\mu}{\ln (\mu / \nu)} $, given by our Theorem~\ref{thm-bound-c2} (our main contribution) to be presented on Page~\pageref{thm-bound-c2}. The brown line is from Theorem~\ref{thm-alpha} (our secondary contribution) on Page~\pageref{thm-alpha}. The green line shows Claim~\ref{thm-kiffer2018better-Theorem-4.4a} on Page~\pageref{thm-alpha}, which is Theorem~4.4 on Page~8 of~\cite{kiffer2018better} after we correct \mbox{$\mu \cdot p$} to $\alpha$ in many places of~\cite{kiffer2018better} and perform some computations presented in Appendix~\ref{app-thm-kiffer2018better-Theorem-4.4a} of the online full version~\cite{fullpaper} (see Table~\ref{table-Notation} for the notation's meanings).
%  For simplicity of explanation, we just say Claim~\ref{thm-kiffer2018better-Theorem-4.4a} as~\cite{kiffer2018better}'s result below.
 Yet,~\cite{kiffer2018better}'s result as well as its induced Claim~\ref{thm-kiffer2018better-Theorem-4.4a} has a minor issue that the probability that \textit{only one} honest miner succeeds in solving a puzzle is computed as the probability that \textit{at least one} honest miner succeeds in solving a puzzle in one round, so we fix it to obtain Theorem~\ref{thm-alpha} (our effort of obtaining Theorem~\ref{thm-alpha} to fix~\cite{kiffer2018better} are based on new Markov chains, to present a more detailed explanation than~\cite{kiffer2018better}). In Claim~\ref{thm-kiffer2018better-Theorem-4.4a} originating from \cite{kiffer2018better}'s result, Inequality~(\ref{eq-thm-alpha-delta_3a-complex}) as the condition on $c$ for consistency is quite complex. In our Theorem~\ref{thm-alpha} fixing \cite{kiffer2018better}, the condition on $c$ for consistency is also quite complex after the expressions of $\alpha$ and $\alpha_1$  are plugged in. In contrast,  Theorem~\ref{thm-bound-c2} as our main contribution presents the neat condition $c > \frac{2\mu}{\ln (\mu / \nu)} $ to ensure consistency. Since we obtain Theorem~\ref{thm-bound-c2} based on Theorem~\ref{thm-alpha}, the neat condition of Theorem~\ref{thm-bound-c2} is sufficient but not necessary to get the condition of Theorem~\ref{thm-alpha}. Yet, these two conditions are almost the same since the red and brown lines almost overlap in Figure~\ref{figcompre}. This shows that we almost do not lose any  tightness of the result in the move from Theorem~\ref{thm-alpha} to Theorem~\ref{thm-bound-c2} for seeking a neater condition.

Since the red, brown, and green lines almost overlap in Figure~\ref{figcompre}, we now focus on the red, blue, and magenta lines. As stated, the red line of Figure~\ref{figcompre} shows our consistency result in Theorem~\ref{thm-bound-c2} on Page~\pageref{thm-bound-c2}. From the condition $c > \frac{2\mu}{\ln (\mu / \nu)} = \frac{2(1-\nu)}{\ln \frac{1-\nu}{\nu} }$,   our maximal $\nu_{\max}$ can be solved numerically given $c$ (strictly speaking, $\nu_{\max}$ cannot be achieved due to the strict inequality sign). This gives the red line. 

The blue line of Figure~\ref{figcompre} is from the consistency analysis of~\cite{pass2017analysis}. The consistency condition of~\cite{pass2017analysis} is $\alpha[1-(2\Delta+2)\alpha]>\beta$, where $\alpha := 1 - (1-p)^{\mu n}$ and $\beta := \nu n p$. Roughly speaking, $\alpha \approx \mu n p$ and $2\Delta+2\approx 2\Delta $, so $\alpha[1-(2\Delta+2)\alpha]>\beta$ is approximately $1-2\Delta \mu n p > \frac{\nu}{1-\nu}$, where we note $\mu = 1- \nu$. Then we further obtain   $p< \frac{1-2\nu}{ 2(1-\nu)^2\Delta n }   $ and hence $c:=\frac{1}{p n \Delta} > \frac{ 2(1-\nu)^2}{1-2\nu}$. This implies \mbox{$\nu < \frac{1}{2} (2-c + \sqrt{c^2-2c}) $}, where $c > 2$. The blue line of Figure~\ref{figcompre} shows this. 

The magenta line of Figure~\ref{figcompre} illustrates an attack of~\cite{pass2017analysis} which breaks consistency. Remark~8.5 of~\cite{pass2017analysis} presents an attack  which works when $\frac{1}{c} > \frac{1}{\nu} - \frac{1}{1-\nu}$. This inequality means $\nu > \frac{2c+1 - \sqrt{4c^2+1}}{2} $.

From Figure~\ref{figcompre}, the red line illustrating our consistency result is strictly above the blue line for consistency of~\cite{pass2017analysis}. Hence, our consistency result is much stronger than that of~\cite{pass2017analysis} in the sense that our result tolerates much more fraction of adversarial computational power. A future direction is to see whether it is possible to reduce the gap between the red line for our consistency result and the magenta line representing an attack on consistency from~\cite{pass2017analysis}.

% We formulate two Markov chains, called the suffix-of-previous-and-current-states Markov chain $\mathcal{C}_{\boldsymbol{F}}$ and the 
% suffix-of-far-previous-states-and-previous-states Markov chain $\mathcal{C}_{\boldsymbol{F}||\boldsymbol{P}}$. As the name suggests, 

% The analysis and algorithms for solving the problem are elaborated in Section~\ref{sec-Analysis-Algorithms}. Section~\ref{sec-Simulation} gives simulation results. In Section~\ref{sec-Discussion-Future}, we discuss limitations of the paper identify several directions for future research.
%   Finally, Section~\ref{sec-Conclusion} concludes the paper.

\subsection{Additional related work} \label{sec-relatedother}

% Blockchain were originally designed to be public systems, where every computing node can join.

The essence of blockchain is a consensus protocol to achieve agreement among distributed nodes.  The seminal blockchain protocol by Nakamoto~\cite{nakamoto2008bitcoin} leads to the popular application of Bitcoin. Bitcoin is a cryptocurrency whose ledger is maintained by the public instead of trusted authorities.
% That is so-called decentralization. 

Nakamoto's blockchain protocol is built on the proof of work (POW)~\cite{nakamoto2008bitcoin}. When a node creates a block, the node should provide a solution of a cryptographic puzzle based on hash functions. Every node  maintains its own chain and accepts the longest chain of the ones it receives from the network.

Recently, formal analyses of blockchain protocols have received considerable attention~\cite{garay2015bitcoin,pass2017analysis,pass2017fruitchains,shi2018analysis}. Three commonly analyzed properties are consistency, chain growth, and chain quality.

In~\cite{nakamoto2008bitcoin,garay2015bitcoin}, \textit{consistency} is defined as the property that  with overwhelming probability in $T$, at any round, the chains of two honest
players can differ only in the last $T$ blocks. Pass, Seeman, and Shelat~\cite{pass2017analysis} identify that this definition is not sufficient for consensus, since it does not exclude a protocol which oscillates between  
different chains. Hence, they require an additional property, referred to as \textit{future self-consistence}: with overwhelming probability in $T$, at any two rounds $r$ and $s$, the chains
of any honest player at $r$ and $s$ differs only in blocks within the last $T$ blocks. The consistency notion used in~\cite{kiffer2018better} and our current paper combines the consistency definition of~\cite{nakamoto2008bitcoin,garay2015bitcoin} and future self-consistence of~\cite{pass2017analysis}. Specifically, by consistency, we mean that with overwhelming probability in $T$, for any two rounds $r$ and $s$ with $r < s$, all but the last $T$ blocks in the chain of any honest player $i$ at round $r$ is a prefix of the chain of any honest player $j$ at round $s$.

In addition to consistency analyzed by~\cite{nakamoto2008bitcoin,garay2015bitcoin,pass2017analysis,kiffer2018better}, chain growth and chain quality for Nakamoto's blockchain protocol are also studied in the literature~\cite{pass2017analysis,pass2017fruitchains,kiayias2015speed,garay2017bitcoin,zhang2019lay}. The chain growth is at least $g$ if with overwhelming probability in $T$, the
chain of honest players grew by at least $T$ blocks in the last $T/g$ rounds. The chain quality is at least $q$ if with overwhelming probability in $T$, for any $T$ consecutive blocks in
any chain held by some honest player, the fraction of blocks contributed by
honest players is at least $q$. In this paper, we analyze only consistency. A future direction is to investigate how to use our proof methods for the analyses of chain growth and chain quality.

% In order to achieve consensus among the nodes, there cannot be many nodes who create a block at the same time. Therefore, the difficulty of the puzzle is somewhat high, which leads to low efficiency and high consumption of energy. 

After POW, blockchain protocols based on an alternative paradigm called  the Proof of Stake (POS) have also been proposed~\cite{kiayias2017ouroboros,david2018ouroboros,badertscher2018ouroboros,li2017securing}. POS typically consumes less computation power than POW. The ingenious Algorand protocol~\cite{gilad2017algorand} combines POS and the classical practical Byzantine fault tolerance (PBFT) protocol of~\cite{castro1999practical}. We refer interested readers to recent surveys~\cite{wang2019survey,li2017survey} for more details of POW, POS, and other types of blockchain protocols.

% Although GHOST (used in Ethereum), an example of POW besides Bitcoin, reduces the energy consumption to some degree, the nodes still need high computation power.

\section{The Model for Nakamoto's Blockchain Protocol} \label{sec-model}

As in many blockchain studies, we adopt the formalization of Garay, Kiayas and Leonardos~\cite{garay2015bitcoin} and Pass, Seeman, and Shelat~\cite{pass2017analysis} for Nakamoto's blockchain protocol. We will mostly follow the notation of~\cite{kiffer2018better}, which presents a clear explanation of the formalization.

% \junzhao{More details here}

A blockchain is a pair of algorithms $(\Pi, \verb|ext|)$. The stateful
algorithm $\Pi$ maintains a local state variable $\mathcal{C}$ and also receives a
security parameter $\kappa$ as an input. The variable $\mathcal{C}$ is commonly referred to as the chain, since it contains a set of blocks. A block is an abstract record containing a message. The algorithm $\verb|ext|(\kappa,\mathcal{C})$ outputs an ordered sequence of messages.

The execution of a blockchain protocol $(\Pi, \verb|ext|)$ is directed by an environment $Z(1^{\kappa})$. It activates each of $n$ players as either \textit{honest} or \textit{corrupt}. For simplicity, all $n$ players are assumed to have identical computing power. Each honest player has a current view of the blockchain and aims to build blocks at the end of the chain. Each corrupted player is controlled by an adversary $\mathcal{A}$. We assume that at any point, $\mathcal{A}$ can corrupt an honest party or uncorrupt a corrupted player, but the fraction of corrupted players is at most $\nu$. For ease of analysis, we can just consider the worst case where $\mathcal{A}$ controls $\nu$ fraction of corrupted players at each round.

We consider the network to be asynchronous, and allow the adversary $\mathcal{A}$ to have the following capabilities: 
\begin{myitemize}
\item[\ding{172}] $\mathcal{A}$ can delay and/or reorder all messages up to a delay of~$\Delta$ rounds, but $\mathcal{A}$ cannot modify messages sent by honest players.
\item[\ding{173}] $\mathcal{A}$ fully controls all corrupted players; i.e., $\mathcal{A}$ reads all their inputs/messages and sets their outputs/messages to be sent.
\end{myitemize} 
Strategies taken by the adversary $\mathcal{A}$ can be letting all corrupted players work on the same block or different ones. 

All players have access to a random function \mbox{$H:\{0,1\}^* \to \{0,1\}^{\kappa}$}  through the following two oracles. First, $\verb|H|(x)$ simply outputs $H(x)$. Second, the verification oracle $\verb|H|.\verb|ver|(x,y)$ outputs $1$ if and only if $H(x)=y$ and $0$ otherwise. How $\verb|H|$ and $\verb|H|.\verb|ver|$ can be accessed is specified as follows:
\begin{myitemize}
\item In each round, the players, as well as the adversary $\mathcal{A}$, make \textit{any} number of queries to $\verb|H|.\verb|ver|$.
\item In each round, each honest player can make  only a single query $\verb|H|$ and the queries made by honest players are \textit{parallel} so that even if they manage to mine several blocks, their longest chain can increase by at most $1$. In contrast, the adversary $\mathcal{A}$ controlling $q$ players can make $q$ \textit{sequential} queries to $\verb|H|$. 
%  , so that the adversary may extend the chain by multiple blocks in one round.
\end{myitemize} 
The above model captures that we account for  only the effort of
\textit{finding} a solution to a ``proof of work'', and consider that \textit{checking} the validity of a solution is negligible. A ``proof of work'' given the block $h_{-1}$ and message $m$ is to find a string $\eta$ such that $\verb|H|(h_{-1},\eta,m) \leq D_p$, where the blockchain protocol sets $D_p$ such that the probability of finding $\eta$ to satisfy the above relation is $p$. This quantity $p$ is referred to as the hardness of the proof of work.

Given the above, we now describe an execution of a blockchain protocol. At the beginning, the environment $Z(1^{\kappa})$ instantiate $n$ players, which have identical computing power. The protocol proceeds in rounds  as follows. At each round, each player $i$ does the following:
\begin{myitemize}
\item $i$ receives blocks created by other players and includes the blocks in its chain based on the protocol $\Pi$;
\item $i$ can make at most one query to the oracle $\verb|H|$ and creates a block with probability $p$; and
\item $i$ receives some message from $Z(1^{\kappa})$ and includes the message in the block that $i$ tries to publish, where the message contains transactions to be included
in the blockchain. 
\end{myitemize}

As already noted, $\nu$ denotes the fraction of corrupted players controlled by the adversary. With $\mu$ being the fraction of honest players, we have 
\begin{align}
\mu + \nu= 1 . \label{eq-mu-nu-sum-to-1}
\end{align}
Throughout the paper, we enforce 
\begin{align}
0 < \nu < \frac{1}{2} < \mu, \label{ineq-nu-mu}
\end{align}
and the trivial condition
\begin{align}
n \geq 4. \label{ineq-n-geq-4}
\end{align} 
From Eq.~(\ref{eq-mu-nu-sum-to-1}), Inequality~(\ref{ineq-nu-mu}) simply means the following two conditions:
\begin{myitemize}
\item[i)]  the fraction of computational power controlled by benign miners is greater than that controlled by the adversary; and 
\item[ii)]  the adversary controls non-zero fraction of computational power.
\end{myitemize}

With $n, p, \mu$, and $\nu$ introduced above, we now define $\alpha$, $\overline{\alpha}$, and $\alpha_1$, which will be used in our theorems to be presented in Section~\ref{sec-results-consistency}. All these notation are given in Table~\ref{table-Notation} on Page~\pageref{table-Notation}. The meanings of $\alpha$, $\overline{\alpha}$, and $\alpha_1$ are as follows: 
% In our theorems to be presented in Section~\ref{sec-results-consistency}, we will use $\alpha$, $\overline{\alpha}$, and $\alpha_1$, which are defined as follows:
\begin{align}
\alpha :~& \text{the probability that \textit{at least one} honest miner}, \nonumber \\ & \text{succeeds in solving a puzzle in one round}, \label{eq-alpha} \\ \overline{\alpha} :~& \text{the probability that \textit{no} honest miner}, \nonumber \\ & \text{succeeds in solving a puzzle in one round}, \label{eq-overline-alpha}   \\ \alpha_1  :~& \text{the probability that \textit{only one} honest miner}, \nonumber \\ & \text{succeeds in solving a puzzle in one round}. \label{eq-overline-alpha1}
\end{align}
Next, we derive the expressions of $\alpha$, $\overline{\alpha}$, and $\alpha_1$.
% Let $\alpha$ be the probability that \textit{some} honest miner succeeds in solving a puzzle in one round. Let $\overline{\alpha}$ be the probability that \textit{no} honest miner succeeds in solving a puzzle in one round.
 Since each honest node mines a block independently with probability $p$ in a round, $X$ denoting the number of blocks mined by the $\mu n$ honest nodes in each round follows $\text{binom}( \mu n, p)$, which denotes a binomial distribution with $ \mu n$ being the number of trials and $p$ being the success probability for each trial. Hence, we have
\begin{align}
\alpha & = \bp{X>0} = 1 - (1- p)^{\mu n}, \label{eq-define-alpha} \\ \overline{\alpha} &  = \bp{X=0} = 1 - \alpha =   (1- p)^{\mu n}, \label{eq-define-alpha-bar} \\ \alpha_1 & = \bp{X=1}  = p \mu n  \times (1-p)^{\mu n - 1}. \label{eq-define-alpha-1}
\end{align}
Let $\beta$ be the expected number of blocks mined in each round by the adversary controlling $\nu n$ miners. Then it holds that
\begin{align}
\beta =p \nu n. 
\end{align} 

\section{Our Results for the Consistency
Property of Nakamoto's Blockchain Protocol} \label{sec-results-consistency}

Our results for the consistency
property of Nakamoto's blockchain protocol are presented as Theorems~\ref{thm-bound-c2} and~\ref{thm-alpha} below.

From~\cite{pass2017analysis,kiffer2018better}, blockchain consistency is defined as follows.

% We will show the consistency of Nakamoto's blockchain protocol by proving that in a window of $T$ slots, there is at least \mbox{$1-O(1)\cdot\exp\left(-\Omega\left(T\right)\right)$}  probability for the event that the number of convergence opportunities is greater than the number of blocks mined by the adversary.

\begin{defn}[Blockchain consistency] \label{def-Blockchain-consistency}
{Nakamoto's blockchain protocol satisfies consistency} if
for any positive integer $T$, with at least \mbox{$1-O(1)\cdot\exp\left(-\Omega\left(T\right)\right)$}  probability, for any two rounds $r$ and $s$ with $r < s$, all but the last $T$ blocks in the chain of any honest player $i$ at round $r$ is a prefix of the chain of any honest player $j$ at round $s$.   
\end{defn}

The asymptotic notation in this paper such as $O \left(\cdot\right)$ and $\Omega \left(\cdot\right)$ is standard\footnote{Given two positive
sequences $f_T$ and $g_T$ indexed by $T$, we have
\begin{myitemize}
    \item $f_T = O \left(g_T\right)$ means that there exist positive
  constants $c_1$ and $T_1$ such that $f_T \leq c_1 g_T$ for all $t \geq
  T_1$.
\item $f_T = \Omega \left(g_T\right)$ means that there exist positive
  constants $c_2$ and $T_2$ such that $f_T \geq c_2 g_T$ for all $t \geq
  T_2$. %Namely, $f_T = \Omega \left(g_T\right)$ holds if and only if
  %$g_T = O \left(f_T\right)$ holds. %
%    \item $f(n)\lesssim g(n)$ means that there exists a function $z(n)$ such that $f(n) \leq z(n)$ for any sufficiently large $n$ and $g(n)\sim z(n)$ both hold;
%    i.e., $f(n)\lesssim g(n)$ means $f(n) \leq g(n) \cdot [1+o(1)]$ for any sufficiently large $n$.\\
\end{myitemize} \label{ft1}
}; see Footnote~\ref{ft1}.
The term\footnote{Actually \mbox{$1-O(1)\cdot\exp\left(-\Omega\left(T\right)\right)$} can be simplified as $1-\exp\left(-\Omega\left(T\right)\right)$ since $O(1)\cdot\exp\left(-\Omega\left(T\right)\right) = \exp\left(\ln O(1)-\Omega\left(T\right)\right)$ and $\ln O(1)-\Omega\left(T\right)$ can also be written $ -\Omega\left(T\right)$.} \mbox{$O(1)\cdot\exp\left(-\Omega\left(T\right)\right)$}  above decays at least exponentially with respect to $T$. 
% An intuitive understanding of the above consistency notion is that it implies that
 Intuitively, the above consistency notion implies that
there is at least \mbox{$1-O(1)\cdot\exp\left(-\Omega\left(T\right)\right)$}  probability for the event that
honest players agree on the current chain, except
for $T$ ``unconfirmed'' blocks at the end of the chain.

Based on Definition~\ref{def-Blockchain-consistency}, Lemma~\ref{def-Blockchain-consistency} below presents a sufficient condition for consistency which we will use to prove our theorems.

\begin{lem}[Blockchain consistency] \label{lem-Blockchain-consistency}
{Nakamoto's blockchain protocol satisfies consistency} if
for any positive integer $T$, in a window of $T$ slots, there is at least \mbox{$1-O(1)\cdot\exp\left(-\Omega\left(T\right)\right)$}  probability for the event that the number of convergence opportunities is greater than the number of blocks mined by the adversary, where a \underline{\textbf{convergence opportunity}} is an event which results in all honest players to agree on a single longest chain.\vspace{-2pt} 
\end{lem}

% The intuition of Lemma~\ref{lem-Blockchain-consistency} is as follows.

Our main contribution on the consistency of Nakamoto's blockchain protocol is given as Theorem~\ref{thm-bound-c2} below.\vspace{-2pt} 

% \begin{defn}
% In this paper, we say that
% \begin{quote}
% \mbox{\textbf{Nakamoto's blockchain protocol satisfies consistency}}    
% \end{quote} 
% if for any positive integer $T$, in a window of $T$ slots, there is at least \mbox{$1-O(1)\cdot\exp\left(-\Omega\left(T\right)\right)$}  probability for the event that the number of convergence opportunities is greater than the number of blocks mined by the adversary.   
% \end{defn}

% \begin{defn}
% In this paper, we say that
% \begin{quote}
% \mbox{\textbf{Nakamoto's blockchain protocol satisfies consistency}}    
% \end{quote} if 
%   for any positive integer $T$, in a window of $T$ slots, there is at least \mbox{$1-O(1)\cdot\exp\left(-\Omega\left(T\right)\right)$}  probability for the event that the number of convergence opportunities is greater than the number of blocks mined by the adversary.   
% \end{defn}

% how does it compare with related work?

\begin{thm} \label{thm-bound-c2} %\label{thm-bound-c2-condition-thm-Delta}
% If
% \begin{align}
% \text{$ \Delta > \frac{1+ \ln \frac{\mu}{\nu}}{4 \epsilon_2_1}   $ for a constant $\epsilon_2_1$ satisfying $0 < \epsilon_2_1 \iffalse_0 \fi  < 1$,}  \label{eq-condition-thm-Delta}
% \end{align}
% where we will explain in ? later that in ?, $\epsilon_2_1$ of ? can be very small (e.g.,
Nakamoto's blockchain protocol satisfies consistency when there exist constants $\epsilon_1$ and $\epsilon_2 $  satisfying \mbox{$0 < \epsilon_1 \iffalse_0 \fi  < 1$} and $\epsilon_2 > 0$ such that $c$ denoting $\frac{1}{p n \Delta}$ satisfies
\begin{align}
c  \geq \max\left\{\left( \frac{2\mu}{\ln \frac{\mu}{\nu}} +  \frac{1}{\Delta}\right)    \frac{1 + \epsilon_2 }{1-\epsilon_1} ,~ \frac{(\ln \frac{\mu}{\nu}+1)\mu}{\epsilon_1 \Delta \ln \frac{\mu}{\nu} } \right\}  .  \label{eq-thm2-c-v1-condition-thm-Delta}
\end{align}
To better understand Inequality~(\ref{eq-thm2-c-v1-condition-thm-Delta}), we present the following result, which we will use in Remark~\ref{thm-alpha-rem1} to show that Inequality~(\ref{eq-thm2-c-v1-condition-thm-Delta}) specifies $c$ to be just slightly greater than $\frac{2\mu}{\ln (\mu / \nu)}$.

If there exist positive constants $\delta_1$ and $\delta_2$ satisfying \mbox{$\delta_1+\delta_2<1$}  such that 
\begin{align}
\frac{1}{1+\exp(\Delta^{\delta_1})}  \leq  \nu \leq \frac{1}{1+\exp\left(\frac{1}{\Delta^{\delta_2}-1}\right)}, \label{nu-delta-1-delta-2}
\end{align}
we can write Inequality~(\ref{eq-thm2-c-v1-condition-thm-Delta}) as
\begin{align}
c  \geq \frac{2\mu}{\ln (\mu / \nu)} \cdot \left(1 + \epsilon_2  \right)\cdot  \frac{1+\Delta^{\delta_1-1}}{1-\Delta^{\delta_1+\delta_2-1}} .  \label{c-mu-nu}
\end{align} 
In Remark~\ref{thm-alpha-rem1}, we will explain that under Inequality~(\ref{nu-delta-1-delta-2}), the condition on $c$ as Inequality~(\ref{c-mu-nu}) enforces
\begin{align}
\textup{$c$ to be just slightly greater than $\frac{2\mu}{\ln (\mu / \nu)}$} . \nonumber
\end{align} 
\end{thm}
% We present two special cases \ding{172} and \ding{173} as follows, where the fraction $\nu$ of nodes controlled by the adversary is non-zero or zero, respectively. In Remark~\ref{thm-alpha-rem1}, we will use Result~\ding{172} to show that Inequality~(\ref{eq-thm2-c-v1-condition-thm-Delta}) means that $c$ just needs to be slightly greater than $\frac{2\mu}{\ln (\mu / \nu)}$ for most $\nu \in (0, \frac{1}{2})$.
% \begin{myitemize}
% \item[\ding{172}] \textbf{Case of $\nu \in (0, \frac{1}{2})$:} Consistency of Nakamoto's blockchain protocol holds in a window of $T$ rounds with probability at least \mbox{$1-O(1)\cdot\exp\left(-\Omega\left(T\right)\right)$} if there exist positive constants $\delta_1$ and $\delta_2$ satisfying $\delta_1+\delta_2<1$ such that 
% \begin{align}
% \frac{1}{1+\exp(\Delta^{\delta_1})} < \nu < \frac{1}{1+\exp\left(\frac{1}{\Delta^{\delta_2}-1}\right)}, \label{nu-delta-1-delta-2}
% \end{align}
% and
% \begin{align}
% c  \geq \frac{2\mu}{\ln (\mu / \nu)} \cdot \left(1 + \epsilon_2  \right)\cdot  \frac{1+\Delta^{\delta_1-1}}{1-\Delta^{\delta_1+\delta_2-1}} .  \label{c-mu-nu}
% \end{align}
% \item[\ding{173}] \textbf{Case of zero $\nu$:} If $\mu=1$ and $\nu = 0$, consistency of Nakamoto's blockchain protocol holds in a window of $T$ rounds with probability at least \mbox{$1-O(1)\cdot\exp\left(-\Omega\left(T\right)\right)$}
% \end{myitemize}

% \begin{rem}
% Our result applies to any $\nu \in (0, \frac{1}{2})$, since .
% \end{rem}

\begin{rem} \label{thm-alpha-rem1} We now explain that Inequality~(\ref{c-mu-nu}) enforces $c$ to be just slightly greater than $\frac{2\mu}{\ln (\mu / \nu)}$ for $\nu$ satisfying Inequality~(\ref{nu-delta-1-delta-2}), which will be shown to cover almost all $\nu \in (0, \frac{1}{2})$. Here we consider $\Delta = 10^{13}$ which is used in Figure 1 of Pass~\emph{et~al.}~\cite{pass2017analysis}, a seminal work on the consistency
property of Nakamoto's blockchain protocol, but our discussions readily apply to other values of $\Delta$. We consider two sets of $\delta_1$ and $\delta_2$ values which cover slightly different ranges of $\nu$.\vspace{-2pt} 
\begin{myitemize}
\item For $\Delta = 10^{13}$ of~\cite{pass2017analysis}, we let $\delta_1 = \frac{1}{6} $ and $\delta_2  = \frac{1}{2}$ so that Inequalities~(\ref{nu-delta-1-delta-2}) and~(\ref{c-mu-nu}) become %10^(13/6)/(ln 10)
\begin{align}
10^{-63}  \leq  \nu  \leq  0.5 - 10^{-7} , \label{nu-delta-1-delta-2-case1}
\end{align}
and
\begin{align}
c  \geq \frac{2\mu}{\ln (\mu / \nu)} \cdot \left(1 + \epsilon_2  \right)\cdot \left(1 + 5 \times 10^{-5} \right).  \label{c-mu-nu-case1}
\end{align}
Inequalities~(\ref{nu-delta-1-delta-2-case1}) and~(\ref{c-mu-nu-case1}) mean that $c$ just needs to be slightly greater than $\frac{2\mu}{\ln (\mu / \nu)}$ for $10^{-63}  \leq  \nu  \leq  0.5 - 10^{-7}$, since the positive  constant $\epsilon_2$ in Inequality~(\ref{c-mu-nu-case1}) can be arbitrarily small.
\item Inequality~(\ref{nu-delta-1-delta-2-case1}) in the above case considers $10^{-63}  \leq  \nu  \leq  0.5 - 10^{-7}$. Below we increase the upper bound  for $\nu$ from $0.5 - 10^{-7}$ in Inequality~(\ref{nu-delta-1-delta-2-case1}) to $0.5 - 10^{-9}$ in Inequality~(\ref{nu-delta-1-delta-2-case2}) by increasing $\delta_2$ from $\frac{1}{2}$ above to $\frac{2}{3}$ here. After increasing $\delta_2$, to ensure that the term $\frac{1+\Delta^{\delta_1-1}}{1-\Delta^{\delta_1+\delta_2-1}}$ in Inequality~(\ref{c-mu-nu}) is still just slightly greater than $1$, we slightly decrease $\delta_1$ from $\frac{1}{6} $ above to $\frac{1}{8} $\vspace{1.5pt} here, which increases the lower bound for $\nu$ from $10^{-63}$ in Inequality~(\ref{nu-delta-1-delta-2-case1})  to $10^{-18}$ in Inequality~(\ref{nu-delta-1-delta-2-case2}). Specifically, for $\Delta = 10^{13}$ of~\cite{pass2017analysis}, we let $\delta_1 = \frac{1}{8} $ and $\delta_2  = \frac{2}{3}$ so that Inequalities~(\ref{nu-delta-1-delta-2}) and~(\ref{c-mu-nu}) become %10^(13/8)/(ln 10)
\begin{align}
10^{-18}  \leq  \nu  \leq  0.5 - 10^{-9}, \label{nu-delta-1-delta-2-case2}
\end{align}
and
\begin{align}
c  \geq \frac{2\mu}{\ln (\mu / \nu)} \cdot \left(1 + \epsilon_2  \right)\cdot \left(1 + 2 \times 10^{-3} \right).   \label{c-mu-nu-case2}
\end{align} %(1+10^(-13*7/8))/((1-10^(-13*5/24))
Inequalities~(\ref{nu-delta-1-delta-2-case2}) and~(\ref{c-mu-nu-case2}) mean that $c$ just needs to be slightly greater than $\frac{2\mu}{\ln (\mu / \nu)}$ for $10^{-18}  \leq  \nu  \leq  0.5 - 10^{-9}$, since the positive  constant $\epsilon_2$ in Inequality~(\ref{c-mu-nu-case1}) can be arbitrarily small.
\end{myitemize}

% Theorems~\ref{thm-bound-c2} and~\ref{thm-alpha} are proved in Sections~\ref{sec-Proof-Theorem-thm-bound-c2} and~\ref{sec-Proof-Theorem-thm-alpha}, respectively. Below, we discuss the novelties of Theorems~\ref{thm-bound-c2} and~\ref{thm-alpha}.

The proof of Theorem~\ref{thm-bound-c2} will be explained in Section~\ref{sec-Proof-Theorem-thm-bound-c2}. Below, we discuss the novelty of Theorem~\ref{thm-bound-c2}.

\noindent \textbf{Novelty of our Theorem~\ref{thm-bound-c2}.} The analysis and results of our Theorem~\ref{thm-bound-c2} are both novel. Moreover, with Inequality~(\ref{nu-delta-1-delta-2-case1}) considering $10^{-63}  \leq  \nu  \leq  0.5 - 10^{-7}$ and Inequality~(\ref{nu-delta-1-delta-2-case2}) considering $10^{-18}  \leq  \nu  \leq  0.5 - 10^{-9}$, we summarize Inequalities~(\ref{nu-delta-1-delta-2-case1})--(\ref{c-mu-nu-case2}) to know that 
\begin{align}
& \text{to ensure the consistency
property of Nakamoto's} \nonumber  \\ & \text{blockchain protocol, $c$ denoting $\frac{1}{p n \Delta}$ just needs to be} \nonumber  \\ & \text{slightly greater than $\frac{2\mu}{\ln (\mu / \nu)}$ for most $\nu \in (0, \frac{1}{2})$.}  \nonumber  
\end{align}
Our paper is the first one in the literature to derive such a neat expression $\frac{2\mu}{\ln (\mu / \nu)}$.
\end{rem}

Our secondary contribution is the following Theorem~\ref{thm-alpha}, which fixes an issue of~\cite{kiffer2018better} (details later). We also use Theorem~\ref{thm-alpha} to prove Theorem~\ref{thm-bound-c2} above.

% Inequality~(\ref{eq-thm2-c-v1-condition-thm-Delta}) of our  implies Corollary~\ref{cor-bound-c2-slightly} which for the first time gives the neat result of $c$ being slightly greater than $\frac{2\mu}{\ln (\mu / \nu)}$ to ensure consistency
% property of Nakamoto's blockchain protocol.

\begin{thm} \label{thm-alpha} Nakamoto's blockchain protocol satisfies consistency if there exists a positive constant $\delta_1$ such that
\begin{align}
{\overline{\alpha}}^{2\Delta} \alpha_1 \geq (1+\delta_1) \beta, \textup{ for $\beta:=p \nu n$} , \label{eq-thm-alpha}
\end{align}
where $\overline{\alpha}$ (resp.,~$\alpha_1$) denotes the probability that \textit{no} (resp.,~\textit{only one}) honest miner succeeds in solving a puzzle in one round, and is given by  Eq.~(\ref{eq-define-alpha-bar})  (resp.,~Eq.~(\ref{eq-define-alpha-1})), while $\beta$ denotes the expected number of blocks mined in each round by the adversary controlling $\nu n$ miners.
% From the expressions of $\alpha$ and $\overline{\alpha}$ in Eq.~(\ref{eq-define-alpha}) and Eq.~(\ref{eq-define-alpha-bar}) as well as $\beta =p \nu n$ and $c=\frac{1}{p n \Delta}$, we can show that Inequality~(\ref{eq-thm-alpha}) means the following complex condition involving $c$:
% \begin{align}
% \mu \bigg(1-\frac{1}{cn\Delta}\bigg)^{2\mu n\Delta}   \bigg(1-\frac{1}{cn\Delta}\bigg)^{\mu n-1} \geq (1+\delta_1)  \nu. \label{eq-thm-alpha-complex}
% \end{align}
% More specifically, if Inequality~(\ref{eq-thm-alpha}) holds, then
%  is at least \mbox{$1 - c \exp\left(-\frac{{\delta_2}^2 T \hspace{1pt} {\overline{\alpha}}^{2\Delta} \alpha_1}{72\tau(\alpha, \Delta)}\right)  - \exp\left(- T \nu n \cdot d(\delta_4, p)  \right)$,}  where
% \begin{myitemize}
% \item $c$ is a positive constant independent of $T,n,p,\mu,\Delta$,
% \item $\delta_2:=1-(1+\delta_1)^{-1/3}$ and  $\delta_4:=(1+\delta_1)^{1/3}-1$,
% \item $\tau(\alpha, \Delta)$ is a positive function of $\alpha$ and $\Delta$, and hence independent of $T$,
% \item $d(\delta_4, p)$ is a positive function of $\delta_4$ and $p$, and hence independent of $T$.
% \end{myitemize} 
\end{thm}

The proof of Theorem~\ref{thm-alpha} will be explained in Section~\ref{sec-Proof-Theorem-thm-alpha}. Below, we discuss the novelty of Theorem~\ref{thm-alpha}.

\noindent \textbf{Novelty of our Theorem~\ref{thm-alpha}.} Our Theorem~\ref{thm-alpha} is also novel in the sense its result as Inequality~(\ref{eq-thm-alpha}) has not been presented in any related work. 
%  Although a recent study by Kiffer~\emph{et~al.}~\cite{kiffer2018better} presents a seemingly similar result, their result is actually incorrect, as will be explained in the next paragraph. In addition, although~\cite{kiffer2018better}
 Although a recent study by Kiffer~\emph{et~al.}~\cite{kiffer2018better} also adopts a Markov-chain based approach that our Theorem~\ref{thm-alpha} uses,
 %to analyze ?,
 our Theorem~\ref{thm-alpha} differentiates from~\cite{kiffer2018better} in the following aspects as we will discuss:
\begin{myitemize}
\item[\ding{202}] First,~\cite{kiffer2018better} does not use the following two Markov chains which we propose for the first time and use to prove our Theorem~\ref{thm-alpha}: 
\begin{myitemize}
\item[\ding{172}] a Markov chain which models the transition of a variable denoting the suffix of the concatenation of the previous states and the current state,
\item[\ding{173}] a Markov chain modeling the transition of a variable which denotes the concatenation of i) the suffix of previous states before the $\Delta$ to last state, ii) the previous $\Delta$ states, and iii) the current state.  
\end{myitemize} 
\item[\ding{203}] Second, the analysis of~\cite{kiffer2018better} has minor errors. In~\cite{kiffer2018better}, the computations of $\ell_{11}$ and $\ell_{10}$ (defined on Page~7 of~\cite{kiffer2018better})  are incorrect. Specifically, $\frac{1}{\mu p}$ therein should be $\frac{1}{\alpha}$ (i.e., $\frac{1}{1 - (1- p)^{\mu n}} $).
\item[\ding{204}]  Third, even after we correct \mbox{$\mu \cdot p$} to $\alpha$ in many places of~\cite{kiffer2018better} and perform some computations to obtain Claim~\ref{thm-kiffer2018better-Theorem-4.4a} below from Theorem~4.4 on Page~8 of~\cite{kiffer2018better}, the result of~\cite{kiffer2018better} (and hence Claim~\ref{thm-kiffer2018better-Theorem-4.4a}) still has a minor issue. In~\cite{kiffer2018better}, to compute the convergence opportunities, one subevent is that  \textit{at least one} honest miner succeeds in solving a puzzle in one round (which happens with probability $\alpha$ in Eq.~(\ref{eq-define-alpha})), while the correct subevent should be that \textit{only one} honest miner succeeds in solving a puzzle in one round (which happens with probability $\alpha_1$ in Eq.~(\ref{eq-define-alpha-1})). Although~\cite{kiffer2018better} mentions ``a single honest mined block'', but its calculation actually uses ``at least one honest mined block'' (this leads to no appearance of $\alpha_1$ in~\cite{kiffer2018better}'s consistency condition). Our Theorem~\ref{thm-alpha} fixes the above issue of~\cite{kiffer2018better} (as noted in ``\ding{202}'' above, we also introduce novel Markov chains to present a clearer proof).

% ~\cite{kiffer2018better} does not present the Inequality result~(\ref{eq-thm-alpha}) of our Theorem~\ref{thm-alpha}. A result given as Inequality~(1) in~\cite{kiffer2018better} looks similar to our~(\ref{eq-thm-alpha}), but is incorrect due to the incorrect computations of $\ell_{11}$ and $\ell_{10}$ noted above. Also, the Markov chain in Figure 2 on Page~6 of~\cite{kiffer2018better} has only two states and cannot cover all possible states.
% \item Second, the analysis of~\cite{kiffer2018better} has minor errors and is more difficult to understand than our proof for Theorem~\ref{thm-alpha}.
% \item Third,~\cite{kiffer2018better} does not present the Inequality result~(\ref{eq-thm-alpha}) of our Theorem~\ref{thm-alpha}. A result in~\cite{kiffer2018better} looks similar to our~(\ref{eq-thm-alpha}), but is incorrect. 
\end{myitemize}

Here we give the reason why we present a detailed proof for Theorem~\ref{thm-alpha} instead of just replacing $\alpha$  with $\alpha_1$ in Inequality~(\ref{eq-thm-alpha-delta_3a}), a condition based on the analysis of~\cite{kiffer2018better}. We find the proof~\cite{kiffer2018better} not intuitive to understand. For instance, Page~6 of~\cite{kiffer2018better} uses the Markov chain \mbox{$ \rotatecurvearrowright \hspace{-2pt} S_{0} \rightleftarrows S_{1} \hspace{-2pt} \rotatecurvearrowleft$} to analyze consistency, with $S_{0}$ denoting the ``messy'' state  where honest mined blocks occur in less than $\Delta$ rounds from one another, and $S_{1}$ denoting the state where quiet periods between honest
mined blocks is at least $\Delta$ rounds. Taking the transition $S_{1} \to S_{1}$ as an example, it happens after a honest mined block followed by a quiet period
of at least $\Delta$ rounds. As the occurrence of the transition $S_{1} \to S_{1}$ needs multiple rounds, the Markov chain \mbox{$ \rotatecurvearrowright \hspace{-2pt} S_{0} \rightleftarrows S_{1} \hspace{-2pt} \rotatecurvearrowleft$} of~\cite{kiffer2018better} cannot characterize the states in the middle of the transition. Due to this, we present a proof of Theorem~\ref{thm-alpha} from scratch using more detailed Markov chains.  We emphasize again that the detailed proof of Theorem~\ref{thm-alpha} involving the novel Markov chains is our secondary contribution while Theorem~\ref{thm-bound-c2} presenting a neat condition to ensure   consistency is our major contribution. After obtaining an inequality (Inequality~(\ref{eq-thm-alpha-delta_3b}) in~\cite{fullpaper}) to ensure consistency,~\cite{kiffer2018better} does not analyze the inequality to provide a more understandable bound for $c$ as our Theorem~\ref{thm-bound-c2} does. Our proof of moving from Theorem~\ref{thm-alpha} to Theorem~\ref{thm-bound-c2} in Section~\ref{sec-Proof-Theorem-thm-bound-c2} is quite involved.

% \noindent \textbf{Comparing our Theorem~\ref{thm-alpha} with Claim~\ref{thm-kiffer2018better-Theorem-4.4a} below resulting from~\cite{kiffer2018better} (after partially fixing~\cite{kiffer2018better}).} 

% \subsection{Result of~\cite{kiffer2018better} after our corrections}

%after fixing its minor errors

We now state Claim~\ref{thm-kiffer2018better-Theorem-4.4a} on Page~\pageref{thm-alpha}, which is Theorem~4.4 on Page~8 of~\cite{kiffer2018better} after we correct \mbox{$\mu \cdot p$} to $\alpha$ in many places of~\cite{kiffer2018better} and perform some computations presented in Appendix~\ref{app-thm-kiffer2018better-Theorem-4.4a} of the online full version~\cite{fullpaper}.

\begin{Claima}[Theorem~4.4 on Page~8 of~\cite{kiffer2018better} after we correct \mbox{$\mu \cdot p$} to $\alpha$ in many places of~\cite{kiffer2018better} and perform some computations] \label{thm-kiffer2018better-Theorem-4.4a} Nakamoto's blockchain protocol satisfies consistency if there exists a positive constant $\delta_3$ such that
\begin{align}
{\overline{\alpha}}^{2\Delta}  \alpha  \geq (1+\delta_3) \beta.  \label{eq-thm-alpha-delta_3a}
\end{align}
From the expressions of $\alpha$ and $\overline{\alpha}$ in Eq.~(\ref{eq-define-alpha}) and Eq.~(\ref{eq-define-alpha-bar}) as well as $\beta =p \nu n$ and $c=\frac{1}{p n \Delta}$,   Inequality~(\ref{eq-thm-alpha-delta_3a}) means the following complex condition involving $c$:
\begin{align}
\textstyle{\bigg(1-\frac{1}{cn\Delta}\bigg)^{2\mu n\Delta} \bigg(1- \bigg(1-\frac{1}{cn\Delta}\bigg)^{\mu n}\bigg) \geq (1+\delta_3) \frac{\nu}{c \Delta}} . \label{eq-thm-alpha-delta_3a-complex}
\end{align}
\end{Claima}

How we rewrite Theorem~4.4 as Claim~\ref{thm-kiffer2018better-Theorem-4.4a} is presented in Appendix~\ref{app-thm-kiffer2018better-Theorem-4.4a} of the online full version~\cite{fullpaper}. We present the result as the claim due to the issue mentioned in ``\ding{204}'' above.

\linespread{0.98}\selectfont

\setlength{\belowdisplayskip}{1pt plus 0pt minus 1pt}%
\setlength{\belowdisplayshortskip}{1pt plus 0pt minus 1pt}
 \setlength{\abovedisplayskip}{1pt plus 0pt minus 1pt}
\setlength{\abovedisplayshortskip}{0pt minus 0.0pt}
   \parskip=0pt plus 1pt
 \setlist{nosep}

 \section{Proof of Theorem~\ref{thm-bound-c2} Given Theorem~\ref{thm-alpha}} \label{sec-Proof-Theorem-thm-bound-c2}

% We decompose Inequality~(\ref{eq-thm2-c-v1-condition-thm-Delta}) of Theorem~\ref{thm-bound-c2} into two parts (specifically, Inequalities~(\ref{eq-condition-pn}) and~(\ref{eq-thm2-c-v1}) given soon) and present Theorem~\ref{simpler-form-thm-bound-c2} below. Afterwards,  Section~\ref{subsec-simpler-form-thm-bound-c2} will present the proof of Theorem~\ref{simpler-form-thm-bound-c2}, while we use Theorem~\ref{simpler-form-thm-bound-c2} to show Theorem~\ref{thm-bound-c2} in Section~\ref{subsec-thm-bound-c2}.

We decompose Inequality~(\ref{eq-thm2-c-v1-condition-thm-Delta}) of Theorem~\ref{thm-bound-c2} into Inequalities~(\ref{eq-condition-pn}) and~(\ref{eq-thm2-c-v1}), to present Theorem~\ref{simpler-form-thm-bound-c2} below. 

\begin{thm} \label{simpler-form-thm-bound-c2} %\label{simpler-form-thm-bound-c2-condition-thm-Delta}
% If
% \begin{align}
% \text{$ \Delta > \frac{1+ \ln \frac{\mu}{\nu}}{4 \epsilon_2_1}   $ for a constant $\epsilon_2_1$ satisfying $0 < \epsilon_2_1 \iffalse_0 \fi  < 1$,}  \label{simpler-form-eq-condition-thm-Delta}
% \end{align}
% where we will explain in ? later that in ?, $\epsilon_2_1$ of ? can be very small (e.g.,
Consistency of Nakamoto's blockchain protocol holds in a window of $T$ rounds with probability at least \mbox{$1-O(1)\cdot\exp\left(-\Omega\left(T\right)\right)$}, when there exist constants $\epsilon_1$ and $\epsilon_2 $  satisfying \mbox{$0 < \epsilon_1 \iffalse_0 \fi  < 1$} and $\epsilon_2 > 0$ such that we have
\begin{align}
\text{$p n  \leq \frac{\epsilon_1 \ln \frac{\mu}{\nu}}{(\ln \frac{\mu}{\nu} + 1) \mu}$,
%for a constant $\epsilon_1$ satisfying \mbox{$0 < \epsilon_1 \iffalse_0 \fi  < 1$}.
}  \label{eq-condition-pn}
\end{align}
and $c$ denoting $\frac{1}{p n \Delta}$ satisfies
\begin{align}
c  \geq \left[ \frac{2\mu}{\ln (\mu / \nu)} +  \frac{1}{\Delta}\right]    \frac{1 + \epsilon_2 }{1-\epsilon_1}.  \label{eq-thm2-c-v1}
\end{align}
\end{thm}

Since $c$ denotes $\frac{1}{p n \Delta}$, it is straightforward to show that a combination of Inequalities~(\ref{eq-condition-pn}) and~(\ref{eq-thm2-c-v1}) is the same as Inequality~(\ref{eq-thm2-c-v1-condition-thm-Delta}), which is a condition of Theorem~\ref{thm-bound-c2}. 

%   Section~\ref{subsec-simpler-form-thm-bound-c2} below   presents the proof of Theorem~\ref{simpler-form-thm-bound-c2} using Theorem~\ref{thm-alpha}. In Section~\ref{subsec-thm-bound-c2}, we use Theorem~\ref{simpler-form-thm-bound-c2} to show Theorem~\ref{thm-bound-c2}.

Below we  present the proof of Theorem~\ref{simpler-form-thm-bound-c2} using Theorem~\ref{thm-alpha}. In Appendix~\ref{subsec-thm-bound-c2} of the online full version~\cite{fullpaper}, we use Theorem~\ref{simpler-form-thm-bound-c2} to show Theorem~\ref{thm-bound-c2}.

\subsection{Proof of Theorem~\ref{simpler-form-thm-bound-c2} using Theorem~\ref{thm-alpha}} \label{subsec-simpler-form-thm-bound-c2}

% To prove Theorem~\ref{thm-bound-c2}, we will first analyze how ?. To ?, we define

% \subsection{An Overview of the Proof}

% Intuitively, ?. Formally,
%  we will establish the following Theorem~\ref{thm-alpha}.

% \begin{lem} \label{prop-alpha}
% If
% \begin{align}
% {\overline{\alpha}} \geq \left( \frac{1+\delta_1}{1 -  p \mu n } \cdot  \frac{\nu}{\mu} \right)^{1/(2\Delta)},  \label{prop-alpha-eq}
% \end{align}
% then Inequality~(\ref{eq-thm-alpha}) holds.
% \end{lem}

% We will prove Theorem~\ref{thm-alpha} in Section~\ref{sec-thm-alpha}?.
 
To prove Theorem~\ref{simpler-form-thm-bound-c2} based on Theorem~\ref{thm-alpha},  we will show that given Inequality~(\ref{eq-condition-pn}), Inequality~(\ref{eq-thm2-c-v1}) implies Inequality~(\ref{eq-thm-alpha}). To this end, we analyze Inequality~(\ref{eq-thm-alpha}) through a series of transformations. Before stating the transformations, we note that in the rest of the paper,  ``$\Longleftarrow$'', ``$\Longrightarrow$'', and ``$\Longleftrightarrow$'' represent ``is implied by'', ``implies'', and ``is equivalent to'', respectively.  To prove Theorem~\ref{simpler-form-thm-bound-c2}, we will convert Inequality~(\ref{eq-thm-alpha}) in a number of steps and obtain the following results, where we will explain soon how to set $\delta_1$ and $\delta_5$. 
\begin{align}
& \text{Nakamoto's blockchain protocol satisfies consistency} \nonumber \\
& \xLeftarrow{\text{Theorem~\ref{thm-alpha}}} \left\{{\overline{\alpha}}^{2\Delta} \alpha_1 \geq (1+\delta_1) p \nu n\right\} \label{eq-lem-thm-alpha} \\  & \xLeftarrow{\textup{Lemma~\ref{prop-eq-bound-c-step-4}}} \left\{{\overline{\alpha}} \geq \left( \frac{1+\delta_1}{1 - p \mu n } \cdot  \frac{\nu}{\mu} \right)^{1/(2\Delta)}\right\}  \label{eq-bound-c-step-4} \\
& \xLeftarrow{\textup{Lemma~\ref{prop1}}} \left\{{\overline{\alpha}} \geq \left(  1+ \frac{\delta_5 \iffalse_7 \fi }{2\Delta} \right) \cdot \left(   \frac{\nu}{\mu} \right)^{1/(2\Delta)}\right\} \label{eq-lem-prop1}
% \text{~and~Inequality~(\ref{eq-condition-pn-complex})}
\\
% \left\{{\overline{\alpha}} \geq \left( \frac{1+\delta_1}{1 - \frac{1}{2} p \mu n } \cdot  \frac{\nu}{\mu} \right)^{1/(2\Delta)}\right\}
& \xLeftarrow{\textup{Lemma~\ref{prop-eq-bound-c-step-3}}} \left\{c \geq  \frac{1}{  n \Delta \left\{1-\left[ \left(1+ \frac{\delta_5 \iffalse_7 \fi }{2\Delta}\right) \left( \frac{\nu}{\mu} \right)^{1/(2\Delta)} \right]^{1/(\mu n)}\right\}}\right\} \label{eq-bound-c-step-3} \\
% \left\{c \geq \frac{1}{ n \Delta \left\{1-\left[ \left(1+ \frac{\delta_5 \iffalse_7 \fi }{2\Delta}\right) \left( \frac{\nu}{\mu} \right)^{1/(2\Delta)} \right]^{1/(\mu n)}\right\}}\right\}
& \xLeftarrow{\textup{Lemma~\ref{prop-eq-bound-c-step-2}}} \left\{c
%\geq   \frac{\mu}{\Delta \left[ 1 - \left(1+ \frac{\delta_5 \iffalse_7 \fi }{2\Delta}\right) \left( \frac{\nu}{\mu} \right)^{1/(2\Delta)} \right]}
\geq \frac{\mu}{\Delta \left[ 1 - \left(1+ \frac{\delta_5 \iffalse_7 \fi }{2\Delta}\right) \left( \frac{\nu}{\mu} \right)^{1/(2\Delta)} \right]}   \right\}
%\label{eq-c-C}
 \label{eq-bound-c-step-2} \\
% \left\{c \geq \frac{\mu}{\Delta \left[ 1 - \left(1+ \frac{\delta_5 \iffalse_7 \fi }{2\Delta}\right) \left( \frac{\nu}{\mu} \right)^{1/(2\Delta)} \right]}  \geq  \frac{1}{  n \Delta \left\{1-\left[ \left(1+ \frac{\delta_5 \iffalse_7 \fi }{2\Delta}\right) \left( \frac{\nu}{\mu} \right)^{1/(2\Delta)} \right]^{1/(\mu n)}\right\}}\right\}
& \xLeftarrow{\textup{Lemma~\ref{prop-eq-bound-c-step-1}}} \left\{c \geq \frac{\mu}{\Delta \left[ 1 - \left( \frac{\nu}{\mu} \right)^{1/(2\Delta)}\right]} \cdot \left(1+\frac{\delta_5 \iffalse_7 \fi }{\ln \frac{\mu}{\nu} - \delta_5 \iffalse_7 \fi }\right) \right\} \label{eq-bound-c-step-1}  \\
% \left\{c \geq \frac{\mu}{\Delta \left[ 1 - \left( \frac{\nu}{\mu} \right)^{1/(2\Delta)}\right]} \cdot \left(1+\frac{\delta_5 \iffalse_7 \fi }{\ln \frac{\mu}{\nu} - \delta_5 \iffalse_7 \fi }\right) \right\}
& \xLeftarrow{\textup{Lemma~\ref{prop-eq-bound-c-step-0}}} \left\{c  \geq  \left[ \frac{2\mu}{\ln (\mu / \nu)} +  \frac{\mu}{\Delta}\right]  \cdot \left(1+\frac{\delta_5 \iffalse_7 \fi }{\ln \frac{\mu}{\nu} - \delta_5 \iffalse_7 \fi }\right) \right\}
% \\ & ~~~~~~~~~~~~\text{~(a condition which we will show as Inequality~(\ref{eq-thm2-c-v1}) of Theorem~\ref{thm-bound-c2})}.
\label{eq-bound-c-step-0}
% \end{align}
% \begin{align}
% & \left\{c  \geq  \left[ \frac{2\mu}{\ln (\mu / \nu)} +  \frac{\mu}{\Delta}\right]  \cdot \left(1+\frac{\delta_5 \iffalse_7 \fi }{\ln \frac{\mu}{\nu} - \delta_5 \iffalse_7 \fi }\right) \right\} \text{~for $\delta_5$ to be given by Eq.~(\ref{eq-define-delta_2}) with $\epsilon_1$ in Inequality~(\ref{eq-thm2-c-v1})}
\\
& \xLeftarrow{\textup{Lemma~\ref{lem-c-simplified}}} \left\{c  \geq  \left[ \frac{2\mu}{\ln (\mu / \nu)} +  \frac{1}{\Delta}\right]  \cdot \frac{1 + \epsilon_2 }{1-\epsilon_1}\right\}\nonumber  \\ &~~~~~~~~~\text{~~\,(i.e., Inequality~(\ref{eq-thm2-c-v1}) of Theorem~\ref{simpler-form-thm-bound-c2}))}.  \label{eq-c-geq-delta-1-simplified}
\end{align} 
The statements of
\mbox{Lemmas~\ref{prop-eq-bound-c-step-4}--\ref{lem-c-simplified}} used above  are deferred to the end of this subsection for clarity, while their proofs will be presented in the Appendicies of the online full version~\cite{fullpaper}.

% As shown above, the results~(\ref{eq-lem-thm-alpha})--(\ref{eq-c-geq-delta-1-simplified}) are presented as \mbox{Lemmas~\ref{prop-eq-bound-c-step-4}--\ref{lem-c-simplified}}, whose statements are deferred to the end of this subsection for clarity with proofs  to be presented in Sections~\ref{sec-thm-alpha}--\ref{subse-lem-c-simplified}.

 \mbox{Lemmas~\ref{prop-eq-bound-c-step-4}--\ref{lem-c-simplified}} also involve extra conditions on $ pn $, $\delta_1$, and $\delta_5$, which are not explicitly stated in (\ref{eq-lem-thm-alpha})--(\ref{eq-c-geq-delta-1-simplified}). We will show on Page~\pageref{lem-conditions} that these conditions on $ pn $ are implied by Inequality~(\ref{eq-condition-pn}) of Theorem~\ref{simpler-form-thm-bound-c2}. To satisfy conditions on $\delta_1$ and $\delta_5$  in \mbox{Lemmas~\ref{prop-eq-bound-c-step-4}--\ref{lem-c-simplified}} for proving Theorem~\ref{simpler-form-thm-bound-c2} (the conditions will be discussed in detail on Page~\pageref{lem-conditions}), we will set $\delta_5$ and $\delta_1$ as follows:
 \begin{align}
\delta_5  & = \textstyle{\frac{(\epsilon_1 + \epsilon_2) \ln \frac{\mu}{\nu}}{\epsilon_1 + \epsilon_2 + (1 - \epsilon_1) \cdot (\ln \frac{\mu}{\nu}+1)}, }\text{~and} \label{eq-define-delta_2}  \\
\delta_1 & = \textstyle{( 1+ \delta_5) \cdot \left( 1 - \frac{\epsilon_1 \ln \frac{\mu}{\nu}}{\ln \frac{\mu}{\nu} + 1}\right) - 1}  \text{ with the above $\delta_5$.}  
% \nonumber  \\ &= \left[ 1 + \frac{(\epsilon_1 + \epsilon_2) \ln \frac{\mu}{\nu}}{(\epsilon_1 + \epsilon_2) + (1 - \epsilon_1) \cdot (\ln \frac{\mu}{\nu}+1)}\right]
% \nonumber  \\ & \quad \times \left( 1 - \frac{\epsilon_1 \ln \frac{\mu}{\nu}}{\ln \frac{\mu}{\nu} + 1}\right) - 1.
\label{eq-define-delta_1}
\end{align}
We note that $\delta_5$ and $\delta_1$ in Eq.~(\ref{eq-define-delta_2}) and Eq.~(\ref{eq-define-delta_1})  are both positive  for $ 0 <\epsilon_1 \iffalse_0 \fi  < 1$ and $ \epsilon_2 \iffalse_0 \fi > 0 $. The details are given in Appendix~\ref{app-delta_5-delta_1} of the online full version~\cite{fullpaper}.  

Below, we give \textit{intuitive}  explanations for a) how we obtain the condition  on $pn$ in Inequality~(\ref{eq-condition-pn}) of Theorem~\ref{simpler-form-thm-bound-c2}, and b) why we set  $\delta_5$ and $\delta_1$ according to Eq.~(\ref{eq-define-delta_2}) and~(\ref{eq-define-delta_1}) in order to have (\ref{eq-lem-thm-alpha})--(\ref{eq-c-geq-delta-1-simplified}) get through. The explanations are just intuitive since some steps come from necessity arguments while  some other steps result from   sufficiency arguments. On Page~\pageref{lem-conditions}, we will formally explain that enforcing the condition on $pn$ in Inequality~(\ref{eq-condition-pn}) and setting constants $\delta_5$ and $\delta_1$ according to Eq.~(\ref{eq-define-delta_2}) and~(\ref{eq-define-delta_1}) will ensure that all conditions of \mbox{Lemmas~\ref{prop-eq-bound-c-step-4}--\ref{lem-c-simplified}} are satisfied.

\textbf{How do we obtain the condition on $pn$ in Inequality~(\ref{eq-condition-pn}) of Theorem~\ref{simpler-form-thm-bound-c2}? }

In~(\ref{eq-bound-c-step-1}) and ~(\ref{eq-bound-c-step-0}), we observe the expression $\ln \frac{\mu}{\nu} - \delta_5 \iffalse_7 \fi $, which requires $\delta_5$ to be smaller than $\ln \frac{\mu}{\nu} $, as will become clear in Lemmas~\ref{prop-eq-bound-c-step-1} and~\ref{prop-eq-bound-c-step-0}. From~(\ref{eq-lem-prop1}), we see that Lemma~\ref{prop1} is used to provide $\left(\frac{1+\delta_1}{1 - p \mu n }\right)^{1/(2\Delta)} \leq  1+ \frac{\delta_5 \iffalse_7 \fi }{2\Delta} $. A necessary condition for this is $\left(\frac{1}{1 - p \mu n }\right)^{1/(2\Delta)} <  1+ \frac{\delta_5 \iffalse_7 \fi }{2\Delta} $, for which a sufficient condition is $\frac{1}{1 - p \mu n } < 1 + \delta_5$ since we know from the binomial series that $ 1+ \delta_5 \iffalse_7 \fi < \left(1+ \frac{\delta_5 \iffalse_7 \fi }{2\Delta}\right)^{2\Delta}  $. For $\delta_5<\ln \frac{\mu}{\nu} $, this implies $\frac{1}{1 - p \mu n } < 1 + \ln \frac{\mu}{\nu}$, for which a sufficient condition is
\begin{align}
\text{$p n  \leq  \frac{\epsilon_1 \ln \frac{\mu}{\nu}}{(\ln \frac{\mu}{\nu} + 1) \mu}  $ for a  constant $0<\epsilon_1 \iffalse_0 \fi  < 1$.}  \label{eq-condition-pn-complex}
\end{align}
% $p n  < \frac{\epsilon_1 \iffalse_0 \fi}{\mu} \left(1 - \frac{1}{1 + \ln \frac{\mu}{\nu}} \right) = \frac{\epsilon_1 \ln \frac{\mu}{\nu}}{(\ln \frac{\mu}{\nu} + 1) \mu}  $ for a positive constant $\epsilon_1 \iffalse_0 \fi  < 1$.
This Inequality~(\ref{eq-condition-pn-complex}) is stronger than $pn < \frac{1}{\mu}$ used in Lemma~\ref{prop-eq-bound-c-step-4}. Hence, our condition on $pn $ is just Inequality~(\ref{eq-condition-pn-complex}), which is exactly Inequality~(\ref{eq-condition-pn}) of Theorem~\ref{simpler-form-thm-bound-c2}.

%  Constraining $\epsilon_1$ to be smaller than $1$ rather than just $2$, we can simplify Inequality~(\ref{eq-condition-pn-complex}). Specifically, given $\frac{\ln \frac{\mu}{\nu}}{\ln \frac{\mu}{\nu} + 1}<1$,
%  %  prove $\frac{\epsilon_1 \ln \frac{\mu}{\nu}}{(\ln \frac{\mu}{\nu} + 1) \mu} < 1$ by showing that $\frac{ \ln \frac{\mu}{\nu}}{(\ln \frac{\mu}{\nu} + 1) \mu}=\frac{ \ln \frac{1-\nu}{\nu}}{(1 + \ln \frac{1-\nu}{\nu}) \cdot (1-\nu)}  $ is a decreasing function for $0 < \nu < \frac{1}{2} $ and converges to its supremum $1$ as $\nu \to 0 $. Hence,
%   a sufficient (but not necessary) condition for Inequality~(\ref{eq-condition-pn-complex}) is that $p n  < \frac{\epsilon_1 \ln \frac{\mu}{\nu}}{(\ln \frac{\mu}{\nu} + 1) \mu} $ for a positive constant $\epsilon_1 \iffalse_0 \fi  < 1$, which is exactly Inequality~(\ref{eq-condition-pn}) of Theorem~\ref{simpler-form-thm-bound-c2}. We can also prove that if  Inequality~(\ref{eq-condition-pn}) is replaced by the slightly more general  Inequality~(\ref{eq-condition-pn-complex}), Theorem~\ref{simpler-form-thm-bound-c2} still holds.

%  We use Inequality~(\ref{eq-condition-pn}) instead of Inequality~(\ref{eq-condition-pn-complex})

\textbf{How do we set constants $\delta_5$ and $\delta_1$ according to Eq.~(\ref{eq-define-delta_2}) and~(\ref{eq-define-delta_1}) to have (\ref{eq-lem-thm-alpha})--(\ref{eq-c-geq-delta-1-simplified}) get through?}

As discussed above, from~(\ref{eq-lem-prop1}), we see that Lemma~\ref{prop1} is used to provide $\left(\frac{1+\delta_1}{1 - p \mu n }\right)^{1/(2\Delta)} \leq  1+ \frac{\delta_5 \iffalse_7 \fi }{2\Delta} $, for which a sufficient condition is $\frac{1+\delta_1}{1 - p \mu n } \leq 1+ \delta_5 \iffalse_7 \fi  $ since we know from the binomial series that $ 1+ \delta_5 \iffalse_7 \fi < \left(1+ \frac{\delta_5 \iffalse_7 \fi }{2\Delta}\right)^{2\Delta}  $. We have just explained above the reasoning behind enforcing Inequality~(\ref{eq-condition-pn}) of Theorem~\ref{simpler-form-thm-bound-c2}; i.e., $p n  \leq \frac{\epsilon_1 \ln \frac{\mu}{\nu}}{(\ln \frac{\mu}{\nu} + 1) \mu} $ for a positive constant $\epsilon_1 \iffalse_0 \fi  < 1$. Then a sufficient condition to ensure the existence of positive $\delta_1$ satisfying  $\frac{1+\delta_1}{1 - p \mu n } \leq 1+ \delta_5 \iffalse_7 \fi  $ discussed just above is $\frac{1}{1 - \frac{\epsilon_1 \ln \frac{\mu}{\nu}}{1 + \ln \frac{\mu}{\nu}}  } <  1+ \delta_5 \iffalse_7 \fi   $, which gives $\delta_5 > \frac{\epsilon_1 \ln \frac{\mu}{\nu}}{1 + (1 - \epsilon_1) \ln \frac{\mu}{\nu}} $. 
% (to be presented as Inequality~(\ref{eq-define-delta_2-bound}) in Lemma~\ref{prop1} on Page~\pageref{prop1}).
 For such $\delta_5$, the expression $\left(1+\frac{\delta_5 \iffalse_7 \fi }{\ln \frac{\mu}{\nu} - \delta_5 \iffalse_7 \fi }\right)$ appearing in~(\ref{eq-bound-c-step-0})
%and hence in Theorem~\ref{simpler-form-thm-bound-c2}
 is greater than $  \left[ 1+\frac{\frac{\epsilon_1 \ln \frac{\mu}{\nu}}{1 + (1 - \epsilon_1) \ln \frac{\mu}{\nu}}}{\ln \frac{\mu}{\nu} - \frac{\epsilon_1 \ln \frac{\mu}{\nu}}{1 + (1 - \epsilon_1) \ln \frac{\mu}{\nu}} }\right] = \left[1+ \frac{\epsilon_1 \iffalse_0 \fi   }{(1-\epsilon_1 \iffalse_0 \fi ) \cdot (\ln \frac{\mu}{\nu}+1) } \right]$. Then we can select $\delta_5$ such that $\left(1+\frac{\delta_5 \iffalse_7 \fi }{\ln \frac{\mu}{\nu} - \delta_5 \iffalse_7 \fi }\right)$ equals $\left[1+ \frac{\epsilon_1 + \epsilon_2 \iffalse_0 \fi   }{(1-\epsilon_1 \iffalse_0 \fi ) \cdot (\ln \frac{\mu}{\nu}+1) } \right]$ for a positive constant $\epsilon_2$. This gives $\delta_5$ by Eq.~(\ref{eq-define-delta_2}).

Recalling $\frac{1+\delta_1}{1 -   p \mu n } \leq  1+ \delta_5 \iffalse_7 \fi  $ discussed above to produce Lemma~\ref{prop1}, we have $1+\delta_1 \leq ( 1+ \delta_5) \cdot (1 - p \mu n) $, for which we know from Inequality~(\ref{eq-condition-pn})  of Theorem~\ref{simpler-form-thm-bound-c2} \vspace{1pt} (i.e., $p n \leq \frac{\epsilon_1 \ln \frac{\mu}{\nu}}{(\ln \frac{\mu}{\nu} + 1) \mu} $ for a positive constant $\epsilon_1 \iffalse_0 \fi  < 1$) that a sufficient condition is $1+\delta_1 \leq ( 1+ \delta_5) \cdot \left( 1 - \frac{\epsilon_1 \ln \frac{\mu}{\nu}}{\ln \frac{\mu}{\nu} + 1}\right)  $. \vspace{1pt} Taking ``$\leq$'' here as ``$=$'' for simplicity, we set $\delta_1$ by Eq.~(\ref{eq-define-delta_1}).
%  The first equality result in~(\ref{eq-define-delta_1}) will be presented as Eq.~(\ref{eq-define-delta_1-copy}) in Lemma~\ref{prop1} on Page~\pageref{prop1}.
 
%   We can show that $\delta_5$ and $\delta_1$ in Eq.~(\ref{eq-define-delta_2}) and Eq.~(\ref{eq-define-delta_1})  are both positive  for $ 0 <\epsilon_1 \iffalse_0 \fi  < 1$ and $ \epsilon_2 \iffalse_0 \fi > 0 $. ?

%   Formal explanations for $\delta_5>0$ and $\delta_1>0$ will be given in the proof of Lemma~\ref{prop1} for clarity.

% this can be further implied by (note that again this is a sufficient but not necessary) condition

We now state \mbox{Lemmas~\ref{prop-eq-bound-c-step-4}--\ref{lem-c-simplified}}, which are proved in the Appendicies of the online full version~\cite{fullpaper}.

\begin{lem} \label{prop-eq-bound-c-step-4}
Under  
\begin{align}
0<p \mu n<1, \label{eq_0-pmun-1}
\end{align}
if
\begin{align} \textstyle{{\overline{\alpha}} \geq \left( \frac{1+\delta_1}{1 -  p \mu n } \cdot  \frac{\nu}{\mu} \right)^{1/(2\Delta)},}
 \label{eq_alpha_delta_1_D}
\end{align}
then Inequality~(\ref{eq-thm-alpha}) of Theorem~\ref{thm-alpha}  follows; i.e., \mbox{${\overline{\alpha}}^{2\Delta} \alpha_1 \geq (1+\delta_1) p \nu n$.} 
\end{lem}
\begin{rem}
The above result shows~(\ref{eq-bound-c-step-4}) under~(\ref{eq_0-pmun-1}), where (\ref{eq-bound-c-step-4}) is 
\begin{align}
 \left\{{\overline{\alpha}}^{2\Delta} \alpha_1 \geq (1+\delta_1) p \nu n\right\}  \xLeftarrow{\textup{Lemma~\ref{prop-eq-bound-c-step-4}}}  \textup{Inequality~(\ref{eq_alpha_delta_1_D})}. \label{eq-bound-c-step-4-repeat}
\end{align}
\end{rem}

\begin{lem} \label{prop1}
If Inequality~(\ref{eq-condition-pn}) of Theorem~\ref{simpler-form-thm-bound-c2} holds; i.e., if there exists a positive constant $0<\epsilon_1 \iffalse_0 \fi  < 1$ such that Inequality~(\ref{eq-condition-pn}) of Theorem~\ref{simpler-form-thm-bound-c2} holds,
% \begin{align}
% \text{$p n  \leq  \frac{\epsilon_1 \ln \frac{\mu}{\nu}}{(\ln \frac{\mu}{\nu} + 1) \mu}  $}, \label{eq-condition-pn}
% \end{align} 
% \begin{align}
% \text{$p n  < \min \left\{  \frac{\epsilon_1 \ln \frac{\mu}{\nu}}{(\ln \frac{\mu}{\nu} + 1) \mu},~\frac{1}{\mu}\right\}  $ for a constant $0 <\epsilon_1 \iffalse_0 \fi  < 2$,}  \label{eq-condition-pn-weaker}
% \end{align}
%  where Inequality~(\ref{eq-condition-pn-weaker}) holds under Inequality~(\ref{eq-condition-pn}) of Theorem~\ref{simpler-form-thm-bound-c2} (note \mbox{$0 < \epsilon_1 \iffalse_0 \fi  < 1$} in~(\ref{eq-condition-pn}) is weakened as $0 < \epsilon_1 \iffalse_0 \fi  < 2$ in~(\ref{eq-condition-pn-weaker})),
   then for \begin{align} \textstyle{\delta_5 > \frac{\epsilon_1 \ln \frac{\mu}{\nu}}{1 + (1 - \epsilon_1) \ln \frac{\mu}{\nu}} ,}
 \label{eq-define-delta_2-bound}
\end{align}and $\delta_1 $ given by
\begin{align} \textstyle{\delta_1 = ( 1+ \delta_5) \cdot \left( 1 - \frac{\epsilon_1 \ln \frac{\mu}{\nu}}{\ln \frac{\mu}{\nu} + 1}\right) - 1  ,}
 \label{eq-define-delta_1-copy}
\end{align}
we have $  \delta_5 > 0$, $\delta_1 > 0$, and
\begin{align} \textstyle{\left(\frac{1+\delta_1}{1 - p \mu n }\right)^{1/(2\Delta)} \leq  1+ \frac{\delta_5 \iffalse_7 \fi }{2\Delta} .}
 \label{eq-delta-1-delta-4}
\end{align}

\begin{rem}
Inequality~(\ref{eq-delta-1-delta-4}) means that  under
  \begin{align}
\textstyle{{\overline{\alpha}} \geq \left( 1+ \frac{\delta_5 \iffalse_7 \fi }{2\Delta} \right) \cdot \left(  \frac{\nu}{\mu} \right)^{1/(2\Delta)},}   \label{eq-c-lower-bound-prop1}
\end{align}
Inequality~(\ref{eq_alpha_delta_1_D}) of Lemma~\ref{prop-eq-bound-c-step-4} follows. Thus, under~(\ref{eq-condition-pn}) and~(\ref{eq-define-delta_2-bound}), we have
\begin{align}
\textup{Inequality~(\ref{eq_alpha_delta_1_D})} \xLeftarrow{\textup{Lemma~\ref{prop1}}}\textup{Inequality~(\ref{eq-c-lower-bound-prop1})} . \label{eq-lem-prop1-repeat}
\end{align}
\end{rem}

% \begin{rem}
% For proving Theorem~\ref{simpler-form-thm-bound-c2}, we set $\delta_5$ and $\delta_1$ according to Eq.~(\ref{eq-define-delta_2}) and Eq.~(\ref{eq-define-delta_1}) with constants \mbox{$0 < \epsilon_1 \iffalse_0 \fi  < 1$} and $  \epsilon_2 > 0$ from Theorem~\ref{simpler-form-thm-bound-c2}. Then our~(\ref{eq-define-delta_2-bound-repeat}) induces~(\ref{eq-define-delta_2-bound}), and (\ref{eq-define-delta_1}) is the same as~(\ref{eq-define-delta_1-copy}).  
% % that~(\ref{eq-define-delta_2-bound}) and~(\ref{eq-define-delta_1-copy}) are satisfied, where we have proved~(\ref{eq-define-delta_2-bound}) via~(\ref{eq-define-delta_2-bound}) and .
% \end{rem}

% defining $\delta_5 \iffalse_7 \fi  : = \frac{1}{2} [\frac{2\mu}{ 2\mu - \epsilon_1 \iffalse_0 \fi (\mu - \nu)} + \frac{\mu}{\nu}] - 1 $ and $\delta_1 :  =
% % (1+\delta_5 \iffalse_7 \fi ) [1 - \frac{\epsilon_1 \iffalse_0 \fi (\mu - \nu)}{2\mu}] - 1   = \frac{1}{2} [\frac{2\mu}{ 2\mu - \epsilon_1 \iffalse_0 \fi (\mu - \nu)} + \frac{\mu}{\nu}] [1 - \frac{\epsilon_1 \iffalse_0 \fi (\mu - \nu)}{2\mu}] - 1  =
% \frac{\mu}{2\nu} [1 - \frac{\epsilon_1 \iffalse_0 \fi (\mu - \nu)}{2\mu}] - \frac{1}{2} $, we have $0<\delta_5 \iffalse_7 \fi < \frac{\mu-\nu}{\nu}$, $\delta_1 > 0$, and $\left(\frac{1+\delta_1}{1 - \frac{1}{2} p \mu n }\right)^{1/(2\Delta)} \leq  1+ \frac{\delta_5 \iffalse_7 \fi }{2\Delta} $.
\end{lem}

\begin{lem} \label{prop-eq-bound-c-step-3}

%Under Inequality~(\ref{eq-condition-pn-weaker}), 

% For $\delta_1$ and $\delta_5$ satisfying~(\ref{eq-define-delta_1-copy})~(\ref{eq-define-delta_2-bound}) and
Under
\begin{align}
0<\delta_5 \iffalse_7 \fi < \ln \frac{\mu}{\nu}, \label{eq-define-delta_2-upperbound}
\end{align}
  if $c$ denoting $\frac{1}{p n \Delta}$ satisfies
  \begin{align}
\textstyle{c \geq \frac{1}{ n \Delta \left\{1-\left[ \left(1+ \frac{\delta_5 \iffalse_7 \fi }{2\Delta}\right) \left( \frac{\nu}{\mu} \right)^{1/(2\Delta)} \right]^{1/(\mu n)}\right\}},} \label{eq-c-lower-bound-1}
\end{align}
 then we have Inequality~(\ref{eq-c-lower-bound-prop1}). Note that  under Inequality~(\ref{eq-define-delta_2-upperbound}), the denominator in Inequality~(\ref{eq-c-lower-bound-1}) is positive  from Proposition~\ref{prop-delta-2-upper-bound} to be presented soon.
%  Again, when we set $\delta_5$ and $\delta_1$ according to Eq.~(\ref{eq-define-delta_2}) and Eq.~(\ref{eq-define-delta_1}) with constants \mbox{$0 < \epsilon_1 \iffalse_0 \fi  < 1$} and $  \epsilon_2/\epsilon_1 > 0$ given in Theorem~\ref{simpler-form-thm-bound-c2}, we can show that~(\ref{eq-define-delta_2-bound})~(\ref{eq-define-delta_1-copy}) and~(\ref{eq-define-delta_2-upperbound}) are all satisfied.
% For $0< \nu < \mu$ and $p n  < \max\left\{ \frac{1}{\mu},~ \frac{\epsilon_1 \iffalse_0 \fi (\mu - \nu)}{\mu^2}  \right\}$ for a positive constant $\epsilon_1 \iffalse_0 \fi  < 2$, if $\delta_5 \iffalse_7 \fi  > \frac{2\mu}{ 2\mu - \epsilon_1 \iffalse_0 \fi (\mu - \nu)} - 1 $ and $c \geq \frac{1}{ n \Delta \left\{1-\left[ \left(1+ \frac{\delta_5 \iffalse_7 \fi }{2\Delta}\right) \left( \frac{\nu}{\mu} \right)^{1/(2\Delta)} \right]^{1/(\mu n)}\right\}}$, then setting $\delta_1 $ as $ \left[ 1 - \frac{\epsilon_1 \iffalse_0 \fi (\mu - \nu)}{2\mu} \right] \cdot \left(1+\delta_5 \iffalse_7 \fi \right) - 1$,
\end{lem}

\begin{rem}
From the above result, under Inequality~(\ref{eq-define-delta_2-upperbound}), we have
\begin{align}
\textup{Inequality~(\ref{eq-c-lower-bound-prop1})} \xLeftarrow{\textup{Lemma~\ref{prop-eq-bound-c-step-3}}}\textup{Inequality~(\ref{eq-c-lower-bound-1})} . \label{eq-bound-c-step-3-repeat}
\end{align}
\end{rem}

\begin{rem}
For proving Theorem~\ref{simpler-form-thm-bound-c2}, we set $\delta_5$ and $\delta_1$ according to Eq.~(\ref{eq-define-delta_2}) and Eq.~(\ref{eq-define-delta_1}) with constants \mbox{$0 < \epsilon_1 \iffalse_0 \fi  < 1$} and $  \epsilon_2 > 0$,  so that~(\ref{eq-define-delta_2-bound})~(\ref{eq-define-delta_1-copy}) and~(\ref{eq-define-delta_2-upperbound}) of Lemmas~\ref{prop1} and~\ref{prop-eq-bound-c-step-3} are satisfied, as explained below. First, the result that Eq.~(\ref{eq-define-delta_2}) implies~(\ref{eq-define-delta_2-bound}) has been shown in~(\ref{eq-define-delta_2-bound-repeat}). Second, Eq.~(\ref{eq-define-delta_1}) is the same as (\ref{eq-define-delta_1-copy}). Finally, for $\delta_5$ in Eq.~(\ref{eq-define-delta_2}) with \mbox{$0 < \epsilon_1 \iffalse_0 \fi  < 1$}, we have $\delta_5  = \frac{(\epsilon_1 + \epsilon_2) \ln \frac{\mu}{\nu}}{\epsilon_1 + \epsilon_2 + (1 - \epsilon_1) \cdot (\ln \frac{\mu}{\nu}+1)} <  \frac{(\epsilon_1 + \epsilon_2) \ln \frac{\mu}{\nu}}{\epsilon_1 + \epsilon_2  } =  \ln \frac{\mu}{\nu}$, which gives~(\ref{eq-define-delta_2-upperbound}).
\end{rem}

\begin{prop} \label{prop-delta-2-upper-bound}
Under Inequality~(\ref{eq-define-delta_2-upperbound}), we have \begin{align}
\textstyle{1 - \left(1+ \frac{\delta_5 \iffalse_7 \fi }{2\Delta}\right) \left( \frac{\nu}{\mu} \right)^{1/(2\Delta)} > 0.} \nonumber
\end{align}
\end{prop}

\begin{lem} \label{prop-eq-bound-c-step-2}
Under Inequality~(\ref{eq-define-delta_2-upperbound}), we have 
\begin{align}
&\textstyle{\frac{\mu}{\Delta \left[ 1 - \left(1+ \frac{\delta_5 \iffalse_7 \fi }{2\Delta}\right) \left( \frac{\nu}{\mu} \right)^{1/(2\Delta)} \right]} }  \geq \textstyle{\frac{1}{  n \Delta \left\{1-\left[ \left(1+ \frac{\delta_5 \iffalse_7 \fi }{2\Delta}\right) \left( \frac{\nu}{\mu} \right)^{1/(2\Delta)} \right]^{1/(\mu n)}\right\}},}   \label{eq-c-lower-bound-prop-eq-bound-c-step-2-1-ineq}
\end{align} 
where the denominators in both sides of Inequality~(\ref{eq-c-lower-bound-prop-eq-bound-c-step-2-1-ineq}) are positive from Proposition~\ref{prop-delta-2-upper-bound} above.
\end{lem}
\begin{rem}
Inequality~(\ref{eq-c-lower-bound-prop-eq-bound-c-step-2-1-ineq}) means that if $c$ denoting $\frac{1}{p n \Delta}$ satisfies
  \begin{align}
\textstyle{c \geq \frac{\mu}{\Delta \left[ 1 - \left(1+ \frac{\delta_5 \iffalse_7 \fi }{2\Delta}\right) \left( \frac{\nu}{\mu} \right)^{1/(2\Delta)} \right]} ,} \label{eq-c-lower-bound-prop-eq-bound-c-step-2-1}
\end{align}
then
Inequality~(\ref{eq-c-lower-bound-1}) of Lemma~\ref{prop-eq-bound-c-step-3} follows. Thus, under Inequality~(\ref{eq-define-delta_2-upperbound}),  we have
\begin{align}
\textup{Inequality~(\ref{eq-c-lower-bound-1})} \xLeftarrow{\textup{Lemma~\ref{prop-eq-bound-c-step-2}}}\textup{Inequality~(\ref{eq-c-lower-bound-prop-eq-bound-c-step-2-1})} . \label{eq-bound-c-step-2-repeat}
\end{align}
\end{rem} 

\begin{lem} \label{prop-eq-bound-c-step-1}
Under Inequality~(\ref{eq-define-delta_2-upperbound}), we have 
\begin{align}
\textstyle{\frac{1}{ 1 - \left( \frac{\nu}{\mu} \right)^{1/(2\Delta)}} \cdot \left(1+\frac{\delta_5 \iffalse_7 \fi }{\ln \frac{\mu}{\nu} - \delta_5 \iffalse_7 \fi }\right) >   \frac{1}{ 1 - \left(1+ \frac{\delta_5 \iffalse_7 \fi }{2\Delta}\right) \left( \frac{\nu}{\mu} \right)^{1/(2\Delta)}}.} \label{eq-prop-eq-bound-c-step-1} 
\end{align} 
\end{lem}
\begin{rem}
Inequality~(\ref{eq-prop-eq-bound-c-step-1}) means that if $c$ denoting $\frac{1}{p n \Delta}$ satisfies
  \begin{align}
\textstyle{c \geq \frac{\mu}{\Delta \left[ 1 - \left( \frac{\nu}{\mu} \right)^{1/(2\Delta)}\right]} \cdot \left(1+\frac{\delta_5 \iffalse_7 \fi }{\ln \frac{\mu}{\nu} - \delta_5 \iffalse_7 \fi }\right) , }\label{eq-c-lower-bound-prop-eq-bound-c-step-1}
\end{align}
then
Inequality~(\ref{eq-c-lower-bound-prop-eq-bound-c-step-2-1}) follows. Thus, under Inequality~(\ref{eq-define-delta_2-upperbound}),  we have
\begin{align}
\textup{Inequality~(\ref{eq-c-lower-bound-prop-eq-bound-c-step-2-1})} \xLeftarrow{\textup{Lemma~\ref{prop-eq-bound-c-step-1}}}\textup{Inequality~(\ref{eq-c-lower-bound-prop-eq-bound-c-step-1})} . \label{eq-bound-c-step-1-repeat}
\end{align} 
\end{rem} 

\begin{lem} \label{prop-eq-bound-c-step-0}
We have
   \begin{align}
\textstyle{\frac{2}{\ln (\mu / \nu)} \leq  \frac{1}{\Delta \left[ 1 - \left( \frac{\nu}{\mu} \right)^{1/(2\Delta)}\right]}    \leq  \frac{2}{\ln (\mu / \nu)} + \frac{1}{\Delta}.} \label{eq-prop-eq-bound-c-step-0}
\end{align}
\end{lem}

\begin{rem}
Inequality~(\ref{eq-prop-eq-bound-c-step-0}) means that if $c$ denoting $\frac{1}{p n \Delta}$ satisfies
  \begin{align}
\textstyle{c  \geq  \left[ \frac{2\mu}{\ln (\mu / \nu)} +  \frac{\mu}{\Delta}\right]  \cdot \left(1+\frac{\delta_5 \iffalse_7 \fi }{\ln \frac{\mu}{\nu} - \delta_5 \iffalse_7 \fi }\right) ,} \label{eq-c-lower-bound-prop-eq-bound-c-step-0}
\end{align}
where $\delta_5$ satisfies $0<\delta_5 \iffalse_7 \fi < \ln \frac{\mu}{\nu}$ (i.e., Inequality~(\ref{eq-define-delta_2-upperbound})), then Inequality~(\ref{eq-c-lower-bound-prop-eq-bound-c-step-1}) follows. Thus, under Inequality~(\ref{eq-define-delta_2-upperbound}),  we have
\begin{align}
\textup{Inequality~(\ref{eq-c-lower-bound-prop-eq-bound-c-step-1})} \xLeftarrow{\textup{Lemma~\ref{prop-eq-bound-c-step-0}}}\textup{Inequality~(\ref{eq-c-lower-bound-prop-eq-bound-c-step-0})} . \label{eq-bound-c-step-0-repeat}
\end{align}  
\end{rem} 

\begin{lem}  \label{lem-c-simplified}
For constants \mbox{$0 < \epsilon_1 \iffalse_0 \fi  < 1$} and $  \epsilon_2 > 0$, with $\delta_5$ given by Eq.~(\ref{eq-define-delta_2}), 
%  ; i.e., with 
% \begin{align}
% \delta_5 = \frac{(\epsilon_1 + \epsilon_2) \ln \frac{\mu}{\nu}}{(\epsilon_1 + \epsilon_2) + (1 - \epsilon_1) \cdot (\ln \frac{\mu}{\nu}+1)} \label{eq-define-delta_2}
% \end{align} 
 we have
 \begin{align}
\textstyle{1+\frac{\delta_5 \iffalse_7 \fi }{\ln \frac{\mu}{\nu} - \delta_5 \iffalse_7 \fi } <   \frac{1 + \epsilon_2 }{1-\epsilon_1}.} \label{eq-lem-c-simplified}
\end{align}
\end{lem}

\begin{rem}
Inequality~(\ref{eq-lem-c-simplified}) means that under Inequality~(\ref{eq-define-delta_2}), if $c$ denoting $\frac{1}{p n \Delta}$\vspace{1pt} satisfies  Inequality~(\ref{eq-thm2-c-v1}) of Theorem~\ref{simpler-form-thm-bound-c2} (i.e., \mbox{$c  \geq  \left[ \frac{2\mu}{\ln (\mu / \nu)} +  \frac{\mu}{\Delta}\right]  \cdot  \frac{1 + \epsilon_2 }{1-\epsilon_1}$}),
%   \begin{align}
% c  \geq  \left[ \frac{2\mu}{\ln (\mu / \nu)} +  \frac{\mu}{\Delta}\right]  \cdot  \frac{1 + \epsilon_2 }{1-\epsilon_1} , \label{eq-thm2-c-v1}
% \end{align}
 then
Inequality~(\ref{eq-c-lower-bound-prop-eq-bound-c-step-0}) follows. Thus, under Inequality~(\ref{eq-define-delta_2}),  we have
\begin{align}
\textup{Inequality~(\ref{eq-c-lower-bound-prop-eq-bound-c-step-0})} \xLeftarrow{\textup{Lemma~\ref{lem-c-simplified}}}\textup{Inequality~(\ref{eq-thm2-c-v1})} . \label{eq-c-geq-delta-1-simplified-repeat}
\end{align}  
\end{rem}

\textbf{Putting All Things Together to Prove Theorem~\ref{simpler-form-thm-bound-c2}.} The above results~(\ref{eq-bound-c-step-4-repeat})~(\ref{eq-lem-prop1-repeat})~(\ref{eq-bound-c-step-3-repeat})~(\ref{eq-bound-c-step-2-repeat})~(\ref{eq-bound-c-step-1-repeat})~(\ref{eq-bound-c-step-0-repeat})~(\ref{eq-c-geq-delta-1-simplified-repeat}) are exactly (\ref{eq-bound-c-step-4})~(\ref{eq-lem-prop1})~(\ref{eq-bound-c-step-3})~(\ref{eq-bound-c-step-2})~(\ref{eq-bound-c-step-1})~(\ref{eq-bound-c-step-0})~(\ref{eq-c-geq-delta-1-simplified})  discussed earlier, which along with~(\ref{eq-lem-thm-alpha}) implies the desired result of Theorem~\ref{simpler-form-thm-bound-c2} that consistency of Nakamoto's blockchain protocol follows if Inequality~(\ref{eq-thm2-c-v1}) holds, under the assumption that we enforce all conditions of~(\ref{eq-bound-c-step-4-repeat})~(\ref{eq-lem-prop1-repeat})~(\ref{eq-bound-c-step-3-repeat})~(\ref{eq-bound-c-step-2-repeat})~(\ref{eq-bound-c-step-1-repeat})~(\ref{eq-bound-c-step-0-repeat})~(\ref{eq-c-geq-delta-1-simplified-repeat}). Now we discuss these conditions: \label{lem-conditions}
\begin{myitemize}
\item (\ref{eq-bound-c-step-4-repeat}) needs the condition~(\ref{eq_0-pmun-1}) that Lemma~\ref{prop-eq-bound-c-step-4} requires,
\item (\ref{eq-lem-prop1-repeat}) needs the condition~(\ref{eq-condition-pn}) and~(\ref{eq-define-delta_2-bound}) that Lemma~\ref{prop1} requires, and
\item  (\ref{eq-bound-c-step-3-repeat}) (resp.~(\ref{eq-bound-c-step-2-repeat})~(\ref{eq-bound-c-step-1-repeat}) and~(\ref{eq-bound-c-step-0-repeat})) needs the condition~(\ref{eq-define-delta_2-upperbound}) that Lemma~\ref{prop-eq-bound-c-step-3} (resp.~Lemmas~\ref{prop-eq-bound-c-step-2}, \ref{prop-eq-bound-c-step-1}, and~\ref{prop-eq-bound-c-step-0}) requires,
\item (\ref{eq-c-geq-delta-1-simplified-repeat}) needs the condition~(\ref{eq-define-delta_2}) that Lemma~\ref{lem-c-simplified} requires.
% \item (\ref{eq-bound-c-step-2-repeat}) needs the condition~(\ref{eq-define-delta_2-upperbound}), 
% \item (\ref{eq-bound-c-step-1-repeat}) needs the condition~(\ref{eq-define-delta_2-upperbound}), 
% \item (\ref{eq-bound-c-step-0-repeat}) needs the condition~(\ref{eq-define-delta_2-upperbound}), 
% \item (\ref{eq-c-geq-delta-1-simplified-repeat}) needs the condition~(\ref{eq-define-delta_2-upperbound}), 
\end{myitemize}
Hence, to complete proving Theorem~\ref{simpler-form-thm-bound-c2}, we just need to enforce~(\ref{eq_0-pmun-1})~(\ref{eq-define-delta_2-bound})~(\ref{eq-define-delta_2-upperbound}) and~(\ref{eq-define-delta_2}) given Inequality~(\ref{eq-condition-pn}) (i.e., $p n  \leq \frac{\epsilon_1 \ln \frac{\mu}{\nu}}{(\ln \frac{\mu}{\nu} + 1) \mu}$) with \mbox{$0 < \epsilon_1 \iffalse_0 \fi  < 1$} and $\epsilon_2 > 0$ from Theorem~\ref{simpler-form-thm-bound-c2}. To this end, we have the following:
\begin{myitemize}
\item We obtain Inequality~(\ref{eq_0-pmun-1}) from Inequality~(\ref{eq-condition-pn}) with  \mbox{$0 < \epsilon_1 \iffalse_0 \fi  < 1$}, in view of $p n  \leq \frac{\epsilon_1 \ln \frac{\mu}{\nu}}{(\ln \frac{\mu}{\nu} + 1) \mu} <\frac{1}{\mu}$.
\item After we define $\delta_5$ according to (\ref{eq-define-delta_2}), we obtain Inequality~(\ref{eq-define-delta_2-bound}) in view of~(\ref{eq-define-delta_2-bound-repeat}), and obtain Inequality~(\ref{eq-define-delta_2-upperbound}) in view of $\delta_5 = \frac{(\epsilon_1 + \epsilon_2) \ln \frac{\mu}{\nu}}{(\epsilon_1 + \epsilon_2) + (1 - \epsilon_1) \cdot (\ln \frac{\mu}{\nu}+1)} \iffalse_7 \fi < \frac{(\epsilon_1 + \epsilon_2) \ln \frac{\mu}{\nu}}{(\epsilon_1 + \epsilon_2) } \iffalse_7 \fi = \ln \frac{\mu}{\nu}$.  
\end{myitemize}

Summarizing the above, we have shown Theorem~\ref{simpler-form-thm-bound-c2} using (\ref{eq-bound-c-step-4-repeat})~(\ref{eq-lem-prop1-repeat})~(\ref{eq-bound-c-step-3-repeat})~(\ref{eq-bound-c-step-2-repeat})~(\ref{eq-bound-c-step-1-repeat})~(\ref{eq-bound-c-step-0-repeat})~(\ref{eq-c-geq-delta-1-simplified-repeat}), which hold respectively after we prove \mbox{Lemmas~\ref{prop-eq-bound-c-step-4}--\ref{lem-c-simplified}}  in Appendices~\ref{subse-prop-eq-bound-c-step-4}--\ref{subse-lem-c-simplified} of the online full version~\cite{fullpaper}. In Appendix~\ref{subsec-thm-bound-c2} of~\cite{fullpaper}, we use Theorem~\ref{simpler-form-thm-bound-c2} to show Theorem~\ref{thm-bound-c2}. \qeda

% Lemma~\ref{prop-eq-bound-c-step-4}

% Since $\delta_5 \iffalse_7 \fi $ is set as $\frac{(\epsilon_1 + \epsilon_2) \ln \frac{\mu}{\nu}}{\epsilon_1 + \epsilon_2 + (1 - \epsilon_1) \cdot (\ln \frac{\mu}{\nu}+1)}$ from Eq.~(\ref{eq-define-delta_2}), we compute the term
% $ \frac{\delta_5 \iffalse_7 \fi }{\ln \frac{\mu}{\nu} - \delta_5 \iffalse_7 \fi } $ in~(\ref{eq-bound-c-step-0}) as $ \frac{\epsilon_1 + \epsilon_2 \iffalse_0 \fi   }{(1-\epsilon_1 \iffalse_0 \fi ) \cdot (\ln \frac{\mu}{\nu}+1) }$. Hence, the results~(\ref{eq-bound-c-step-4})--(\ref{eq-bound-c-step-0}) above together imply that
% \begin{align}
%  &\left\{ c  \geq  \left[ \frac{2\mu}{\ln (\mu / \nu)} +  \frac{1}{\Delta}\right]  \cdot \frac{1 + \epsilon_2 }{1-\epsilon_1}\right\} \nonumber  \\ &\Longrightarrow \left\{{\overline{\alpha}}^{2\Delta} \alpha_1 \geq (1+\delta_1) p \nu n\right\}, \label{eq-c-geq-delta-1}
% \end{align}
% if all conditions of \mbox{Lemmas~\ref{prop-eq-bound-c-step-4}--\ref{lem-c-simplified}} are satisfied, which we will explain soon. Then (\ref{eq-c-geq-delta-1}) along with Theorem~\ref{thm-alpha} induces the desired Theorem~\ref{simpler-form-thm-bound-c2}.

% As noted, we now explain that all conditions of \mbox{Lemmas~\ref{prop-eq-bound-c-step-4}--\ref{lem-c-simplified}} are satisfied. To this end, we first present the conditions used in the lemmas.
% Lemma~\ref{prop-eq-bound-c-step-4} uses $0<p \mu n<1$ and the trivial condition $n \geq 4$. Lemma~\ref{prop-eq-bound-c-step-3}

% ===

\begin{figure*}[!t]
 \centering
\includegraphics[scale=0.45]{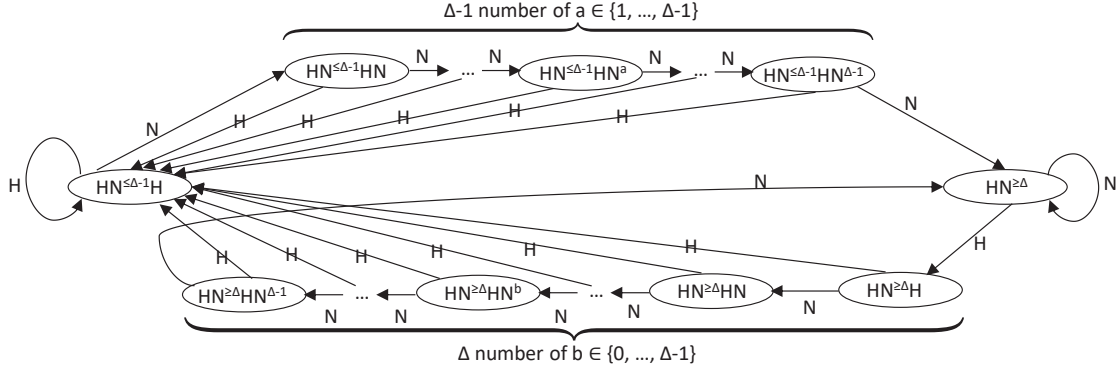} \vspace{-5pt}
\caption{The suffix-of-previous-and-current-states Markov chain $\mathcal{C}_{\boldsymbol{F}}$, which models the transition of a variable denoting the suffix of the concatenation of the previous states and the current state.\vspace{-20pt}} \label{fig-Markov-F}
\end{figure*}

\section{Proof of Theorem~\ref{thm-alpha}} \label{sec-Proof-Theorem-thm-alpha}

% fixing other parameters

We use $A(t_0, t_0+T-1)$ to denote the number of blocks mined the adversary in the $T$ rounds from round $t_0$ to $t_0+T-1$, and use $C(t_0, t_0+T-1)$ to denote the number of times that $HN^{\geq \Delta} || H_1 N^{\Delta}$ is visited (i.e., the number of convergence opportunities) in the $T$ rounds from round $t_0$ to $t_0+T-1$. Then 
we will show in Section~\ref{sec-eq-E-C-E-A} that Inequality~(\ref{eq-thm-alpha}) of Theorem~\ref{thm-alpha} is the same as
\begin{align}
 \bE{C(t_0, t_0+T-1)}\geq (1+\delta_1) \cdot \bE{A(t_0, t_0+T-1)}. \label{eq-thm-alpha-repeat}
\end{align}

Here we discuss the intuition of requiring Inequality~(\ref{eq-thm-alpha-repeat}), which then gives Inequality~(\ref{eq-thm-alpha}). First, we will prove that  the probability of \mbox{$C(t_0, t_0+T-1)$}
 being a constant factor smaller than its expectation $\bE{C(t_0, t_0+T-1)}$ is exponentially small in $T$. Formally, for any positive constant $\delta_2<1$, we have
\begin{align}
 &\bp{C(t_0, t_0+T-1) \leq (1-\delta_2) \cdot \bE{C(t_0, t_0+T-1)}}  \nonumber  \\ & \leq O(1)\cdot\exp\left(-\Omega\left(T\right)\right) . \label{prop-event-1-prob-eq-simple}
\end{align}
Second, we will prove that the probability of \mbox{$A(t_0, t_0+T-1)$}
 being a constant factor greater than its expectation $\bE{A(t_0, t_0+T-1)}$ is exponentially small in $T$. Formally, for any positive constant $\delta_4$, we have
 \begin{align}
 &\bp{A(t_0, t_0+T-1) \geq (1+\delta_4) \cdot \bE{A(t_0, t_0+T-1)}}  \nonumber  \\ & \leq O(1)\cdot\exp\left(-\Omega\left(T\right)\right) . \label{prop-event-2-prob-eq-simple}
\end{align}
In (\ref{prop-event-1-prob-eq-simple}) and~(\ref{prop-event-2-prob-eq-simple}), the term $O(1)$ is with respect to~$T$. 

% As will become clear, a more specific expression for the bound in the right hand side (RHS) of Inequality~(\ref{prop-event-1-prob-eq-simple}) (resp., Inequality~(\ref{prop-event-2-prob-eq-simple})) is $c \exp\left(-\frac{{\delta_2}^2 T \hspace{1pt} {\overline{\alpha}}^{2\Delta} \alpha_1}{72\tau(\alpha, \Delta)}\right) $ (resp., $\exp\left(- T \nu n \cdot d(\delta_4, p)  \right)$) appearing in the statement of Theorem~\ref{thm-alpha}. 

Via a union bound to combine Inequalities~(\ref{prop-event-1-prob-eq-simple}) and~(\ref{prop-event-2-prob-eq-simple}), $\left(C(t_0, t_0+T-1) \leq (1-\delta_2) \cdot \bE{C(t_0, t_0+T-1)} \right)\lor \left( A(t_0, t_0+T-1) \geq (1+\delta_4) \cdot \bE{A(t_0, t_0+T-1)}\right)$ happens with probability no greater than the result of summing the bounds in the right hand side (RHS) of Inequalities~(\ref{prop-event-1-prob-eq-simple}) and~(\ref{prop-event-2-prob-eq-simple}), which can also be written as $O(1)\cdot\exp\left(-\Omega\left(T\right)\right)$. Then we have (at least) \mbox{$1-O(1)\cdot\exp\left(-\Omega\left(T\right)\right)$} probability for the above union event's complement, $\left(C(t_0, t_0+T-1) > (1-\delta_2) \cdot \bE{C(t_0, t_0+T-1)} \right)\land \left( A(t_0, t_0+T-1) < (1+\delta_4) \cdot \bE{A(t_0, t_0+T-1)}\right)$, implying that \mbox{$C(t_0, t_0+T-1) - A(t_0, t_0+T-1) $}  is greater than
\begin{align}
 &(1\hspace{-1.5pt}-\hspace{-1.5pt}\delta_2) \cdot \bE{C(t_0, t_0\hspace{-1.5pt}+\hspace{-1.5pt}T\hspace{-1.5pt}-\hspace{-1.5pt}1)} - (1\hspace{-1.5pt}+\hspace{-1.5pt}\delta_4) \cdot \bE{A(t_0, t_0\hspace{-1.5pt}+\hspace{-1.5pt}T\hspace{-1.5pt}-\hspace{-1.5pt}1)} . \label{eq-delta-2-delta-3}
\end{align}
From Inequality~(\ref{eq-thm-alpha-repeat}), we bound the term in~(\ref{eq-delta-2-delta-3}) by
\begin{align}
\text{(\ref{eq-delta-2-delta-3})} \geq [(1-\delta_2) \cdot  (1+\delta_1)- (1+\delta_4)] \cdot \bE{A(t_0, t_0+T-1)}. \label{eq-delta-2-delta-3v2}
\end{align} 
Then to obtain the desired result that \mbox{$C(t_0, t_0+T-1) - A(t_0, t_0+T-1) $} is $\Omega(T)$ with probability \mbox{$1-O(1)\cdot\exp\left(-\Omega\left(T\right)\right)$}, we select positive constants $\delta_2 < 1$ and $\delta_4$ such that the term in~(\ref{eq-delta-2-delta-3v2}) is $\Omega(T)$. It will become clear from Eq.~(\ref{eq-Expectation-A-t-0-t-0-T-1-simple}) that $A(t_0, t_0+T-1) $ can be written as $\Omega(T)$, so we select positive constants $\delta_2 < 1$ and $\delta_4$ such that $[(1-\delta_2) \cdot  (1+\delta_1)- (1+\delta_4)]$ appearing in~(\ref{eq-delta-2-delta-3v2}) is a positive constant. To this end, we set
\begin{align}
& \delta_2:=1-(1+\delta_1)^{-1/3},~~~ \delta_4:=(1+\delta_1)^{1/3} - 1 , 
\end{align}
so that Inequality~(\ref{eq-delta-2-delta-3v2}) becomes
\begin{align}
\hspace{-20pt}\text{(\ref{eq-delta-2-delta-3})} \geq  \big[(1+\delta_1)^{2/3}-(1+\delta_1)^{1/3}\big] \cdot \bE{A(t_0, t_0+T-1)}. \label{eq-delta-2-delta-3v3}
\end{align} 
Summarizing the above, we have
\begin{align}
 &\text{If Inequalities~(\ref{eq-thm-alpha-repeat})~(\ref{prop-event-1-prob-eq-simple}) and~(\ref{prop-event-2-prob-eq-simple}) hold, then}  \nonumber  \\ &\text{\mbox{$C(t_0, t_0+T-1) - A(t_0, t_0+T-1) $} is greater than}  \nonumber  \\ &\text{$\big[(1+\delta_1)^{2/3}-(1+\delta_1)^{1/3}\big] \cdot \bE{A(t_0, t_0+T-1)}$}  \nonumber  \\ &\text{with probability \mbox{$1-O(1)\cdot\exp\left(-\Omega\left(T\right)\right)$}}   . \label{eq-delta-2-delta-3v5}
\end{align}

In the rest of this section, we will first prove in Section~\ref{sec-eq-E-C-E-A} that Inequality~(\ref{eq-thm-alpha}) of Theorem~\ref{thm-alpha} is the same as Inequality~(\ref{eq-thm-alpha-repeat}). It will also become clear in Section~\ref{sec-eq-E-C-E-A} that \mbox{$A(t_0, t_0+T-1) $}  can be written as $\Omega(T)$. We prove Inequalities~(\ref{prop-event-1-prob-eq-simple}) and~(\ref{prop-event-2-prob-eq-simple}) in Appendices~\ref{sec-prop-event-1-prob-eq-simple} and \ref{sec-prop-event-2-prob-eq-simple} of the online full version~\cite{fullpaper}. In Section~\ref{sec-thm-1-Putting-things-together}, we combine  the results of Appendices~\ref{sec-prop-event-1-prob-eq-simple} and \ref{sec-prop-event-2-prob-eq-simple} of~\cite{fullpaper} with~(\ref{eq-delta-2-delta-3v5}) to complete the proof of to Theorem~\ref{thm-alpha}.

% which further induces Inequality~(\ref{eq-E-C-E-A}) given Inequality~(\ref{eq-delta-2-delta-3v5}).
 %means that there exist positive constants $\delta_2<1$ and $\delta_4$ such that Inequality~(\ref{eq-E-C-E-A}) holds.

% \subsection{Proving that Inequality~(\ref{eq-thm-alpha}) implies Inequalities~(\ref{eq-thm-alpha-repeat}) and~(\ref{eq-E-C-E-A})} \label{sec-eq-E-C-E-A}

\subsection{Proving that Inequality~(\ref{eq-thm-alpha}) is the same as Inequality~(\ref{eq-thm-alpha-repeat})} \label{sec-eq-E-C-E-A}

  Inequality~(\ref{eq-thm-alpha}) of Theorem~\ref{thm-alpha} is \mbox{${\overline{\alpha}}^{2\Delta} \alpha_1 \geq (1+\delta_1) p \nu n$.}  To show Inequality~(\ref{eq-thm-alpha-repeat}), we will explain
\begin{align}
\bE{C(t_0, t_0+T-1)} = T \hspace{1pt} {\overline{\alpha}}^{2\Delta} \alpha_1, \label{eq-Expectation-C-t-0-t-0-T-1-simple}
\end{align}
and
\begin{align}
\bE{A(t_0, t_0+T-1)} = T p \nu n, \label{eq-Expectation-A-t-0-t-0-T-1-simple}
\end{align}

We first show Eq.~(\ref{eq-Expectation-A-t-0-t-0-T-1-simple}). Since the adversary controls $\nu n$ nodes and each node mines a block independently with probability $p$ in each round, the number of blocks mined (by $\nu n$ nodes controlled) by the adversary in each round follows $\text{binom}( \nu n, p)$, which denotes a binomial distribution with $ \nu n$ being the number of trials and $p$ being the success probability for each trial. Then $A(t_0, t_0+T-1)$ denoting the number of blocks mined the adversary in the $T$ rounds from round $t_0$ to $t_0+T-1$ is the sum of $T$ indepdendent random variables, each of which obeys $\text{binom}( \nu n, p)$. Hence,  $A(t_0, t_0+T-1)$ follows $\text{binom}(T \nu n, p)$. Then Eq.~(\ref{eq-Expectation-A-t-0-t-0-T-1-simple}) clearly follows.

% , which denotes a binomial distribution with $T \nu n$ being the number of trials and $p$ being the success probability for each trial. 

We now present the proof of Eq.~(\ref{eq-Expectation-C-t-0-t-0-T-1-simple}.

In each round, one of the following events will happen: i) $H$, which means that at least one block is mined by the benign (i.e., honest) nodes, and ii) $N$, which means that no block is mined by the benign nodes. By a round's state, we refer to whether $H$ or $N$ happens, and we know from the definitions of $\alpha$ and $\overline{\alpha}$ in Eq.~(\ref{eq-define-alpha}) and Eq.~(\ref{eq-define-alpha-bar}) that~$H$ (resp.,~$N$) happens with probability $\alpha$ (resp.,~$\overline{\alpha}$). Then we define \text{State-Set} to characterize the possible values that a round's state can take:
\begin{align}
\text{State-Set}   & := \{H,~N\} . \label{eq-State-Set}
\end{align} 
Let $S_t$ be the random variable representing the state at round $t$. We will use $s_t \in \text{State-Set}$ as an instantiation of $S_t$.

We consider a Markov chain $\mathcal{C}_{\boldsymbol{F}}$ for the suffix of all the states in all rounds up to round $t$, where ``$\boldsymbol{F}$'' means suffix. We will explain that Figure~\ref{fig-Markov-F} can represent this Markov chain. To avoid confusion, we use ``$\boldsymbol{F}$'' instead of ``$\boldsymbol{S}$'' since the symbol $S$ is used to represent the state at a round. We will call $\mathcal{C}_{\boldsymbol{F}}$ as the suffix-of-previous-and-current-states Markov chain. At round $t$, let  random variable $\boldsymbol{F}_t$ represent the suffix of the states in all rounds up to round $t$; i.e., $\boldsymbol{F}_t$ represents the vertex visited at round $t$ in the Markov chain $\mathcal{C}_{\boldsymbol{F}}$.  
% We formulate two Markov chains, called  suffix-of-previous-and-current-states Markov chain $\mathcal{C}_{\boldsymbol{F}}$ and 
% suffix-of-far-previous-states-and-previous-states Markov chain $\mathcal{C}_{\boldsymbol{F}||\boldsymbol{P}}$.

% $\boldsymbol{F}_t$ denotes the random variable 
After at least two $H$ have happened by round $t$ (which holds for sufficiently large $t$), we will explain below that we can characterize all possible $\boldsymbol{F}_t$ by the following $2\Delta+1$ values which form the $\text{Suffix-Set}$:
%consists of $2\Delta+1$ values
\begin{align}
\text{Suffix-Set}   & : =  \left\{\begin{subarray}{l}HN^{\leq \Delta-1}H,~HN^{\leq \Delta-1}HN^{a},\\[2pt] HN^{\geq \Delta},~HN^{\geq \Delta}HN^{b} \end{subarray}~\Big|~\begin{subarray}{l}a \in \{1,\ldots, \Delta-1\},\\[2pt] b \in \{0,\ldots, \Delta-1\} \end{subarray}\right\}  . \label{eq-Suffix-Set}
\end{align}
In~(\ref{eq-Suffix-Set}), the term $N^{\leq \Delta-1}$ means a series of  $N$ which has at most $\Delta-1$ number of consecutive $N$; i.e., zero $N$ (i.e., null), one $N$, $\ldots$, or $\Delta-1$ number of $N$. Similarly, $N^{\geq \Delta}$ means a series of  $N$ which has at least $\Delta$ number of consecutive $N$, while $N^{a}$ (resp.,~$N^{b}$) means $a$ (resp.,~$b$) number of consecutive $N$. Supposing $\Delta=3$ for the purpose of giving an example (practical $\Delta$ is much larger) and the states from round $1$ to round $10$ are $H,N,H,H,N,N,H,N,N,N$, then the corresponding $\boldsymbol{F}_7, \boldsymbol{F}_8, \boldsymbol{F}_9$, and $\boldsymbol{F}_{10}$ (i.e., $\boldsymbol{F}_{t}$ at time $t=7,8,9$, and $10$) are $HN^{\leq \Delta-1}H$, $HN^{\leq \Delta-1}HN^{a}$ with $a=1$, $HN^{\leq \Delta-1}HN^{a}$ with $a=2$, and $HN^{\geq \Delta}$ ($HN^{\geq 3}$ covers $HN^{3}$), respectively.

To see why we can characterize all possible $\boldsymbol{F}_t$ by~(\ref{eq-Suffix-Set}), we discuss the following cases, where we recall that  $S_i$ represents the state at round $i$: 
\begin{myitemize}
\item If $S_t$ is $H$ and $S_{t-1}$ is $H$, then we can set $\boldsymbol{F}_t$ as $HN^{\leq \Delta-1}H$ which covers $HH$ when ``$N^{\leq \Delta-1}$'' becomes $0$ number of $N$ (i.e., null); 
\item If $S_t$ is $H$ and $S_{t-1}$ is $N$, as we consider that at least two $H$ have happened by round $t$ (which holds for sufficiently large $t$), suppose the previous $H$ closest to round $t$ happens at round $t-c$ for some $c>0$. In other words, $S_{t-c}$ and $S_t$ are $H$ while $S_i$ for each $i \in \{t-c+1, \ldots, t-1\}$ is $N$ so that the series $S_{t-c} \ldots S_{t}$ can be written as $HN^{c-1}H$. Then if $c-1\leq \Delta-1$,  we can set $\boldsymbol{F}_t$ as $HN^{\leq \Delta-1}H$; if $c-1\geq \Delta$,  we can set $\boldsymbol{F}_t$ as $HN^{\geq \Delta}HN^{b}$ which covers $HN^{\geq \Delta}H$ when $b$ takes $0$.
\item If $S_t$ is $N$, as we consider that at least two $H$ have happened by round $t$ (which holds for sufficiently large $t$), suppose the $H$ closest to and before round $t$   happens at round $t-d$ for some $d>0$. In other words, $S_{t-d}$ is $H$ while $S_i$ for each $i \in \{t-d+1, \ldots, t\}$ is $N$ so that the series $S_{t-d} \ldots S_{t}$ can be written as $HN^{d}$. Then we have two subcases:
\begin{myitemize}
\item If $d\geq \Delta$,  we can set $\boldsymbol{F}_t$ as $HN^{\geq \Delta}$. 
\item If $d\leq \Delta-1$, given that $S_{t-d} \ldots S_{t}$ is $HN^{d}$, we further discuss the states before round $t-d$. Again, since we consider that at least two $H$ have happened by round $t$ (which holds for sufficiently large $t$), suppose the previous $H$ closest to round $t-d$ happens at round $t-f$ for some $f>d$. In other words, $S_{t-f}$ is $H$ while $S_i$ for each $i \in \{t-f+1, \ldots, t-d-1\}$ is $N$ so that the series $S_{t-f} \ldots S_{t}$ can be written as $HN^{f-d-1}HN^{d}$. Recalling this subcase discusses $d\leq \Delta-1$, if \mbox{$f-d-1 \leq \Delta-1$,}  we can set $\boldsymbol{F}_t$ as $HN^{\leq \Delta-1}HN^{a}$ for $a = d \in \{1,\ldots, \Delta-1\}$; if \mbox{$f-d-1 \geq \Delta$,}  we can set $\boldsymbol{F}_t$ as $HN^{\geq \Delta}HN^{b}$ for $b = d \in \{1,\ldots, \Delta-1\}$. 
\end{myitemize} 
% \item If , then we can set 
% \item If , then we can set 
% \item If , then we can set 
% \item If , then we can set 
\end{myitemize}

% \junzhao{From our Figure~\ref{fig-Markov-F} and~\cite{kiffer2018better}'s Figure 3, we see the differences between our suffix-of-previous-and-current-states Markov chain and~\cite{kiffer2018better}'s Markov chain are:}

In Figure~\ref{fig-Markov-F} on Page~\pageref{fig-Markov-F}, we plot the transition  of $\boldsymbol{F}_t$ in the suffix-of-previous-and-current-states Markov chain $\mathcal{C}_{\boldsymbol{F}}$, which is \textit{time-homogeneous}, \textit{irreducible}, and \textit{ergodic}. In particular, from~\cite{levin2017markov,kiffer2018better}, time-homogeneous means that the transition does not depend on the time; irreducible means getting to any state from any other state has non-zero probability; and ergodic means that each state has a positive mean recurrence time and is aperiodic (i.e., the period is $1$).

As illustrated in Figure~\ref{fig-Markov-F}, the transition rules in the Markov chain $\mathcal{C}_{\boldsymbol{F}}$ are as follows:
\begin{myitemize}
\item[\ding{172}] First, for any $a \in \{1,\ldots, \Delta-1\}$, the event that $\boldsymbol{F}_t$ at time $t$ is $HN^{\leq \Delta-1}HN^{a}$ can only result from that $\boldsymbol{F}_{t-1}$ at time $t-1$ is $HN^{\leq \Delta-1}HN^{a-1}$, which by a recursive argument can only result from that $\boldsymbol{F}_{t-a}$ at time $t-a$ is $HN^{\leq \Delta-1}H$. Moreover, moving from $\boldsymbol{F}_{t-a} = HN^{\leq \Delta-1}H$ to $\boldsymbol{F}_t = HN^{\leq \Delta-1}HN^{a}$ requires that $S_i$ for each $i \in \{ t-a+1, \ldots, t\}$ is $N$.
\item[\ding{173}] Second, for any $b \in \{0,\ldots, \Delta-1\}$, the event that $\boldsymbol{F}_t$ at time $t$ is $HN^{\geq \Delta}HN^{b}$ can only result from that $\boldsymbol{F}_{t-1}$ at time $t-1$ is $HN^{\geq \Delta}HN^{b-1}$, which by a recursive argument can only result from that $\boldsymbol{F}_{t-b}$ at time $t-b$ is $HN^{\geq \Delta}H$, and also $\boldsymbol{F}_{t-b-1}$ at time $t-b-1$ is $HN^{\geq \Delta}$. Moreover, moving from $\boldsymbol{F}_{t-b-1} = HN^{\geq \Delta}$ to $\boldsymbol{F}_t = HN^{\geq \Delta}HN^{b}$ requires that $S_{t-a}$ is $H$, and   $S_i$ for each $i \in \{ t-a+1, \ldots, t\}$ is $N$.
\item[\ding{174}] Third, the event that $\boldsymbol{F}_t$ at time $t$ is $HN^{\leq \Delta-1}H$ can result from the combination of the following two events: i) $\boldsymbol{F}_{t-1}$ at time $t-1$ is $HN^{\leq \Delta-1}H$ or $HN^{\leq \Delta-1}HN^{a}$ for $a \in \{1,\ldots, \Delta-1\}$ or $HN^{\geq \Delta}HN^{b}$ for $b \in \{0,\ldots, \Delta-1\}$; and ii) $S_t$ is $H$.
\item[\ding{175}] Fourth, the event that $\boldsymbol{F}_t$ at time $t$ is $HN^{\geq \Delta}$ can result from the combination of the following two events: i) $\boldsymbol{F}_{t-1}$ at time $t-1$ is $HN^{\geq \Delta}$ or $HN^{\leq \Delta-1}HN^{\Delta-1}$ or $HN^{\geq \Delta}HN^{\Delta-1}$; and ii) $S_t$ is $N$. 
\end{myitemize}

% $\boldsymbol{f}_t \in \text{Suffix-Set}$
In Appendix~\ref{app-CF-stationary} of the online full version~\cite{fullpaper},
we derive the stationary distribution of the suffix-of-previous-and-current-states Markov chain $\mathcal{C}_{\boldsymbol{F}}$ as follows: 
\begin{subnumcases}{}
\pi_{\boldsymbol{F}}(HN^{\leq \Delta-1}H) = \alpha\cdot  (1-{\overline{\alpha}}^{\Delta}), \label{eq-pi-F-1}  \\ \pi_{\boldsymbol{F}}(HN^{\leq \Delta-1}HN^{a}) = \alpha\cdot  (1-{\overline{\alpha}}^{\Delta}) \cdot  {\overline{\alpha}}^{a}, \label{eq-pi-F-2} \\ ~~~~~~~~~~~~~~~~~~\forall a \in \{1,\ldots, \Delta-1\}, \nonumber  \\ \pi_{\boldsymbol{F}}(HN^{\geq \Delta}) =  {\overline{\alpha}}^{\Delta},  \label{eq-pi-F-3}  \\ \pi_{\boldsymbol{F}}(HN^{\geq \Delta}HN^{b}) =   \alpha \cdot  {\overline{\alpha}}^{\Delta+b}, \label{eq-pi-F-4} \\ ~~~~~~~~~~~~~~~~~~\forall b \in \{0,\ldots, \Delta-1\}. \nonumber
\end{subnumcases}

% $\alpha$ and $\Delta$ together completely specify the Markov chain $\mathcal{C}_{\boldsymbol{F}||\boldsymbol{P}}$ (note that given $\alpha$, $\overline{\alpha}  : = 1 - \alpha$ is also given).

% \begin{subnumcases}{a}
% b\\
% c
% \end{subnumcases}

% \begin{subnumcases}{a}
% b\\
% c
% \end{subnumcases}

We now use the suffix-of-previous-and-current-states Markov chain $\mathcal{C}_{\boldsymbol{F}}$ to construct another Markov chain. For notational purpose, we let $\boldsymbol{P}$ stand for $S_{t-\Delta} \ldots S_{t}$, which are  states in the  previous $\Delta$ rounds and the  state in the current round $t$.  Then we consider a Markov chain to represent the transition of $\boldsymbol{F}_{t-\Delta-1} S_{t-\Delta} \ldots S_{t}$, and denote this Markov chain by $\mathcal{C}_{\boldsymbol{F}||\boldsymbol{P}}$, where ``$||$'' intuitively means concatenation. The random variable $\boldsymbol{F}_{t-\Delta-1}$ represents the suffix of the states in all rounds up to round $t-\Delta-1$, so $\boldsymbol{F}_{t-\Delta-1} S_{t-\Delta} \ldots S_{t}$ means the concatenation of i) the suffix of previous states before the $\Delta$ to last state, ii) the previous $\Delta$ states, and iii) the current state. We can see that the Markov chain $\mathcal{C}_{\boldsymbol{F}||\boldsymbol{P}}$ is \textit{time-homogeneous}, \textit{irreducible}, and \textit{ergodic}.

As it will become clear, to analyze $S_{t}$ of $\boldsymbol{F}_{t-\Delta-1} S_{t-\Delta} \ldots S_{t}$, knowing whether $S_{t}$ is $H$ or $N$ is not enough, and we need to know   the exact number of blocks mined by the honest nodes at round $t$ in the case of $S_{t}$ being $H$ (i.e., when at least one block is mined by the honest nodes at round $t$). To this end, we let $H_h$ be the event that the honest nodes mine $h$ number of block at round $t$. Then the values that $S_{t}$ can take is given by the following set:
\begin{align}
\text{Detailed-State-Set}   & := \{H_h,~N~|~1\leq h \leq \mu n\} . \label{eq-Detailed-State-Set}
\end{align} 
Clearly, the $H$ state comprises all $H_h$ states for \mbox{$1\leq h \leq \mu n$.}

% In each round, one of the following events will happen: i) $H$, which means that at least one block is mined by the benign (i.e., honest) nodes, and ii) $N$, which means that no block is mined by the benign nodes. By a round's state, we refer to whether $H$ or $N$ happens. Then we define \text{State-Set} to characterize the possible values that a round's state can take:

% Note that all $H_h$ states for $1\leq h \leq \mu n$ in Eq.~(\ref{eq-Detailed-State-Set}) are considered as the $H$ state in Eq.~(\ref{eq-State-Set}).
 
Below we analyze the stationary distribution of the Markov chain  $\mathcal{C}_{\boldsymbol{F}||\boldsymbol{P}}$.
For $\boldsymbol{f} \in \text{Suffix-Set}$, $s^{(1)} \in \text{Detailed-State-Set}$, $\ldots$, $ s^{(\Delta+1)} \in \text{Detailed-State-Set}$, we let $\pi_{\boldsymbol{F}||\boldsymbol{P}}(\boldsymbol{f} s^{(1)} \ldots s^{(\Delta+1)})$ be the stationary probability of vertex $\boldsymbol{f} s^{(1)} \ldots s^{(\Delta+1)}$,  where \mbox{Suffix-Set} and \mbox{Detailed-State-Set} are given by Eq.~(\ref{eq-Detailed-State-Set}) and~(\ref{eq-Suffix-Set}); i.e., 
\begin{align}
 & \pi_{\boldsymbol{F}||\boldsymbol{P}}(\boldsymbol{f} s^{(1)} \ldots s^{(\Delta+1)}) \nonumber  \\ & =  \lim_{t\to \infty} \pr{\boldsymbol{F}_{t-\Delta-1} S_{t-\Delta} \ldots S_{t} = \boldsymbol{f} s^{(1)} \ldots s^{(\Delta+1)}} .\label{eq-G-t-stationary}
\end{align}

% $\boldsymbol{F}_t S_{t+1} \ldots S_{t+\Delta+1}$

% $\boldsymbol{F}_{t-\Delta-1} S_{t-\Delta} \ldots S_{t} = \boldsymbol{f} s^{(1)} \ldots s^{(\Delta+1)}$

% $S_{t+1} = s^{(\Delta+2)}$ 

% $\boldsymbol{F}_{t-\Delta} S_{t-\Delta+1} \ldots S_{t+1} = \text{suffix}(\boldsymbol{f} s^{(1)})s^{(2)} \ldots s^{(\Delta+1)} s^{(\Delta+2)}$

% $\bp{\boldsymbol{F}_{t-\Delta} S_{t-\Delta+1} \ldots S_{t+1} = \text{suffix}(\boldsymbol{f} s^{(1)})s^{(2)}  \ldots s^{(\Delta+1)} s^{(\Delta+2)}} = \bp{S_{t+1} = s^{(\Delta+2)}}\bp{\boldsymbol{F}_{t-\Delta-1} S_{t-\Delta} \ldots S_{t} = \boldsymbol{f} s^{(1)} \ldots s^{(\Delta+1)}} $

% $\text{suffix}()$

% the stationary distribution  $\pi_{\boldsymbol{F}||\boldsymbol{P}}$ of the Markov chain $\mathcal{C}_{\boldsymbol{F}||\boldsymbol{P}}$

Since $\pr{\boldsymbol{F}_{t-\Delta-1} S_{t-\Delta} \ldots S_{t} = \boldsymbol{f} s^{(1)} \ldots s^{(\Delta+1)}}$ equals $\pr{\boldsymbol{F}_{t-\Delta-1} = \boldsymbol{f}} \prod_{i=1}^{\Delta+1} \bp{S_{t-\Delta-1+i} = s^{(i)}}$, we obtain from Eq.~(\ref{eq-F-t-stationary}) and~(\ref{eq-G-t-stationary}) that
% \begin{align}
% & \pi_{\boldsymbol{F}||\boldsymbol{P}}(\boldsymbol{f}_{t-\Delta-1} s_{t-\Delta} \ldots s_{t})  = \pi_{\boldsymbol{F}}(\boldsymbol{f}_{t-\Delta-1})\prod_{i=t-\Delta}^t \pr{S_{i} = s_{i} } . 
% \end{align}
\begin{align}
 & \textstyle{\pi_{\boldsymbol{F}||\boldsymbol{P}}( \boldsymbol{f} s^{(1)} \ldots s^{(\Delta+1)})   =  \pi_{\boldsymbol{F}}(\boldsymbol{f}) \prod_{i=1}^{\Delta+1} \bp{s^{(i)}} .}\label{eq-F-t-to-G-t}
\end{align}
We can also prove Eq.~(\ref{eq-F-t-to-G-t}) by analyzing the Markov chain  $\mathcal{C}_{\boldsymbol{F}||\boldsymbol{P}}$ directly. A proof is deferred to  Appendix~\ref{Appendix-eq-F-t-to-G-t} of the online full version~\cite{fullpaper}.

% \begin{subnumcases}{\bp{s^{(i)}}=}
% \alpha, & if $s^{(i)} = H$,\\
% {\overline{\alpha}}, & if $s^{(i)} = N$.
% \end{subnumcases}

From Eq.~(\ref{eq-F-t-to-G-t}), we can compute the stationary distribution  $\pi_{\boldsymbol{F}||\boldsymbol{P}}$ of the Markov chain $\mathcal{C}_{\boldsymbol{F}||\boldsymbol{P}}$ using expressions of $\pi_{\boldsymbol{F}}$ in
Eq.~(\ref{eq-pi-F-1})--(\ref{eq-pi-F-4}) and the following Eq.~(\ref{eq-s-i-HN}):
\begin{align}
\bp{s^{(i)}}= \begin{cases}
\binom{\mu n}{h} p^h (1-p)^{\mu n - h} , & \text{if $s^{(i)} = H_h$,}   \\
&\hspace{-80pt}\text{for each $h$ satisfying $1\leq h \leq \mu n$,} \\
{\overline{\alpha}}, \text{ for ${\overline{\alpha}}=(1-p)^{\mu n}$}, & \text{if $s^{(i)} = N$}.
\end{cases}  \label{eq-s-i-HN}
\end{align}
Eq.~(\ref{eq-s-i-HN}) follows from the result that since each honest node mines a block independently with probability $p$ in a round, the number of blocks mined by the $\mu n$ honest nodes in each round follows $\text{binom}( \mu n, p)$, which denotes a binomial distribution with $ \mu n$ being the number of trials and $p$ being the success probability for each trial.

We now explain that when we have $\left( \boldsymbol{f} = HN^{\geq \Delta}\right) \land \left( s^{(1)} = H_1\right) \land \left( s^{(2)} = \ldots = s^{(\Delta+1)} = N\right)$, the $\boldsymbol{F}||\boldsymbol{P}$ state $\boldsymbol{f} s^{(1)} \ldots s^{(\Delta+1)}$, which we write as $HN^{\geq \Delta} || H_1 N^{\Delta}$ for notational simplicity, represents a convergence opportunity. Specifically, the pattern of $HN^{\geq \Delta} || H_1 N^{\Delta}$ means the following consecutive events:
\begin{myitemize}
\item[i)]  a benign node mines a block in a round,
\item[ii)]  at least $\Delta$ rounds pass in which no benign node mines a block, which means that at the end of the $\Delta$ rounds, all benign nodes know all benign blocks and hence agree on the maximum length of the chain (they may not agree on the same chain),
\item[iii)]  a benign node mines a block $\mathcal{B}$ in a new round and thus extends a chain by one more block than the longest chain of the previous round, and
\item[iv)] $\Delta$ rounds pass in which no benign node mines a block. Thus, at the end, all honest miners know the new
block $\mathcal{B}$ and agree on the single longest chain as the one having $\mathcal{B}$.
\end{myitemize} 
% In such case, we write $\boldsymbol{f} s^{(1)} \ldots s^{(\Delta+1)}$ as $HN^{\geq \Delta} || H_1 N^{\Delta}$ for notational simplicity.
 Then we compute the stationary probability of the $\boldsymbol{F}||\boldsymbol{P}$ state $\boldsymbol{f} s^{(1)} \ldots s^{(\Delta+1)}$ state being $HN^{\geq \Delta} || H_1 N^{\Delta}$ as follows by using Eq.~(\ref{eq-F-t-to-G-t}):
\begin{align}
 & \pi_{\boldsymbol{F}||\boldsymbol{P}}(HN^{\geq \Delta} || H_1 N^{\Delta})     =  \pi_{\boldsymbol{F}}(HN^{\geq \Delta})  \bp{H_1} \left( \bp{N}\right)^{\Delta} . \label{a}
\end{align}
From Eq.~(\ref{eq-s-i-HN}), it holds that
\begin{align}
 & \bp{H_1}   = \alpha_1 \text{ for $\alpha_1 := p \mu n  \times (1-p)^{\mu n - 1}$}. \label{eq-H-1}
\end{align} 
From Eq.~(\ref{eq-pi-F-3}) and Eq.~(\ref{eq-H-1}), we obtain
\begin{align}
\pi_{\boldsymbol{F}||\boldsymbol{P}}(HN^{\geq \Delta} || H_1 N^{\Delta})   &   = {\overline{\alpha}}^{\Delta} \cdot  \alpha_1 \cdot {\overline{\alpha}}^{\Delta}  =  {\overline{\alpha}}^{2\Delta} \alpha_1 . \label{eq-F-P-alpha}
\end{align}

We define $f_t$ as the indicator function that the visited vertex at time $t$ is the state $HN^{\geq \Delta} || H_1 N^{\Delta}$. 
%  Then from ?, $f_t$ is a binary variable which takes $1$ with probability ${\overline{\alpha}}^{2\Delta} \alpha_1 $ and  $0$ with probability $1 - {\overline{\alpha}}^{2\Delta} \alpha_1 $.
 For the $T$-step random walk on the Markov chain $\mathcal{C}_{\boldsymbol{F}||\boldsymbol{P}}$ in the $T$ rounds from round $t_0$ to $t_0+T-1$, let the visited vertices be $V_{t_0}, \ldots, V_{t_0+T-1}$. Then from Eq.~(\ref{eq-F-P-alpha}), we have that for $t\in\{t_0,\ldots,t_0+T-1\}$:
\begin{myitemize}
\item $f_t(V_t)$ equals $1$ if $V_t$ is the state $HN^{\geq \Delta} || H_1 N^{\Delta}$, which happens with probability ${\overline{\alpha}}^{2\Delta} \alpha_1 $;
\item $f_t(V_t)$ equals $0$ if $V_t$ is not the state $HN^{\geq \Delta} || H_1 N^{\Delta}$, which happens with probability $1-{\overline{\alpha}}^{2\Delta} \alpha_1 $.
\end{myitemize}

Then the expectation of the binary variable $f_t(V_t)$ is
\begin{align}
\bE{f_t(V_t)} = {\overline{\alpha}}^{2\Delta} \alpha_1  . \label{ftVt-Exp}
\end{align} 
With $C(t_0, t_0+T-1)$ being the number of times that $HN^{\geq \Delta} || H_1 N^{\Delta}$ is visited (i.e., the number of convergence opportunities)  from round $t_0$ to $t_0+T-1$,   we have
\begin{align}
\textstyle{C(t_0, t_0+T-1) = \sum_{t=t_0}^{t_0+T-1} f_t(V_t) . }\label{a}
\end{align}

From the above, the random variables $f_t(V_t)|_{t=t_0}^{t_0+T-1}$ are identically distributed, but are not independent. Since the linearity of expectation holds regardless of whether the random variables are independent, we use Eq.~(\ref{ftVt-Exp}) to obtain
\begin{align}
\textstyle{\bE{C(t_0, t_0+T-1)} = \sum_{t=t_0}^{t_0+T-1} \bE{f_t(V_t)} = T \hspace{1pt} {\overline{\alpha}}^{2\Delta} \alpha_1 ; }\nonumber
\end{align}
i.e., Eq.~(\ref{eq-Expectation-C-t-0-t-0-T-1-simple}) is proved. 
Using Eq.~(\ref{eq-Expectation-C-t-0-t-0-T-1-simple}) and Eq.~(\ref{eq-Expectation-A-t-0-t-0-T-1-simple}) which we have both shown, we know that  Inequality~(\ref{eq-thm-alpha}) is the same as Inequality~(\ref{eq-thm-alpha-repeat}).

% In the next subsection, we will show that \mbox{$C(t_0, t_0+T-1)$}  is in fact concentrated around its expectation with high probability.

% $C(t_0, t_0+T-1)$ has an expectation of

% $A(t_0, t_0+T-1)  $ $\text{binom}(T \nu n, p)$

\subsection{Putting things together to prove Theorem~\ref{thm-alpha}} \label{sec-thm-1-Putting-things-together}

We have proved in Section~\ref{sec-eq-E-C-E-A} that Inequality~(\ref{eq-thm-alpha}) as a condition of Theorem~\ref{thm-alpha} is the same as Inequality~(\ref{eq-thm-alpha-repeat}). Also, Eq.~(\ref{eq-Expectation-A-t-0-t-0-T-1-simple}) in Section~\ref{sec-eq-E-C-E-A} shows that $A(t_0, t_0+T-1) $ can be written as $\Omega(T)$. We prove  Inequalities~(\ref{prop-event-1-prob-eq-simple}) and~(\ref{prop-event-2-prob-eq-simple}) in Appendices~\ref{sec-prop-event-1-prob-eq-simple} and \ref{sec-prop-event-2-prob-eq-simple} of the online full version~\cite{fullpaper} (where~\cite{chung2012chernoff,arratia1989tutorial,zhao2015k,taylor1952hospital} are cited). Then  we combine~(\ref{eq-delta-2-delta-3v5}) and~(\ref{eq-thm-alpha-repeat})~(\ref{prop-event-1-prob-eq-simple})~(\ref{prop-event-2-prob-eq-simple}) with~$A(t_0, t_0+T-1)=\Omega(T)$ to complete proving Theorem~\ref{thm-alpha}.

~ \qeda

 \section{Conclusion} \label{sec-Conclusion}

In this paper, we analyze the consistency  of Nakamoto's blockchain protocol. Let $\mu$ (resp., $\nu$) be the fraction of computational power controlled by benign miners (resp., the adversary), where $\mu + \nu = 1$. With $c$ denoting the expected number of network delays before some block is mined, we prove for the first time that to ensure the consistency property of Nakamoto's blockchain protocol in an asynchronous network, it suffices to have $c$ to be just slightly greater than   $\frac{2\mu}{\ln (\mu/\nu)}$. This expression is both neater and stronger than existing ones. In the proof, we formulate novel Markov chains which characterize the numbers of mined blocks in different rounds.

 \let\OLDthebibliography\thebibliography
\renewcommand\thebibliography[1]{
  \OLDthebibliography{#1}
  \setlength{\parskip}{1pt}
  \setlength{\itemsep}{0pt plus 1pt}
}


% \bibliographystyle{IEEEtran}
% \bibliography{related}

\begin{thebibliography}{10}
\providecommand{\url}[1]{#1}
\csname url@samestyle\endcsname
\providecommand{\newblock}{\relax}
\providecommand{\bibinfo}[2]{#2}
\providecommand{\BIBentrySTDinterwordspacing}{\spaceskip=0pt\relax}
\providecommand{\BIBentryALTinterwordstretchfactor}{4}
\providecommand{\BIBentryALTinterwordspacing}{\spaceskip=\fontdimen2\font plus
\BIBentryALTinterwordstretchfactor\fontdimen3\font minus
  \fontdimen4\font\relax}
\providecommand{\BIBforeignlanguage}[2]{{%
\expandafter\ifx\csname l@#1\endcsname\relax
\typeout{** WARNING: IEEEtran.bst: No hyphenation pattern has been}%
\typeout{** loaded for the language `#1'. Using the pattern for}%
\typeout{** the default language instead.}%
\else
\language=\csname l@#1\endcsname
\fi
#2}}
\providecommand{\BIBdecl}{\relax}
\BIBdecl

\bibitem{nakamoto2008bitcoin}
\BIBentryALTinterwordspacing
S.~Nakamoto, ``Bitcoin: A peer-to-peer electronic cash system,'' 2008.
  [Online]. Available: \url{https://bitcoin.org/bitcoin.pdf}
\BIBentrySTDinterwordspacing

\bibitem{garay2015bitcoin}
J.~Garay, A.~Kiayias, and N.~Leonardos, ``The bitcoin backbone protocol:
  {Analysis} and applications,'' in \emph{Annual International Conference on
  the Theory and Applications of Cryptographic Techniques (EUROCRYPT)}, 2015,
  pp. 281--310.

\bibitem{pass2017analysis}
R.~Pass, L.~Seeman, and A.~Shelat, ``Analysis of the blockchain protocol in
  asynchronous networks,'' in \emph{Annual International Conference on the
  Theory and Applications of Cryptographic Techniques (EUROCRYPT)}, 2017, pp.
  643--673.

\bibitem{pass2017fruitchains}
R.~Pass and E.~Shi, ``Fruitchains: {A} fair blockchain,'' in \emph{ACM
  Symposium on Principles of Distributed Computing (PODC)}, 2017, pp. 315--324.

\bibitem{shi2018analysis}
E.~Shi, ``Analysis of deterministic longest-chain protocols,'' \emph{IACR
  Cryptology ePrint Archive}, vol. 2018, p. 1079, 2018.

\bibitem{kiffer2018better}
L.~Kiffer, R.~Rajaraman, and A.~Shelat, ``A better method to analyze blockchain
  consistency,'' in \emph{ACM SIGSAC Conference on Computer and Communications
  Security (CCS)}, 2018, pp. 729--744.

\bibitem{fullpaper}
J.~Zhao, J.~Tang, Z.~Li, H.~Wang, K.-Y. Lam, and K.~Xue, ``An analysis of
  blockchain consistency in asynchronous networks: Deriving a neat bound,''
  2020, full version of this \mbox{paper}. Available online at \\
  \url{http://www.ntu.edu.sg/home/junzhao/BlockchainConsistency.pdf}.

\bibitem{kiayias2015speed}
A.~Kiayias and G.~Panagiotakos, ``Speed-security tradeoffs in blockchain
  protocols,'' \emph{IACR Cryptology ePrint Archive}, vol. 2015, p. 1019, 2015.

\bibitem{garay2017bitcoin}
J.~Garay, A.~Kiayias, and N.~Leonardos, ``The bitcoin backbone protocol with
  chains of variable difficulty,'' in \emph{Annual International Cryptology
  Conference (CRYPTO)}, 2017, pp. 291--323.

\bibitem{zhang2019lay}
R.~Zhang and B.~Preneel, ``Lay down the common metrics: {Evaluating}
  proof-of-work consensus protocols' security,'' in \emph{IEEE Symposium on
  Security and Privacy (SP)}, 2019.

\bibitem{kiayias2017ouroboros}
A.~Kiayias, A.~Russell, B.~David, and R.~Oliynykov, ``Ouroboros: {A} provably
  secure proof-of-stake blockchain protocol,'' in \emph{Annual International
  Cryptology Conference (CRYPTO)}, 2017, pp. 357--388.

\bibitem{david2018ouroboros}
B.~David, P.~Ga{\v{z}}i, A.~Kiayias, and A.~Russell, ``Ouroboros {Praos}: {An}
  adaptively-secure, semi-synchronous proof-of-stake blockchain,'' in
  \emph{Annual International Conference on the Theory and Applications of
  Cryptographic Techniques (EUROCRYPT)}, 2018, pp. 66--98.

\bibitem{badertscher2018ouroboros}
C.~Badertscher, P.~Ga{\v{z}}i, A.~Kiayias, A.~Russell, and V.~Zikas,
  ``Ouroboros genesis: {Composable} proof-of-stake blockchains with dynamic
  availability,'' in \emph{ACM SIGSAC Conference on Computer and Communications
  Security (CCS)}, 2018, pp. 913--930.

\bibitem{li2017securing}
W.~Li, S.~Andreina, J.-M. Bohli, and G.~Karame, ``Securing proof-of-stake
  blockchain protocols,'' in \emph{Data Privacy Management, Cryptocurrencies
  and Blockchain Technology}.\hskip 1em plus 0.5em minus 0.4em\relax Springer,
  2017, pp. 297--315.

\bibitem{gilad2017algorand}
Y.~Gilad, R.~Hemo, S.~Micali, G.~Vlachos, and N.~Zeldovich, ``Algorand:
  {Scaling} byzantine agreements for cryptocurrencies,'' in \emph{Proceedings
  of the 26th Symposium on Operating Systems Principles (SOSP)}, 2017, pp.
  51--68.

\bibitem{castro1999practical}
M.~Castro, B.~Liskov \emph{et~al.}, ``Practical byzantine fault tolerance,'' in
  \emph{USENIX Symposium on Operating Systems Design and Implementation
  (OSDI)}, vol.~99, no. 1999, 1999, pp. 173--186.

\bibitem{wang2019survey}
W.~Wang, D.~T. Hoang, P.~Hu, Z.~Xiong, D.~Niyato, P.~Wang, Y.~Wen, and D.~I.
  Kim, ``A survey on consensus mechanisms and mining strategy management in
  blockchain networks,'' \emph{IEEE Access}, vol.~7, pp. 22\,328--22\,370,
  2019.

\bibitem{li2017survey}
X.~Li, P.~Jiang, T.~Chen, X.~Luo, and Q.~Wen, ``A survey on the security of
  blockchain systems,'' \emph{Future Generation Computer Systems}, vol.~8, p.
  274, 2017.

\bibitem{levin2017markov}
D.~A. Levin and Y.~Peres, \emph{Markov chains and mixing times}.\hskip 1em plus
  0.5em minus 0.4em\relax American Mathematical Soc., 2017, vol. 107.

\bibitem{chung2012chernoff}
K.-M. Chung, H.~Lam, Z.~Liu, and M.~Mitzenmacher, ``{Chernoff--Hoeffding}
  bounds for {Markov} chains: {Generalized} and simplified,'' \emph{arXiv
  preprint arXiv:1201.0559}, 2012.

\bibitem{arratia1989tutorial}
R.~Arratia and L.~Gordon, ``Tutorial on large deviations for the binomial
  distribution,'' \emph{Bulletin of mathematical biology}, vol.~51, no.~1, pp.
  125--131, 1989.

\bibitem{zhao2015k}
J.~Zhao, O.~Ya{\u{g}}an, and V.~Gligor, ``$ k $-connectivity in random key
  graphs with unreliable links,'' \emph{IEEE Transactions on Information
  Theory}, vol.~61, no.~7, pp. 3810--3836, 2015.

\bibitem{taylor1952hospital}
A.~E. Taylor, ``L'hospital's rule,'' \emph{The American Mathematical Monthly},
  vol.~59, no.~1, pp. 20--24, 1952.

\end{thebibliography}

\newpage

\appendix

\subsection{Proof of Claim~\ref{thm-kiffer2018better-Theorem-4.4a}}
\label{app-thm-kiffer2018better-Theorem-4.4a}

We first present Claim~\ref{thm-kiffer2018better-Theorem-4.4} on Page~\pageref{thm-alpha}, which is Theorem~4.4 on Page~8 of~\cite{kiffer2018better} after we correct \mbox{$\mu \cdot p$} to $\alpha$ in many places of~\cite{kiffer2018better}. Then we   perform some computations to show Claim~\ref{thm-kiffer2018better-Theorem-4.4a} using Claim~\ref{thm-kiffer2018better-Theorem-4.4}.

\begin{Claima}[Theorem~4.4 of~\cite{kiffer2018better} after we correct \mbox{$\mu \cdot p$} to $\alpha$ in many places of~\cite{kiffer2018better}] \label{thm-kiffer2018better-Theorem-4.4} With some notation defined below, Nakamoto's blockchain protocol satisfies consistency if there exists a positive constant $\delta_3$ such that
\begin{align}
\frac{P_{\Delta}^2}{\sum_{i,j\in\{0,1\}}P_{ij}\pi_i \ell_{ij}}  \geq (1+\delta_3) \beta, \textup{ for $\beta:=p \nu n$} , \label{eq-thm-alpha-delta_3b}
\end{align}
where
\begin{itemize}
\item[1):] $\beta$ denotes the expected number of blocks mined in each round by the adversary controlling $\nu n$ miners;
\item[2):] $P_{\Delta}$ (defined on Page~6 of~\cite{kiffer2018better}) denotes the probability of $\Delta$ silent rounds (after we correct \mbox{$\mu \cdot p$} to $\alpha$ on Page~6 of~\cite{kiffer2018better}, it holds that $P_{\Delta} = (1-\alpha)^{\Delta} = {\overline{\alpha}}^{\Delta}$ from the definitions of $\alpha$ and $\overline{\alpha}$ in Eq.~(\ref{eq-define-alpha}) and Eq.~(\ref{eq-define-alpha-bar}));
\item[3):] $\pi_{0}$ and $\pi_{1}$  (denoting stationary probabilities of states $S_{0}$ and $S_{1}$ in the Markov chain \mbox{$ \rotatecurvearrowright \hspace{-2pt} S_{0} \rightleftarrows S_{1} \hspace{-2pt} \rotatecurvearrowleft$} on Page~6 of~\cite{kiffer2018better}) are given as follows:
\begin{itemize}
\item[3a):] $\pi_{0}$ denotes the stationary probability of a ``messy'' state $S_{0}$ where honest mined blocks occur in less than $\Delta$ rounds from one another ($\pi_{0}=1-P_{\Delta}$ from Page~7 of~\cite{kiffer2018better});
\item[3b):] $\pi_{1}$ denotes the stationary probability of the state $S_{1}$ where quiet periods between honest
mined blocks is at least $\Delta$ rounds ($\pi_{1}=P_{\Delta}$ from Page~7 of~\cite{kiffer2018better});
\end{itemize}
\item[4):] $P_{ij}$ for $i,j\in\{0,1\}$ denotes the probability of event $e_{ij}$, which represents the transition from state $S_i$ to state $S_j$; more specifically, 
% s $P_{00}$, $P_{01}$, $P_{11}$, and $P_{10}$  (denoting probabilities of events $e_{00}$, $e_{01}$, $e_{11}$, and $e_{10}$ on Page~6 of~\cite{kiffer2018better}) are given as follows:
\begin{itemize}
\item[4a):] $P_{00}$ denotes the probability of $e_{00}$, meaning one quiet period of less than $\Delta$ rounds, followed by a round with at least one block mined by honest players\footnote{Note the phrase ``a round with at least one block mined by honest players'' in the definitions of $e_{00}$, $e_{11}$, and $e_{10}$  (and hence $P_{00}$, $P_{11}$, and $P_{10}$) of Theorem~\ref{thm-kiffer2018better-Theorem-4.4}. On Page~6 of~\cite{kiffer2018better}, actually the phrase ``single honest mined block'' is used. However, for $e_{ij}$ to exactly mean the transition from state $S_i$ to state $S_j$ for $i,j\in\{0,1\}$ (defined in ``3a)'' and ``3b)'' of the list in Theorem~\ref{thm-kiffer2018better-Theorem-4.4}), there is no reason for requiring ``single honest mined block''. \label{footnote-single-vs-at-least-one}} ($P_{00}=1-P_{\Delta}$ from Page~7 of~\cite{kiffer2018better});
\item[4b):] $P_{01}$ denotes the probability of $e_{01}$, meaning one quiet period that is at least $\Delta$ rounds ($P_{01}=P_{\Delta}$ from Page~7 of~\cite{kiffer2018better});
\item[4c):] $P_{11}$ denotes the probability of $e_{11}$, meaning a single honest mined block$^{\textup{\ref{footnote-single-vs-at-least-one}}}$, followed by a quiet period of at least $\Delta$ rounds ($P_{11}=P_{\Delta}$ from Page~7 of~\cite{kiffer2018better});
\item[4d):] $P_{10}$ denotes the probability of $e_{10}$, meaning a round with at least one block mined by honest players, followed by one quiet period of less than $\Delta$ rounds, followed by a round with at least one block mined by honest players ($P_{10}=1-P_{\Delta}$ from Page~7 of~\cite{kiffer2018better});
\end{itemize}
\item[5):] $\ell_{ij}$ for $i,j\in\{0,1\}$ denoting the expected time spent on the edge $S_i \to S_j$ in the Markov chain \mbox{$ \rotatecurvearrowright \hspace{-2pt} S_{0} \rightleftarrows S_{1} \hspace{-2pt} \rotatecurvearrowleft$} on Page~6 of~\cite{kiffer2018better}:
\begin{itemize}
\item \mbox{With $p_{i | \leq \Delta}$ denoting $\bp{ \textup{hit at time $i$}\,|\,\textup{silience lasted} \leq \Delta }$} (after we correct \mbox{$\mu \cdot p$} to $\alpha$ on Page~7 of~\cite{kiffer2018better}, it holds that $p_{i | \leq \Delta} = \frac{(1-\alpha)^{i-1}\alpha}{\sum_{j=1}^{\Delta}(1-\alpha)^{j-1}\alpha} $), the expressions of $\ell_{00}$, $\ell_{01}$, $\ell_{11}$, and $\ell_{10}$ are as follows after we correct \mbox{$\mu \cdot p$} to $\alpha$ on Page~7 of~\cite{kiffer2018better}: $\ell_{00} = \sum_{i=1}^{\Delta} i p_{i | \leq \Delta}$, $\ell_{01} = \Delta$, $\ell_{11} = \frac{1}{\alpha} + \Delta$, and $\ell_{10} = \frac{1}{\alpha} + \sum_{i=1}^{\Delta} i p_{i | \leq \Delta}$.
\end{itemize}
\end{itemize}
% $\overline{\alpha}$ (resp.,~$\alpha_1$) denotes the probability that \textit{no} (resp.,~\textit{only one}) honest miner succeeds in solving a puzzle in one round, and is given by  Eq.~(\ref{eq-define-alpha-bar})  (resp.,~Eq.~(\ref{eq-define-alpha-1})).  
\end{Claima} 

We now use Claim~\ref{thm-kiffer2018better-Theorem-4.4} to show Claim~\ref{thm-kiffer2018better-Theorem-4.4a}. First, we use the expression of $p_{i | \leq \Delta}$ to compute $\ell_{00}$: 
\begin{align}
\ell_{00} &  = \sum_{i=1}^{\Delta} i p_{i | \leq \Delta}  \nonumber \\ & = \sum_{i=1}^{\Delta} \frac{i(1-\alpha)^{i-1}\alpha}{\sum_{j=1}^{\Delta}(1-\alpha)^{j-1}\alpha}  \nonumber \\ & = \frac{1}{\alpha} - \frac{\Delta(1-\alpha)^{\Delta}}{1-(1-\alpha)^{\Delta}} \nonumber \\ & = \frac{1}{\alpha} - \frac{\Delta P_{\Delta}}{1-P_{\Delta}}
 . \label{eq-p-Q}
\end{align}
Then we calculate the left-hand side in Inequality~(\ref{eq-thm-alpha-delta_3b}):
\begin{align}
& \frac{P_{\Delta}^2}{\sum_{i,j\in\{0,1\}}P_{ij}\pi_i \ell_{ij}}  \nonumber \\ & = \frac{P_{\Delta}^2}{P_{00}\pi_0 \ell_{00} + P_{01}\pi_0 \ell_{01} + P_{11}\pi_1 \ell_{11}+ P_{10}\pi_1 \ell_{10}} \nonumber \\ & = \frac{P_{\Delta}^2}{\left[\begin{array}{l}
  (1-P_{\Delta})(1-P_{\Delta}) (\frac{1}{\alpha} - \frac{\Delta P_{\Delta}}{1-P_{\Delta}}) + P_{\Delta}(1-P_{\Delta}) \Delta \\+ P_{\Delta}P_{\Delta} (\frac{1}{\alpha} + \Delta)+ (1-P_{\Delta})P_{\Delta} (\frac{2}{\alpha} - \frac{\Delta P_{\Delta}}{1-P_{\Delta}})
\end{array}\right]} \nonumber \\ & = P_{\Delta}^2 \alpha \nonumber \\ & = {\overline{\alpha}}^{2\Delta}  \alpha
 . \label{eq-p-Q}
\end{align}
Hence, Claim~\ref{thm-kiffer2018better-Theorem-4.4a} follows from Claim~\ref{thm-kiffer2018better-Theorem-4.4}.

% \begin{prop} \label{prop1}
% If $0< \nu < \mu$ and $p n  < \max\left\{ \frac{1}{\mu},~ \frac{\epsilon_1 \iffalse_0 \fi (\mu - \nu)}{\mu^2}  \right\}$ for a positive constant $\epsilon_1 \iffalse_0 \fi  < 2$, defining $\delta_5 \iffalse_7 \fi  : = \frac{1}{2} [\frac{2\mu}{ 2\mu - \epsilon_1 \iffalse_0 \fi (\mu - \nu)} + \frac{\mu}{\nu}] - 1 $ and $\delta_1 :  =
% % (1+\delta_5 \iffalse_7 \fi ) [1 - \frac{\epsilon_1 \iffalse_0 \fi (\mu - \nu)}{2\mu}] - 1   = \frac{1}{2} [\frac{2\mu}{ 2\mu - \epsilon_1 \iffalse_0 \fi (\mu - \nu)} + \frac{\mu}{\nu}] [1 - \frac{\epsilon_1 \iffalse_0 \fi (\mu - \nu)}{2\mu}] - 1  =
% \frac{\mu}{2\nu} [1 - \frac{\epsilon_1 \iffalse_0 \fi (\mu - \nu)}{2\mu}] - \frac{1}{2} $, we have $0<\delta_5 \iffalse_7 \fi < \frac{\mu-\nu}{\nu}$, $\delta_1 > 0$, and $\left(\frac{1+\delta_1}{1 - \frac{1}{2} p \mu n }\right)^{1/(2\Delta)} \leq  1+ \frac{\delta_5 \iffalse_7 \fi }{2\Delta} $.
% \end{prop}

\subsection{Deriving the stationary distribution of the suffix-of-previous-and-current-states Markov chain $\mathcal{C}_{\boldsymbol{F}}$}
\label{app-CF-stationary}

We now derive the stationary distribution of the suffix-of-previous-and-current-states Markov chain $\mathcal{C}_{\boldsymbol{F}}$. To this end, we first analyze the state transition in $\mathcal{C}_{\boldsymbol{F}}$. 

Let $s_t$ be $\boldsymbol{f}_t$'s state in round $t$. We define a function $\text{suffix}(\cdot)$ such that $\left( \boldsymbol{F}_{t-1} = \boldsymbol{f}_{t-1}\right)\land  \left(S_t = s_t \right)$ produces $\boldsymbol{F}_t = \text{suffix}(\boldsymbol{f}_{t-1}||s_t)$. Then we have 
\begin{align}
& \pr{\boldsymbol{F}_t = \boldsymbol{f}_t}  \nonumber  \\ & = \sum_{\begin{subarray}{l}
\boldsymbol{f}_{t-1}\in \text{Suffix-Set}:\\ \text{suffix}(\boldsymbol{f}_{t-1}||s_t)=\boldsymbol{f}_{t}  
\end{subarray}} \pr{\left( \boldsymbol{F}_{t-1} = \boldsymbol{f}_{t-1}\right)\land  \left(S_t = s_t \right) } \nonumber  \\ &= \sum_{\begin{subarray}{l}
\boldsymbol{f}_{t-1}\in \text{Suffix-Set}:\\ \text{suffix}(\boldsymbol{f}_{t-1}||s_t)=\boldsymbol{f}_{t}  
\end{subarray}}  \left( \pr{\boldsymbol{F}_{t-1} = \boldsymbol{f}_{t-1}} \pr{S_t = s_t} \right) ,\label{eq-F-t-F-t-1}
\end{align}
where the last step uses the independence between $\left( \boldsymbol{F}_{t-1} = \boldsymbol{f}_{t-1}\right) $ and $ \left(S_t = s_t \right)$.

% \subsection{Stationary distribution of the suffix-of-previous-and-current-states Markov chain $\mathcal{C}_{\boldsymbol{F}}$}

Based on Eq.~(\ref{eq-F-t-F-t-1}), we now set $\boldsymbol{f}_t$ as each vertex of Markov chain $\mathcal{C}_{\boldsymbol{F}}$ to obtain the specific transition rules.

\textbf{Case of $\boldsymbol{f}_t$ in Eq.~(\ref{eq-F-t-F-t-1}) being $HN^{\leq \Delta-1}HN^{a}$.}
We obtain from Eq.~(\ref{eq-F-t-F-t-1}) and the above result \ding{172} that  for any $a \in \{1,\ldots, \Delta-1\}$, 
\begin{align}
& \pr{\boldsymbol{F}_t = HN^{\leq \Delta-1}HN^{a}}  \nonumber  \\ & = \pr{\boldsymbol{F}_{t-a} = HN^{\leq \Delta-1}H}  \prod_{i=t-a+1}^{t} \pr{S_{i} = N} \nonumber  \\ &= \pr{\boldsymbol{F}_{t-a} = HN^{\leq \Delta-1}H} \cdot  {\overline{\alpha}}^{a} ,   \label{eq-F-t-a}
\end{align}
where the last step uses $\pr{S_{i} = N} = \overline{\alpha}$.

\textbf{Case of $\boldsymbol{f}_t$ in Eq.~(\ref{eq-F-t-F-t-1}) being $HN^{\geq \Delta}HN^{b}$.}
We obtain from Eq.~(\ref{eq-F-t-F-t-1}) and the above result \ding{173} that  for any $b \in \{0,\ldots, \Delta-1\}$, 
\begin{align}
& \pr{\boldsymbol{F}_t = HN^{\geq \Delta}HN^{b}}  \nonumber  \\ & = \pr{\boldsymbol{F}_{t-b-1} = HN^{\geq \Delta}} \pr{S_{t-b} = H} \prod_{i=t-b+1}^{t} \pr{S_{i} = N} \nonumber  \\ &= \pr{\boldsymbol{F}_{t-b-1} = HN^{\geq \Delta}} \cdot   \alpha \cdot  {\overline{\alpha}}^{b}, \label{eq-F-t-b}
\end{align}
where the last step uses $\pr{S_{t-b} = H} = \alpha$ and $\pr{S_{i} = N} = \overline{\alpha}$.

\textbf{Case of $\boldsymbol{f}_t$ in Eq.~(\ref{eq-F-t-F-t-1}) being $HN^{\leq \Delta-1}H$.}
We obtain from Eq.~(\ref{eq-F-t-F-t-1}) and the above result \ding{174} that 
\begin{align}
& \pr{\boldsymbol{F}_{t} = HN^{\leq \Delta-1}H} \nonumber  \\ & = \pr{S_{t} = H} \cdot \Bigg( \pr{\boldsymbol{F}_{t-1} = HN^{\leq \Delta-1}H} \nonumber  \\ & \quad \quad \quad \quad \quad \quad \quad +\sum_{a=1}^{\Delta-1} \pr{\boldsymbol{F}_{t-1} = HN^{\leq \Delta-1}HN^{a}}\nonumber  \\ & \quad \quad \quad \quad \quad \quad \quad + \sum_{b=0}^{\Delta-1} \pr{\boldsymbol{F}_{t-1} = HN^{\geq \Delta}HN^{b}}\Bigg) \nonumber  \\ & = \alpha \cdot  \Bigg( \pr{\boldsymbol{F}_{t-1} = HN^{\leq \Delta-1}H} \nonumber  \\ &  \quad \quad \quad \quad  + \sum_{a=1}^{\Delta-1} \pr{\boldsymbol{F}_{t-1} = HN^{\leq \Delta-1}HN^{a}}\nonumber  \\ &   \quad \quad \quad \quad + \sum_{b=0}^{\Delta-1} \pr{\boldsymbol{F}_{t-1} = HN^{\geq \Delta}HN^{b}}\Bigg), \label{eq-H-N-Delta-1-H}
\end{align}
where the last step uses $\pr{S_{t} = H} = \alpha$.

\textbf{Case of $\boldsymbol{f}_t$ in Eq.~(\ref{eq-F-t-F-t-1}) being $HN^{\geq \Delta}$.}
We obtain from Eq.~(\ref{eq-F-t-F-t-1}) and the above result \ding{175} that 
\begin{align}
& \pr{\boldsymbol{F}_{t} = HN^{\geq \Delta}} \nonumber  \\ & = \pr{S_{t} = N} \cdot \Bigg( \pr{\boldsymbol{F}_{t-1} = HN^{\geq \Delta}}\nonumber  \\ & \quad \quad \quad \quad \quad \quad \quad +\pr{\boldsymbol{F}_{t-1} = HN^{\leq \Delta-1}HN^{\Delta-1}}\nonumber  \\ & \quad \quad \quad \quad \quad \quad \quad + \pr{\boldsymbol{F}_{t-1} = HN^{\geq \Delta}HN^{\Delta-1}}\Bigg) \nonumber  \\ & = {\overline{\alpha}} \cdot  \Bigg( \pr{\boldsymbol{F}_{t-1} = HN^{\geq \Delta}}\nonumber  \\ &   \quad \quad \quad \quad + \pr{\boldsymbol{F}_{t-1} = HN^{\leq \Delta-1}HN^{\Delta-1}}\nonumber  \\ & \quad   \quad \quad \quad + \pr{\boldsymbol{F}_{t-1} = HN^{\geq \Delta}HN^{\Delta-1}}\Bigg), \label{eq-H-N-geq-Delta}
\end{align}
where the last step uses $\pr{S_{t} = N} = \overline{\alpha}$.

Below we analyze the stationary distribution of the Markov chain  $\mathcal{C}_{\boldsymbol{F}}$.
For $\boldsymbol{f} \in \text{Suffix-Set}$, we let $\pi_{\boldsymbol{F}}(\boldsymbol{f})$ be the stationary probability of vertex $\boldsymbol{f}$,  where \mbox{Suffix-Set} is given by Eq.~(\ref{eq-Suffix-Set}); i.e., 
\begin{align}
 & \pi_{\boldsymbol{F}}(\boldsymbol{f})   =  \lim_{t\to \infty} \pr{\boldsymbol{F}_{t} = \boldsymbol{f}} .\label{eq-F-t-stationary}
\end{align}

Summarizing Eq.~(\ref{eq-F-t-a})--(\ref{eq-H-N-geq-Delta}), to derive Markov chain $\mathcal{C}_{\boldsymbol{F}}$'s stationary distribution denoted by $\pi_{\boldsymbol{F}}$, we obtain 
\begin{subnumcases}{\hspace{-25pt}}
\hspace{-4pt}\pi_{\boldsymbol{F}}(HN^{\leq \Delta-1}HN^{a}) = \pi_{\boldsymbol{F}}(HN^{\leq \Delta-1}H) \cdot  {\overline{\alpha}}^{a},\label{eq-F-t-a-stationary}\\ ~~~~~~~~~~~~~~~~~~\forall a \in \{1,\ldots, \Delta-1\}, \nonumber \\ \hspace{-4pt}\pi_{\boldsymbol{F}}(HN^{\geq \Delta}HN^{b}) = \pi_{\boldsymbol{F}}(HN^{\geq \Delta}) \cdot   \alpha \cdot  {\overline{\alpha}}^{b},\label{eq-F-t-b-stationary}\\ ~~~~~~~~~~~~~~~~~~\forall b \in \{0,\ldots, \Delta-1\},   \nonumber \\ \hspace{-4pt}\pi_{\boldsymbol{F}}(HN^{\leq \Delta-1}H) = \alpha \cdot \Bigg( \pi_{\boldsymbol{F}}(HN^{\leq \Delta-1}H)  \nonumber \\ \hspace{0pt}+\hspace{-1pt} \sum_{a=1}^{\Delta-1} \pi_{\boldsymbol{F}}( HN^{\leq \Delta-1}HN^{a})  \hspace{-1pt}+ \hspace{-1pt}\sum_{b=0}^{\Delta-1} \pi_{\boldsymbol{F}}( HN^{\geq \Delta}HN^{b})\hspace{-1pt}\Bigg)\hspace{-1pt}, \label{eq-H-N-Delta-1-H-stationary} \\ \hspace{-4pt} \pi_{\boldsymbol{F}}( HN^{\geq \Delta}) = {\overline{\alpha}} \cdot \Bigg( \pi_{\boldsymbol{F}}( HN^{\geq \Delta}) \nonumber  \\   \hspace{0pt}+  \pi_{\boldsymbol{F}}( HN^{\leq \Delta-1}HN^{\Delta-1})+ \pi_{\boldsymbol{F}}( HN^{\geq \Delta}HN^{\Delta-1})\Bigg), \label{eq-H-N-geq-Delta-stationary} \\ \hspace{-4pt} \left[\hspace{-5pt} \begin{array}{l}  \pi_{\boldsymbol{F}}(HN^{\leq \Delta-1}H) \hspace{-2pt}+\hspace{-2pt} \sum_{a=1}^{\Delta-1} \pi_{\boldsymbol{F}}(HN^{\leq \Delta-1}HN^{a}) \\[3pt] + \pi_{\boldsymbol{F}}(HN^{\geq \Delta}) \hspace{-2pt}+\hspace{-2pt} \sum_{b=0}^{\Delta-1} \pi_{\boldsymbol{F}}(HN^{\geq \Delta}HN^{b})  \end{array} \hspace{-5pt}\right] \hspace{-2pt} = \hspace{-2pt} 1,\label{eq-F-stationary-sum-to-1}
\end{subnumcases}
% where Eq.~(\ref{eq-F-t-a-stationary})~(\ref{eq-F-t-b-stationary})~(\ref{eq-H-N-Delta-1-H-stationary}) and~(\ref{eq-H-N-geq-Delta-stationary}) are from Eq.~(\ref{eq-F-t-a})~(\ref{eq-F-t-b})~(\ref{eq-H-N-Delta-1-H}) and~(\ref{eq-H-N-geq-Delta}), respectively, and Eq.~(\ref{eq-F-stationary-sum-to-1}) simply means that the stationary probabilities of all the states sum to $1$. 
% From Eq.~(\ref{eq-F-t-a-stationary})~(\ref{eq-F-t-b-stationary})  and~(\ref{eq-F-stationary-sum-to-1}), we derive that 
where Eq.~(\ref{eq-F-t-a-stationary})--(\ref{eq-H-N-geq-Delta-stationary}) are from Eq.~(\ref{eq-F-t-a})--(\ref{eq-H-N-geq-Delta}), respectively, and Eq.~(\ref{eq-F-stationary-sum-to-1}) simply means that the stationary probabilities of all the states sum to $1$.

From Eq.~(\ref{eq-F-t-a-stationary})--(\ref{eq-F-stationary-sum-to-1}), we derive that 
\begin{subnumcases}{}
\pi_{\boldsymbol{F}}(HN^{\leq \Delta-1}H) = \alpha\cdot  (1-{\overline{\alpha}}^{\Delta}),    \\ \pi_{\boldsymbol{F}}(HN^{\leq \Delta-1}HN^{a}) = \alpha\cdot  (1-{\overline{\alpha}}^{\Delta}) \cdot  {\overline{\alpha}}^{a},  \\ ~~~~~~~~~~~~~~~~~~\forall a \in \{1,\ldots, \Delta-1\}, \nonumber  \\ \pi_{\boldsymbol{F}}(HN^{\geq \Delta}) =  {\overline{\alpha}}^{\Delta},     \\ \pi_{\boldsymbol{F}}(HN^{\geq \Delta}HN^{b}) =   \alpha \cdot  {\overline{\alpha}}^{\Delta+b},   \\ ~~~~~~~~~~~~~~~~~~\forall b \in \{0,\ldots, \Delta-1\}. \nonumber
\end{subnumcases}

\subsection{Proving Inequality~(\ref{prop-event-1-prob-eq-simple})} \label{sec-prop-event-1-prob-eq-simple}

Recall from the previous subsection that the $T$-step random walk on the Markov chain $\mathcal{C}_{\boldsymbol{F}||\boldsymbol{P}}$ in the $T$ rounds from round $t_0$ to $t_0+T-1$ visits vertices $V_{t_0}, \ldots, V_{t_0+T-1}$. Let $\phi$ be the initial distribution of the random walk; i.e., $\phi$ represents the distribution at round $t_0$. Also recall that the Markov chain $\mathcal{C}_{\boldsymbol{F}||\boldsymbol{P}}$ is \textit{time-homogeneous}, \textit{irreducible}, and \textit{ergodic}. Let $\tau(\epsilon, \alpha, \Delta)$ be the $\epsilon$-mixing time of $\mathcal{C}_{\boldsymbol{F}||\boldsymbol{P}}$, for $0<\epsilon \leq 1/8$. With $f_t(V_t)$ and $C(t_0, t_0+T-1)$ defined above, we use Theorem 3.1 of Reference~\cite{chung2012chernoff} on the {Chernoff--Hoeffding} bounds for {Markov} chains to obtain the existence of a positive constant $c$ independent of $T,n,p,\mu,\Delta$ such that
\begin{align}
  & \bp{C(t_0, t_0+T-1) \leq (1-\delta_2) \cdot \bE{C(t_0, t_0+T-1)}} \nonumber  \\   &  \leq c \|\phi\|_{\pi} \exp\left(-\frac{{\delta_2}^2 T \hspace{1pt} {\overline{\alpha}}^{2\Delta} \alpha_1}{72\tau(\epsilon, \alpha, \Delta)}\right) , \text{ for constant $0<\delta_2<1$,}  \label{prop-event-1-prob-eq-smaller-than}
\end{align} 
where $\|\phi\|_{\pi}$, denoting the $\pi$-norm of the vector $\phi$, is given by $$\|\phi\|_{\pi}:=\sqrt{\sum_{(\boldsymbol{f}||\boldsymbol{p})\in \text{Domain}(\boldsymbol{F}||\boldsymbol{P})} \frac{(\phi_{\boldsymbol{F}||\boldsymbol{P}}(\boldsymbol{f}||\boldsymbol{p}))^2}{\pi_{\boldsymbol{F}||\boldsymbol{P}}(\boldsymbol{f}||\boldsymbol{p}))}},$$ where $\text{Domain}(\boldsymbol{F}||\boldsymbol{P}): = \text{Suffix-Set} \times \left( \text{Detailed-State-Set} \right)^{\Delta+1} $ since Markov chain $\mathcal{C}_{\boldsymbol{F}||\boldsymbol{P}}$ represents the transition of $\boldsymbol{F}_{t-\Delta-1} S_{t-\Delta} \ldots S_{t}$. The term $ T \hspace{1pt} {\overline{\alpha}}^{2\Delta} \alpha_1$ in Inequality~(\ref{prop-event-1-prob-eq-smaller-than}) comes from Eq.~(\ref{eq-Expectation-C-t-0-t-0-T-1-simple}). 
% \begin{subnumcases}{\hspace{-10pt}}
% \bp{C(t_0, t_0+T-1) \leq (1-\delta_2) \cdot \bE{C(t_0, t_0+T-1)}}  \nonumber  \\  \leq c \|\phi\|_{\pi} \exp\left(-\frac{{\delta_2}^2 T \cdot \bE{C(t_0, t_0+T-1)}}{72\tau(\epsilon, \alpha, \Delta)}\right) , \nonumber  \\  \text{for constant $0<\delta_2<1$} . \label{prop-event-1-prob-eq-smaller-than}
% \end{subnumcases}  
We can also use Theorem 3.1 of Reference~\cite{chung2012chernoff} to compute a bound for the tail probability $\bp{C(t_0, t_0+T-1) \geq (1+\delta_2) \cdot \bE{C(t_0, t_0+T-1)}}$. We do not present the result here since it is not needed.

% Recall ? defined in the previous section; i.e., ?. Consider a random walk in the $T$ rounds from round $t_0$ to $t_0+T-1$,   on  the \textit{time-homogeneous}, \textit{irreducible}, and \textit{ergodic} Markov chain  . Let the 

Proposition~\ref{eq-prop-phi-pi} below provides an upper bound for
$\|\phi\|_{\pi}$.

\begin{prop} \label{eq-prop-phi-pi}
We have $\|\phi\|_{\pi} \leq \frac{1}{\sqrt{\min \pi_{\boldsymbol{F}||\boldsymbol{P}}}} $, where $\min \pi_{\boldsymbol{F}||\boldsymbol{P}}$ denotes the minimal value among $\pi_{\boldsymbol{F}||\boldsymbol{P}}$ and  is given by 
%\min \pi_{\boldsymbol{F}||\boldsymbol{P}}=
 $\alpha\cdot {\overline{\alpha}}^{\Delta-1}  \cdot  \min\left\{ 1-{\overline{\alpha}}^{\Delta}, {\overline{\alpha}}^{\Delta}\right\} \cdot \left(\min \left\{p^{\mu n}, (1-p)^{\mu n}\right\}\right)^{\Delta+1} $.
\end{prop}

We prove Proposition~\ref{eq-prop-phi-pi} in Appendix~\ref{app-eq-prop-phi-pi}.

From Proposition~\ref{eq-prop-phi-pi}, $\|\phi\|_{\pi}$ is upper bounded by a term that depends on $\alpha$ and $\Delta$ (note that when $\alpha$ is given, $\overline{\alpha}  : = 1 - \alpha$ is also given). Also, $\tau(\epsilon, \alpha, \Delta)$ denoting the $\epsilon$-mixing time of the Markov chain $\mathcal{C}_{\boldsymbol{F}||\boldsymbol{P}}$ is clearly a non-increasing function of $\epsilon$ given $\alpha$ and $\Delta$. In view of $0<\epsilon \leq 1/8$, we can select $\epsilon$ as $1/8$ so that the bound in the right hand side of Inequality~(\ref{prop-event-1-prob-eq-smaller-than}) is maximized. Then $\tau(1/8, \alpha, \Delta)$  depends on only $\alpha$ and $\Delta$. Recall from Eq.~(\ref{eq-define-alpha}) that $\alpha$  depends on $n,p,\mu$. Hence, given $n,p,\mu,\Delta$, we use Inequality~(\ref{prop-event-1-prob-eq-smaller-than}) to obtain the desired result (\ref{prop-event-1-prob-eq-simple}) that $\bp{C(t_0, t_0+T-1) \leq (1-\delta_2) \cdot \bE{C(t_0, t_0+T-1)}}$ is upper bounded by $O(1)\cdot\exp\left(-\Omega\left(T\right)\right)$, where $O(1)$ is with respect to $T$.

% \begin{prop} \label{prop-event-1-prob}
% \begin{align}
%  &\bp{C(t_0, t_0+T-1) \leq (1-\delta_2) \cdot \bE{C(t_0, t_0+T-1)}}  \nonumber  \\ & \leq c \|\phi\|_{\pi} \exp\left(-\frac{{\delta_2}^2 T \hspace{1pt} {\overline{\alpha}}^{2\Delta} \alpha_1}{72\tau(\epsilon, \alpha, \Delta)}\right)  . \label{prop-event-1-prob-eq}
% \end{align}
% \end{prop}

\subsection{Proving Inequality~(\ref{prop-event-2-prob-eq-simple})} \label{sec-prop-event-2-prob-eq-simple}

As already explained in Section~\ref{sec-eq-E-C-E-A}, $A(t_0, t_0+T-1)$ follows the binomial distribution $\text{binom}(T \nu n, p)$. 
From~\cite{arratia1989tutorial}, for a positive constant $\delta_4$, with $D\left( (1+\delta_4) p ||p\right)$ denoting the relative entropy between a Bernoulli distribution of parameter $(1+\delta_4) p$ and a Bernoulli distribution of parameter $p$; i.e., defining
\begin{align}
 & D\left( (1+\delta_4) p ||p\right) \nonumber  \\ &: = (1+\delta_4) p \ln (1+\delta_4) + [1-(1+\delta_4) p] \ln \frac{1-(1+\delta_4) p}{1-p}  , \label{a}
\end{align}
we have
\begin{align}
 &\bp{A(t_0, t_0+T-1) \geq (1+\delta_4) \cdot \bE{A(t_0, t_0+T-1)}}  \nonumber  \\ & \leq \exp\left(- T \nu n \cdot D\left( (1+\delta_4) p ||p\right)  \right)  . \label{prop-event-2-prob-eq}
\end{align}
Thus, given $n,p,\nu$, we obtain the desired result (\ref{prop-event-2-prob-eq-simple}) that $\bp{A(t_0, t_0+T-1) \geq (1+\delta_4) \cdot \bE{A(t_0, t_0+T-1)}}$ is upper bounded by $O(1)\cdot\exp\left(-\Omega\left(T\right)\right)$, where $O(1)$ is with respect to $T$.

% \begin{prop} \label{prop-event-2-prob}
% \begin{align}
%  &\bp{A(t_0, t_0+T-1) \geq (1+\delta_4) \cdot \bE{A(t_0, t_0+T-1)}}  \nonumber  \\ & \leq \exp\left(- T \nu n \cdot D\left( (1+\delta_4) p ||p\right)  \right)  . \label{prop-event-2-prob-eq}
% \end{align}
% \end{prop}

\subsection{Using Theorem~\ref{simpler-form-thm-bound-c2} to prove Theorem~\ref{thm-bound-c2}} \label{subsec-thm-bound-c2}

For $c$ denoting $\frac{1}{p n \Delta}$, it is straightforward to show that a combination of Inequalities~(\ref{eq-condition-pn}) and~(\ref{eq-thm2-c-v1}) in Theorem~\ref{simpler-form-thm-bound-c2} is the same as Inequality~(\ref{eq-thm2-c-v1-condition-thm-Delta}) of Theorem~\ref{thm-bound-c2}. Hence, given Theorem~\ref{simpler-form-thm-bound-c2}, we know that if  Inequality~(\ref{eq-thm2-c-v1-condition-thm-Delta}) holds, then the consistency of Nakamoto's blockchain protocol holds in a window of $T$ rounds with probability at least \mbox{$1-O(1)\cdot\exp\left(-\Omega\left(T\right)\right)$}.

To complete the proof of Theorem~\ref{thm-bound-c2}, next we show that under Inequality~(\ref{nu-delta-1-delta-2}), we can write Inequality~(\ref{eq-thm2-c-v1-condition-thm-Delta}) as Inequality~(\ref{c-mu-nu}).

From $\mu = 1-\nu$ and the condition $\nu \geq \frac{1}{1+\exp(\Delta^{\delta_1})} $ of Inequality~(\ref{nu-delta-1-delta-2}), we have
\begin{align}
\ln \frac{\mu}{\nu} = \ln \frac{1-\nu}{\nu} \leq  \ln \frac{~~1-\frac{1}{1+\exp(\Delta^{\delta_1})}~~}{~~\frac{1}{1+\exp(\Delta^{\delta_1})}~~} = \Delta^{\delta_1} .  \label{nu-delta-1-delta-2-part1}
\end{align}

From $\mu = 1-\nu$ and the condition $\nu \leq \frac{1}{1+\exp\left(\frac{1}{\Delta^{\delta_2}-1}\right)}$ of Inequality~(\ref{nu-delta-1-delta-2}), we have
\begin{align}
\ln \frac{\mu}{\nu} = \ln \frac{1-\nu}{\nu} \geq  \ln \frac{~~1-\frac{1}{1+\exp\left(\frac{1}{\Delta^{\delta_2}-1}\right)}~~}{~~\frac{1}{1+\exp\left(\frac{1}{\Delta^{\delta_2}-1}\right)}~~} = \frac{1}{\Delta^{\delta_2}-1} , \label{nu-delta-1-delta-2-part2}
\end{align}
which implies
\begin{align}
\frac{\ln \frac{\mu}{\nu}+1}{ \Delta \ln \frac{\mu}{\nu} } =  \frac{1}{ \Delta}\left( 1 +  \frac{1}{  \ln \frac{\mu}{\nu} }\right) \leq \Delta^{\delta_2-1}. \label{nu-delta-1-delta-2-part3}
\end{align}

Here we set $\epsilon_1$ by
\begin{align}
\epsilon_1: =\Delta^{\delta_1+\delta_2-1}.\label{eq-set-epsilon-1}
\end{align} 

From~(\ref{nu-delta-1-delta-2-part1})~(\ref{nu-delta-1-delta-2-part3})~(\ref{eq-set-epsilon-1}) and the condition \mbox{$\delta_1+\delta_2<1$}, letting $\epsilon_1$ be $\Delta^{\delta_1+\delta_2-1}$, we obtain
\begin{align}
\frac{2\mu}{\ln \frac{\mu}{\nu}} \geq \frac{2\mu}{\Delta^{\delta_1}} = \frac{2\mu}{\epsilon_1} \cdot \Delta^{\delta_2-1} >  \frac{2(\ln \frac{\mu}{\nu}+1)\mu}{\epsilon_1 \Delta \ln \frac{\mu}{\nu} } , \label{nu-delta-1-delta-2-part4}
\end{align}
which means that Inequality~(\ref{eq-thm2-c-v1-condition-thm-Delta}) (i.e., \mbox{$c  \geq \max\left\{\left( \frac{2\mu}{\ln \frac{\mu}{\nu}} +  \frac{1}{\Delta}\right)    \frac{1 + \epsilon_2 }{1-\epsilon_1} ,~ \frac{(\ln \frac{\mu}{\nu}+1)\mu}{\epsilon_1 \Delta \ln \frac{\mu}{\nu} } \right\} $}) becomes
\begin{align}
c  \geq \left[ \frac{2\mu}{\ln (\mu / \nu)} +  \frac{1}{\Delta}\right]    \frac{1 + \epsilon_2 }{1-\epsilon_1}.  \label{eq-thm2-c-v1-new}
\end{align}

From~(\ref{nu-delta-1-delta-2-part1}) and $\mu > \frac{1}{2}  $, we get
\begin{align}
\frac{1}{\Delta} =\Delta^{-\delta_1}  \cdot \Delta^{\delta_1-1} < \frac{2\mu}{\Delta^{\delta_1}}  \cdot \Delta^{\delta_1-1} \leq \frac{2\mu}{\ln \frac{\mu}{\nu} }  \cdot \Delta^{\delta_1-1}, \label{nu-delta-1-delta-2-part5}
\end{align}
which means that a sufficient condition for~(\ref{eq-thm2-c-v1-new}) is
\begin{align}
c & \geq \left[ \frac{2\mu}{\ln (\mu / \nu)} +  \frac{2\mu}{\ln \frac{\mu}{\nu} }  \cdot \Delta^{\delta_1-1}\right]    \frac{1 + \epsilon_2 }{1-\epsilon_1} \nonumber  \\ & = \frac{2\mu}{\ln (\mu / \nu)} \cdot \left(1 + \epsilon_2  \right)\cdot  \frac{1+\Delta^{\delta_1-1}}{1-\Delta^{\delta_1+\delta_2-1}} , \label{nu-delta-1-delta-2-part6}
\end{align}
where the last step uses~(\ref{eq-set-epsilon-1}). 

The above result~(\ref{nu-delta-1-delta-2-part6}) gives Inequality~(\ref{c-mu-nu}). Hence, we have completed proving Theorem~\ref{thm-bound-c2}. \qeda

% \section{Implementation and Experiments}

\subsection{Explaining that $\delta_5$ and $\delta_1$ in Eq.~(\ref{eq-define-delta_2}) and Eq.~(\ref{eq-define-delta_1})  are both positive  for $ 0 <\epsilon_1 \iffalse_0 \fi  < 1$ and $ \epsilon_2 \iffalse_0 \fi > 0 $}
\label{app-delta_5-delta_1}

Clearly, $ \delta_5 > 0 $ since the nominator and denominator of Eq.~(\ref{eq-define-delta_2}) are both positive. In addition, given
\begin{align}
\delta_5  &> \frac{(\epsilon_1 + \epsilon_2) \ln \frac{\mu}{\nu}}{(\epsilon_1 + \epsilon_2) + \frac{\epsilon_1 + \epsilon_2}{\epsilon_1} \cdot (1 - \epsilon_1)\cdot  (\ln \frac{\mu}{\nu}+1)} \nonumber  \\ &=   \frac{\epsilon_1 \ln \frac{\mu}{\nu}}{ \epsilon_1 +   (1 - \epsilon_1)\cdot  (\ln \frac{\mu}{\nu}+1)} = \frac{\epsilon_1 \ln \frac{\mu}{\nu}}{1 + (1 - \epsilon_1) \ln \frac{\mu}{\nu}}  , \label{eq-define-delta_2-bound-repeat}
\end{align}
 we have
 \begin{align}
\delta_1 & = ( 1+ \delta_5) \cdot \left( 1 - \frac{\epsilon_1 \ln \frac{\mu}{\nu}}{\ln \frac{\mu}{\nu} + 1}\right) - 1  \nonumber  \\ &  > \left[ 1 + \frac{\epsilon_1 \ln \frac{\mu}{\nu}}{1 + (1 - \epsilon_1) \ln \frac{\mu}{\nu}} \right] \cdot \left( 1 - \frac{\epsilon_1 \ln \frac{\mu}{\nu}}{\ln \frac{\mu}{\nu} + 1}\right) - 1 = 0.
\end{align}

\subsection{Proof of Proposition~\ref{eq-prop-phi-pi}}
\label{app-eq-prop-phi-pi}

% \noindent\textbf{Proof of Proposition~\ref{eq-prop-phi-pi}.}

The $\pi$-norm of $\phi$ is
\begin{align}
\|\phi\|_{\pi}  & = \sqrt{\sum_{(\boldsymbol{f}||\boldsymbol{p})\in \text{Domain}(\boldsymbol{F}||\boldsymbol{P})} \frac{(\phi_{\boldsymbol{F}||\boldsymbol{P}}(\boldsymbol{f}||\boldsymbol{p}))^2}{\pi_{\boldsymbol{F}||\boldsymbol{P}}(\boldsymbol{f}||\boldsymbol{p}))}}  \nonumber  \\ & \leq  \sqrt{\sum_{(\boldsymbol{f}||\boldsymbol{p})\in \text{Domain}(\boldsymbol{F}||\boldsymbol{P})} \frac{\phi_{\boldsymbol{F}||\boldsymbol{P}}(\boldsymbol{f}||\boldsymbol{p})}{\min \pi_{\boldsymbol{F}||\boldsymbol{P}}}}   \nonumber  \\ & = \frac{1}{\sqrt{\min \pi_{\boldsymbol{F}||\boldsymbol{P}}}} , \label{eq-phi-pi}
\end{align}
where $\min \pi_{\boldsymbol{F}||\boldsymbol{P}}$ denotes the minimal value among $\pi_{\boldsymbol{F}||\boldsymbol{P}}$.

Recall from Eq.~(\ref{eq-F-t-to-G-t}) that
\begin{align}
 & \pi_{\boldsymbol{F}||\boldsymbol{P}}( \boldsymbol{f} s^{(1)} \ldots s^{(\Delta+1)})   =  \pi_{\boldsymbol{F}}(\boldsymbol{f}) \prod_{i=1}^{\Delta+1} \bp{s^{(i)}} .\label{eq-F-t-to-G-t-repeat}
\end{align}
For $s^{(i)} \in \text{Detailed-State-Set}$ for Detailed-State-Set in Eq.~(\ref{eq-Detailed-State-Set}), we have
\begin{align}
\min_{s^{(i)} \in \text{Detailed-State-Set}} \bp{s^{(i)}} = \begin{cases} p^{\mu n}, & \text{if $p \leq \frac{1}{2}$,}  \\  (1-p)^{\mu n}, & \text{if $p > \frac{1}{2}$,} \end{cases} \nonumber
\end{align}
so that we can write
\begin{align}
\min_{s^{(i)} \in \text{Detailed-State-Set}} \bp{s^{(i)}} = \min \left\{p^{\mu n}, (1-p)^{\mu n}\right\}. \label{eq-F-t-to-G-t-repeat-cases2}
\end{align}
Then Eq.~(\ref{eq-F-t-to-G-t-repeat}) implies that 
\begin{align}
\min \pi_{\boldsymbol{F}||\boldsymbol{P}}  & = \left(\min\pi_{\boldsymbol{F}}\right) \cdot \left(\min \left\{p^{\mu n}, (1-p)^{\mu n}\right\}\right)^{\Delta+1}  , \label{eq-phi-pi2}
\end{align}
where the minimal value among $\pi_{\boldsymbol{F}}$ is
\begin{align}
\min\pi_{\boldsymbol{F}}  & = \min\left\{\alpha\cdot  (1-{\overline{\alpha}}^{\Delta}) \cdot  {\overline{\alpha}}^{\Delta-1},\alpha \cdot  {\overline{\alpha}}^{2\Delta-1}\right\}  \nonumber  \\ & = \alpha\cdot {\overline{\alpha}}^{\Delta-1}  \cdot  \min\left\{ 1-{\overline{\alpha}}^{\Delta}, {\overline{\alpha}}^{\Delta}\right\}  . \label{eq-phi-pi3}
\end{align} 
Combining~(\ref{eq-phi-pi})~(\ref{eq-phi-pi2})~(\ref{eq-phi-pi3}), we complete proving Proposition~\ref{eq-prop-phi-pi}.\qeda

\subsection{Proof of Lemma~\ref{prop-eq-bound-c-step-4}} \label{subse-prop-eq-bound-c-step-4}

% Lemma~\ref{prop-eq-bound-c-step-4} and Its Proof

% \noindent\textbf{:}

Recall the expression of $\alpha_1$ in Inequality~(\ref{eq-H-1}); i.e., \mbox{$\alpha_1 = p \mu n  \times (1-p)^{\mu n - 1}$.}  Then given the condition \mbox{$0<p \mu n<1$} and the result $\mu n - 1 > \frac{1}{2} n - 1 \geq 1$ from $\mu > \frac{1}{2}$ and $n\geq 4$, we use Fact 2 on Page~20 of~\cite{zhao2015k} to obtain  
% Given the condition $0<p \mu n<1$, we can prove that $\alpha: = 1 - (1- p)^{\mu n}   $ satisfies
\begin{align}
\alpha_1  & = p \mu n \cdot  (1-p)^{\mu n - 1} \geq  p \mu n \cdot  [1-p\cdot(\mu n - 1)] \nonumber \\ & \geq p \mu n \cdot (1-p \mu n)  .  \label{eq-pmun-alphav1}
\end{align}
% \begin{align}
% \alpha & = 1 - (1- p)^{\mu n} \geq 1 - [ 1 -  p \mu n + \frac{1}{2} (p \mu n )^2 ] \nonumber \\ & = p \mu n - \frac{1}{2} (p \mu n )^2 . \label{eq-pmun-alpha2}
% \end{align}
% To see Inequality~(\ref{eq-pmun-alpha2}), from the Taylor series expansion with Lagrange remainder, there
% exist $ 0 < \theta <1$ such that
% \begin{align}
% (1- p)^{\mu n}   & = 1 - p \mu n + \frac{\mu n(\mu n-1)}{2}p^2   \nonumber \\ & \quad - \frac{\mu n(\mu n-1)(\mu n-2)(1-\theta p)^{\mu n-3}}{6}p^3.
% \label{exy2}
% \end{align}
%  Applying $0< p<1$ and $\mu n \geq n/2 \geq 2$ for $n \geq 4$ (which trivially holds) to (\ref{exy2}), we have $(1- p)^{\mu n}  \leq
% 1 -  p \mu n + \frac{1}{2} (p \mu n )^2 ,$
% % \begin{align}
% % (1- p)^{\mu n} & \leq
% % 1 -  p \mu n + \frac{1}{2} (p \mu n )^2 ,
% % \end{align}
% which implies Inequality~(\ref{eq-pmun-alpha}).

Then Inequality~(\ref{eq-pmun-alphav1})  induces
\begin{align}
&\left\{ p \mu n\cdot \left(1 - p \mu n \right) {\overline{\alpha}}^{2\Delta} \geq (1+\delta_1) p \nu n   \right\} \nonumber \\ &\Longrightarrow \left\{  {\overline{\alpha}}^{2\Delta} \alpha_1 \geq (1+\delta_1) p \nu n \right\} . \label{eq-pmun-alpha}
\end{align}
The statement $ p \mu n\left(1 - p \mu n \right) {\overline{\alpha}}^{2\Delta} \geq (1+\delta_1) p \nu n  $ is equivalent to ${\overline{\alpha}} \geq \left( \frac{1+\delta_1}{1 - p \mu n } \cdot  \frac{\nu}{\mu} \right)^{1/(2\Delta)}$; i.e., Inequality~(\ref{eq_alpha_delta_1_D}). This along with Inequality~(\ref{eq-pmun-alpha}) implies the desired result. \qeda
% \begin{align}
% \left\{   \right\} \Longrightarrow \left\{   \right\}
% \end{align}

% \begin{align}
% \left\{   \right\} \Longrightarrow \left\{   \right\}
% \end{align}

% \begin{align}
% \left\{   \right\} \Longrightarrow \left\{   \right\}
% \end{align}

% \begin{align}
% \left\{   \right\} \Longrightarrow \left\{   \right\}
% \end{align}

% \begin{align}
% \left\{   \right\} \Longrightarrow \left\{   \right\}
% \end{align}

% \begin{align}
% \left\{   \right\} \Longrightarrow \left\{   \right\}
% \end{align}

% \begin{align}
% \overline{\alpha}: = 1 - \alpha =   (1- p)^{\mu n}. \label{eq-over-alpha}
% \end{align}

% \begin{align}
% p \mu n\left(1 - p \mu n \right) {\overline{\alpha}}^{2\Delta} \geq (1+\delta_1) p \nu n
% \end{align}

% \begin{align}
% {\overline{\alpha}}^{2\Delta} \geq   \frac{1+\epsilon_1}{1 - \frac{1}{2} p \mu n } \cdot  \frac{\nu}{\mu}
% \end{align}

% $$\overline{\alpha}: = 1 - \alpha =   (1- p)^{\mu n}$$

% \begin{align}
% {\overline{\alpha}} \geq \left( \frac{1+\delta_1}{1 - \frac{1}{2} p \mu n } \cdot  \frac{\nu}{\mu} \right)^{1/(2\Delta)}
% \end{align}

% \qeda

% To continue proving Lemma~\ref{prop-eq-bound-c-step-3}?, we  upper bound the term $  \left( \frac{1+\delta_1}{1 -  p \mu n } \cdot  \frac{\nu}{\mu} \right)^{1/(2\Delta)}$ via the following lemma.

% Lemma~\ref{prop1} is proved in the Appendix.

\subsection{Proof of Lemma~\ref{prop1}}
\label{app-secprf-prop1}

% \noindent \textbf{Proof of $ 0 < \delta_5 \iffalse_7 \fi < \ln \frac{\mu}{\nu}$:} Given $\delta_5 > \frac{\epsilon_1 \ln \frac{\mu}{\nu}}{1 + (1 - \epsilon_1) \ln \frac{\mu}{\nu}} $ and $0<\epsilon_1 \iffalse_0 \fi <1$, we have $0<\delta_5 \iffalse_7 \fi  < \ln \frac{\mu}{\nu} $.

\noindent \textbf{Proof of $ \delta_5 > 0$:} Given $\delta_5 > \frac{\epsilon_1 \ln \frac{\mu}{\nu}}{1 + (1 - \epsilon_1) \ln \frac{\mu}{\nu}} $ with $0<\epsilon_1 \iffalse_0 \fi <1$ and $\ln \frac{\mu}{\nu}  > 0$ from $0 < \nu  < \mu$, we have $\delta_5 > 0$.

\noindent \textbf{Proof of $\delta_1 > 0$:} Given $\delta_5 > \frac{\epsilon_1 \ln \frac{\mu}{\nu}}{1 + (1 - \epsilon_1) \ln \frac{\mu}{\nu}} $ and  \mbox{$\delta_1 = ( 1+ \delta_5) \cdot \left( 1 - \frac{\epsilon_1 \ln \frac{\mu}{\nu}}{\ln \frac{\mu}{\nu} + 1}\right) - 1$,} we have $$\delta_1 > \left( 1+ \frac{\epsilon_1 \ln \frac{\mu}{\nu}}{1 + (1 - \epsilon_1) \ln \frac{\mu}{\nu}}\right) \cdot \left( 1 - \frac{\epsilon_1 \ln \frac{\mu}{\nu}}{\ln \frac{\mu}{\nu} + 1}\right) - 1  = 0.$$

% Proof of $0<\delta_5 \iffalse_7 \fi < \frac{\mu-\nu}{\nu}$: Given $0< \nu < \mu$ and $\epsilon_1 \iffalse_0 \fi  < 2$, we have $ 0 < \frac{\epsilon_1 \iffalse_0 \fi (\mu - \nu)}{2\mu} < 1-\frac{\nu}{\mu}$. Using these in $\delta_5 \iffalse_7 \fi  : = \frac{1}{2} [\frac{2\mu}{ 2\mu - \epsilon_1 \iffalse_0 \fi (\mu - \nu)} + \frac{\mu}{\nu}] - 1 $, we have $\delta_5 \iffalse_7 \fi  > \frac{1}{2} [1 + \frac{\mu}{\nu}] - 1 = \frac{1}{2} [ \frac{\mu}{\nu}-1] > 0 $ and $\delta_5 \iffalse_7 \fi   <\frac{1}{2} [\frac{2\mu}{ 2\mu - \epsilon_1 \iffalse_0 \fi (\mu - \nu)} + \frac{\mu}{\nu}] - 1 < \frac{1}{2} [\frac{1}{1 - (1-\frac{\nu}{\mu})} + \frac{\mu}{\nu}] - 1 = \frac{\mu-\nu}{\nu} $.

% Proof of $\delta_1 > 0$: Given $0< \nu < \mu$, and $p \mu n <  \frac{\epsilon_1 \iffalse_0 \fi (\mu - \nu)}{\mu}$ for $\epsilon_1 \iffalse_0 \fi  < 2$, we have $0< \frac{\epsilon_1 \iffalse_0 \fi (\mu - \nu)}{2\mu} < 1-\frac{\nu}{\mu} $, which is used in $\delta_1 :  = \frac{\mu}{2\nu} [1 - \frac{\epsilon_1 \iffalse_0 \fi (\mu - \nu)}{2\mu}] - \frac{1}{2} $ to have $\delta_1 > \frac{\mu}{2\nu} \cdot \frac{\nu}{\mu} - \frac{1}{2} = 0$.

\noindent \textbf{Proof of $\left(\frac{1+\delta_1}{1 -  p \mu n }\right)^{1/(2\Delta)} <  1+ \frac{\delta_5 \iffalse_7 \fi }{2\Delta} $:}  Using the conditions $p n  \leq \frac{\epsilon_1 \ln \frac{\mu}{\nu}}{(\ln \frac{\mu}{\nu} + 1) \mu}  $ and $\delta_1 = ( 1+ \delta_5) \cdot \left( 1 - \frac{\epsilon_1 \ln \frac{\mu}{\nu}}{\ln \frac{\mu}{\nu} + 1}\right) - 1  $, we have $1+\delta_1  \leq ( 1+ \delta_5) \cdot \left(1 -  p \mu n\right)$, which means $\frac{1+\delta_1}{1 -  p \mu n }  \leq 1+ \delta_5 \iffalse_7 \fi  $. Moreover, we have $ 1+ \delta_5 \iffalse_7 \fi < \left(1+ \frac{\delta_5 \iffalse_7 \fi }{2\Delta}\right)^{2\Delta}  $ from the binomial series. Summarizing the above results, we obtain $\left(\frac{1+\delta_1}{1 -  p \mu n }\right)^{1/(2\Delta)} <  1+ \frac{\delta_5 \iffalse_7 \fi }{2\Delta} $. \qeda

\subsection{Proof of Lemma~\ref{prop-eq-bound-c-step-3}} \label{subse-prop-eq-bound-c-step-3}

% Lemma~\ref{prop-eq-bound-c-step-3} and Its Proof
% \noindent\textbf{:}

Recalling $\overline{\alpha}  =   (1- p)^{\mu n}$ from Eq.~(\ref{eq-define-alpha-bar}), we have
\begin{align}
&\hspace{-5pt}\left\{  {\overline{\alpha}} \geq \left( 1+ \frac{\delta_5 \iffalse_7 \fi }{2\Delta} \right) \cdot \left(  \frac{\nu}{\mu} \right)^{1/(2\Delta)}  \right\} \nonumber \\ &\hspace{-5pt}\Longleftrightarrow \left\{ (1- p)^{\mu n}   \geq \left( 1+ \frac{\delta_5 \iffalse_7 \fi }{2\Delta} \right) \cdot \left(  \frac{\nu}{\mu} \right)^{1/(2\Delta)} \right\} \nonumber \\  &  \hspace{-5pt}\Longleftrightarrow \left\{ p \leq 1 - \left[  \left( 1+ \frac{\delta_5 \iffalse_7 \fi }{2\Delta} \right)   \left(  \frac{\nu}{\mu} \right)^{1/(2\Delta)}  \right]^{1/(\mu n)}    \right\}  \nonumber \\  &   \hspace{-5pt}\Longleftrightarrow \hspace{-1.5pt}\left\{\hspace{-1.5pt}  c \hspace{-1.5pt}: =\hspace{-1.5pt} \frac{1}{p n \Delta } \hspace{-1.5pt}\geq \hspace{-1.5pt} \frac{1}{ n \Delta \left\{1-\left[ \left(1+ \frac{\delta_5 \iffalse_7 \fi }{2\Delta}\right) \left( \frac{\nu}{\mu} \right)^{1/(2\Delta)} \right]^{1/(\mu n)}\right\}}  \hspace{-1.5pt}\right\} . \nonumber %\label{eq-alpha-overline-transform}
\end{align} 
 \qeda
%  We now use Lemma~\ref{prop1} to analyze the last statement of~(\ref{eq-alpha-overline-transform}). It holds that
% \begin{align}
% & \left\{  c \geq   \frac{1}{ n\Delta\left\{ 1 - \left[  \left( \frac{1+\delta_1}{1 -  p \mu n } \cdot  \frac{\nu}{\mu} \right)^{1/(2\Delta)}   \right]^{1/(\mu n)}\right\} }   \right\}  \nonumber \\ & \xLeftarrow{\textup{Lemma~\ref{prop1}}} \left\{  c \geq  \frac{1}{ n \Delta \left\{1-\left[ \left(1+ \frac{\delta_5 \iffalse_7 \fi }{2\Delta}\right) \left( \frac{\nu}{\mu} \right)^{1/(2\Delta)} \right]^{1/(\mu n)}\right\}}  \right\}   . \label{eq-alpha-overline-transform2}
% \end{align} 
% The combination of~(\ref{eq-alpha-overline-transform}) and~(\ref{eq-alpha-overline-transform2}) completes the proof of Lemma~\ref{prop-eq-bound-c-step-3}.

% $ \frac{1}{  n \Delta \left\{1-\left[ \left(1+ \frac{\delta_5 \iffalse_7 \fi }{2\Delta}\right) \left( \frac{\nu}{\mu} \right)^{1/(2\Delta)} \right]^{1/(\mu n)}\right\}}$

\subsection{Proof of Proposition~\ref{prop-delta-2-upper-bound}} Our goal is to prove
\begin{align}
1 - \left(1+ \frac{\delta_5 \iffalse_7 \fi }{2\Delta}\right) \left( \frac{\nu}{\mu} \right)^{1/(2\Delta)} > 0,   \label{eq-define-C-Delta2}
\end{align}
given the condition $ 0< \delta_5 \iffalse_7 \fi < \ln \frac{\mu}{\nu}$.

Clearly, Inequality~(\ref{eq-define-C-Delta2}) holds once we show 
\begin{align}
1 - \left(1+ \frac{1}{2\Delta} \ln \frac{\mu}{\nu}\right) \left( \frac{\nu}{\mu} \right)^{1/(2\Delta)} > 0 .  \label{eq-define-C-Delta3}
\end{align}

After defining $f(x) : = x^{1/(2\Delta)} - \frac{1}{2\Delta} \ln x - 1 $ \vspace{1pt} for $x \geq 1$, the term $1 - \left(1+ \frac{1}{2\Delta} \ln \frac{\mu}{\nu}\right) \left( \frac{\nu}{\mu} \right)^{1/(2\Delta)}$ in Inequality~(\ref{eq-define-C-Delta3}) becomes $f\big(\frac{\mu}{\nu}\big) \cdot \left( \frac{\nu}{\mu} \right)^{1/(2\Delta)}$, so Inequality~(\ref{eq-define-C-Delta3}) holds once we prove $f\big(\frac{\mu}{\nu}\big)>0$. To this end, we derive $f'(x) : = \frac{1}{2x\Delta} \left( x^{1/(2\Delta)} - 1 \right) > 0$ for $x>1$, so that $f(x)$ is a strictly increasing function for $x \geq 1$. Then given $f(1)=0$, we obtain $f(x)>0$ for $x > 1$ and thus $f\big(\frac{\mu}{\nu}\big)>0$ given $\frac{\mu}{\nu} > 1$. The result $f\big(\frac{\mu}{\nu}\big)>0$ means $\big(\frac{\mu}{\nu}\big)^{1/(2\Delta)} - \frac{1}{2\Delta} \ln \frac{\mu}{\nu} - 1 > 0$, which implies Inequality~(\ref{eq-define-C-Delta3}) and thus Inequality~(\ref{eq-define-C-Delta2}).
 \qeda

\subsection{Proof of Lemma~\ref{prop-eq-bound-c-step-2}} \label{subse-prop-eq-bound-c-step-2}

% Lemma~\ref{prop-eq-bound-c-step-2} and Its Proof
% \noindent\textbf{:}

% \noindent \textbf{Proving Inequality~(\ref{eq-bound-c-step-2}):}

First, we know from Proposition~\ref{prop-delta-2-upper-bound} that   the denominators in both sides of Inequality~(\ref{eq-c-lower-bound-prop-eq-bound-c-step-2-1-ineq}) of Lemma~\ref{prop-eq-bound-c-step-2} are positive.

With $A$ defined by
\begin{align}
A: = 1 - \left(1+ \frac{\delta_5 \iffalse_7 \fi }{2\Delta}\right) \left( \frac{\nu}{\mu} \right)^{1/(2\Delta)}, \label{eq-define-C}
\end{align}
we know $A>0$ from Proposition~\ref{prop-delta-2-upper-bound}. 
% % is the same as 
% % \begin{align}
% % 1 - \left(1+ \frac{1}{2\Delta}\ln \frac{\mu}{\nu}\right) \left( \frac{\nu}{\mu} \right)^{1/(2\Delta)} > 0.  \label{eq-define-C-Delta}
% % \end{align}
% Then Inequality~(\ref{eq-define-C-Delta}) induces that
% We define
% Given $\left( \frac{\mu}{\nu} \right)^{1/(2\Delta)} =\left( 1 + \frac{\mu-\nu}{\nu} \right)^{1/(2\Delta)}  > 1 + \frac{\mu-\nu}{\nu} \cdot \frac{1}{2\Delta} > 1+ \frac{\delta_5 \iffalse_7 \fi }{2\Delta}$, where the last step uses the condition $ \delta_5 \iffalse_7 \fi < \frac{\mu-\nu}{\nu}$,
%  %result $ \delta_5 \iffalse_7 \fi < \frac{\mu-\nu}{\nu}$ from Lemma~\ref{prop1},
%  we obtain $A>0$??. 
  Also, clearly $A<1$.
With $0<A<1$ and $\mu n > \frac{n}{2} \geq 2$ from\vspace{1.5pt} $\mu > \frac{1}{2}$ and $n \geq 4$, we use Fact 2 on Page~20 of~\cite{zhao2015k} to obtain $(1 - \frac{A}{\mu n})^{\mu n} \geq 1 - \frac{A}{\mu n} \cdot \mu n = 1 - A >0 $, which implies $(1 - A)^{1/(\mu n)} \leq 1 - \frac{A}{\mu n}.$ Hence, \begin{align}
\frac{\mu}{A\Delta} =  \frac{1}{  n \Delta [1-   (1-A/(\mu n))]} \geq  \frac{1}{  n \Delta [1- (1 - A)^{1/(\mu n)}]} . \label{eq-defC-result}
\end{align}
We plug Eq.~(\ref{eq-define-C}) (i.e., the expression of $A$) into~(\ref{eq-defC-result}) and complete proving Lemma~\ref{prop-eq-bound-c-step-2}.
\qeda

\subsection{Proof of Lemma~\ref{prop-eq-bound-c-step-1}} \label{subse-prop-eq-bound-c-step-1}

% Lemma~\ref{prop-eq-bound-c-step-1} and Its Proof
% \noindent\textbf{:}

We evaluate $\frac{1}{1 - \left(1+ \frac{\delta_5 \iffalse_7 \fi }{2\Delta}\right) \left( \frac{\nu}{\mu} \right)^{1/(2\Delta)}}$ appearing in the desired result. We have
\begin{align}
&\frac{1}{1 - \left(1+ \frac{\delta_5 \iffalse_7 \fi }{2\Delta}\right) \left( \frac{\nu}{\mu} \right)^{1/(2\Delta)}} = \frac{\left( \frac{\mu}{\nu} \right)^{1/(2\Delta)}}{ \left( \frac{\mu}{\nu} \right)^{1/(2\Delta)}-\left( 1+\frac{\delta_5 \iffalse_7 \fi }{2\Delta}\right)} \nonumber \\ &= \left[ 1+\frac{\frac{\delta_5 \iffalse_7 \fi }{2\Delta}}{ \left( \frac{\mu}{\nu} \right)^{1/(2\Delta)}-\left( 1+\frac{\delta_5 \iffalse_7 \fi }{2\Delta}\right)} \right]  \cdot \frac{1}{ 1 - \left( \frac{\nu}{\mu} \right)^{1/(2\Delta)}}. \label{eq-C-bound}
\end{align}
We further bound the term $ \left( \frac{\mu}{\nu} \right)^{1/(2\Delta)}-\left( 1+\frac{\delta_5 \iffalse_7 \fi }{2\Delta}\right)$ in Eq.~(\ref{eq-C-bound}):
\begin{align}
& \left( \frac{\mu}{\nu} \right)^{1/(2\Delta)}-\left( 1+\frac{\delta_5 \iffalse_7 \fi }{2\Delta}\right) \nonumber \\ & = \exp\left( \frac{1}{2\Delta} \ln \frac{\mu}{\nu} \right) -\left( 1+\frac{\delta_5 \iffalse_7 \fi }{2\Delta}\right) \nonumber \\ &> 1 +  \frac{1}{2\Delta} \ln \frac{\mu}{\nu}  -\left( 1+\frac{\delta_5 \iffalse_7 \fi }{2\Delta}\right) \nonumber \\ & = \frac{ \ln \frac{\mu}{\nu} \iffalse_7 \fi - \delta_5 }{2\Delta},  \label{eq-C-bound2}
\end{align}
where the step of ``$>$'' uses $\exp\left(x\right)> 1+x$ for $x>0$ as well as $\ln \frac{\mu}{\nu}  > 0$ from $0 < \nu  < \mu$.

Applying Inequality~(\ref{eq-C-bound2}) to Eq.~(\ref{eq-C-bound}), we obtain
\begin{align}
 &\frac{1}{1 - \left(1+ \frac{\delta_5 \iffalse_7 \fi }{2\Delta}\right) \left( \frac{\nu}{\mu} \right)^{1/(2\Delta)}} \nonumber \\ & < \left(1+\frac{\frac{\delta_5 \iffalse_7 \fi }{2\Delta}}{\frac{\ln \frac{\mu}{\nu}-\delta_5 \iffalse_7 \fi }{2\Delta}}\right)  \cdot \frac{1}{ 1 - \left( \frac{\nu}{\mu} \right)^{1/(2\Delta)}} \nonumber \\ & \hspace{-10pt} =  \left(1+\frac{\delta_5 \iffalse_7 \fi }{\ln \frac{\mu}{\nu} - \delta_5 \iffalse_7 \fi }\right)  \cdot \frac{1}{ 1 - \left( \frac{\nu}{\mu} \right)^{1/(2\Delta)}}. \label{eq-C-bound3}
\end{align}
% Multiplying $\frac{\mu}{\Delta}$
% on the sides of~(\ref{eq-C-bound3}), we complete proving Lemma~\ref{prop-eq-bound-c-step-1}.
% Using Inequality~(\ref{eq-C-bound2}) in Eq.~(\ref{eq-c-C}), we have
% \begin{align}
% c  \geq  \frac{\mu}{\Delta \left[ 1 - \left( \frac{\nu}{\mu} \right)^{1/(2\Delta)}\right]}  \cdot \left(1+\frac{\delta_5 \iffalse_7 \fi }{\ln \frac{\mu}{\nu} - \delta_5 \iffalse_7 \fi }\right) . \label{eq-thm1-c-v2}
% \end{align}
\qeda

\subsection{Proof of Lemma~\ref{prop-eq-bound-c-step-0}} \label{subse-prop-eq-bound-c-step-0}

% Lemma~\ref{prop-eq-bound-c-step-0} and Its Proof
 %$ \left[ \frac{2\mu}{\ln (\mu / \nu)} +  \frac{\mu}{\Delta}\right]  \cdot \left(1+\frac{\delta_5 \iffalse_7 \fi }{\ln \frac{\mu}{\nu} - \delta_5 \iffalse_7 \fi }\right)   \geq  \frac{\mu}{\Delta \left[ 1 - \left( \frac{\nu}{\mu} \right)^{1/(2\Delta)}\right]} \cdot \left(1+\frac{\delta_5 \iffalse_7 \fi }{\ln \frac{\mu}{\nu} - \delta_5 \iffalse_7 \fi }\right)  $ follows.

% \noindent\textbf{:}

% Lemma~\ref{prop-eq-bound-c-step-0} follows from Proposition~\ref{prop-final} below.

% \qeda

% Note that the expression before ``$\Longleftarrow$'' of~(\ref{eq-bound-c-step-4}) (resp., (\ref{eq-bound-c-step-3}) (\ref{eq-bound-c-step-2}) (\ref{eq-bound-c-step-1})) is the same as the expression after ``$\Longleftarrow'' of~(\ref{eq-bound-c-step-3}) (resp., (\ref{eq-bound-c-step-2}) (\ref{eq-bound-c-step-1}) (\ref{eq-bound-c-step-0})).

% \begin{prop} \label{prop-final}

% For $0< \nu < \mu$,
% $ \frac{2}{\ln (\mu / \nu)} \leq  \frac{1}{\Delta \left[ 1 - \left( \frac{\nu}{\mu} \right)^{1/(2\Delta)}\right]}    \leq  \frac{2}{\ln (\mu / \nu)} + \frac{1}{\Delta}$.
% \end{prop}

% \noindent\textbf{Proof of Proposition~\ref{prop-final}:}

We define
\begin{align}
\lambda: = \frac{\nu}{\mu}  \label{eq-def-lambda}
\end{align}
and  for $0<x\leq 1$,
\begin{align}
f(x): = \frac{x}{1-\lambda^x}   \label{eq-def-f-x}.
\end{align}

Clearly, $0 < \lambda < 1 $ follows from $0< \nu < \mu$. Then the derivative of $f(x)$ is
\begin{align}
f'(x) = \frac{1-\lambda^x-x \cdot (-\ln \lambda)\lambda^x}{(1-\lambda^x)^2} = \frac{g(x)}{(1-\lambda^x)^2}, \label{eq-def-f-prime-x}
\end{align}
where we define $g(x)$ as
\begin{align}
g(x) : = 1-(1-x\ln \lambda)\lambda^x  . \label{eq-def-g-x}
\end{align}
%  and $f''(x): = \frac{1-[1+x\ln (1/\lambda)]\lambda^x}{(1-\lambda^x)^2} $, respectively.

To analyze the sign of $f'(x)$ in Eq.~(\ref{eq-def-f-prime-x}), we discuss the sign of $g(x)$ in Eq.~(\ref{eq-def-g-x}). Hence, we compute
the derivative of $g(x)$ as $g'(x) = (\ln \lambda)^2 \lambda^x  x > 0$ for $0<x\leq 1$, given $0 < \lambda < 1 $. Hence, $g (x)$ strictly increases as $x$ increases for $0<x\leq 1$, implying $g(x)>g(0) = 0$ for $0<x\leq 1$. Using  this in Eq.~(\ref{eq-def-f-prime-x}), we have $f'(x)>0$ for $0<x\leq 1$, so that $f(x)$ strictly increases as $x$ increases for $0<x\leq 1$. Then for any $\epsilon_4 \in (0, \frac{1}{2\Delta})$, we have
\begin{align}
f\left(\frac{1}{2\Delta}\right) >  f(\epsilon_4), \label{eq-f-1-2-Delta-epsilon}
\end{align}
and
\begin{align}
f\left(\frac{1}{2\Delta}\right) - f(\epsilon_4) \leq  \left(\frac{1}{2\Delta} - \epsilon_4 \right) \cdot \max_{x \in [\epsilon_4,  \frac{1}{2\Delta}]} f'(x). \label{eq-f-1-2-Delta-epsilon-v2}
\end{align}
Letting $\epsilon_4 \to 0$ in Inequality~(\ref{eq-f-1-2-Delta-epsilon}), we obtain
\begin{align}
f\left(\frac{1}{2\Delta}\right) \geq \lim_{\epsilon_4 \to 0}   f(\epsilon_4)   = \lim_{\epsilon_4 \to 0}  \frac{\epsilon_4}{1-\lambda^{\epsilon_4}} . \label{eq-f-1-2-Delta-epsilon-v3-old}
\end{align}
To compute $\lim_{\epsilon_4 \to 0}  \frac{\epsilon_4}{1-\lambda^{\epsilon_4}}$ of~(\ref{eq-f-1-2-Delta-epsilon-v3-old}), we note that the nominator and denominator both converge to $0$ as $\epsilon_4 \to 0$, and are also both differentiable for $\epsilon_4 > 0$, so we use L'Hospital's rule (see~\cite{taylor1952hospital}) to obtain
\begin{align}
\lim_{\epsilon_4 \to 0}   f(\epsilon_4) =\lim_{\epsilon_4 \to 0}  \frac{\epsilon_4}{1-\lambda^{\epsilon_4}}   = \lim_{\epsilon_4 \to 0}\frac{1}{-\lambda^{\epsilon_4} \cdot \ln \lambda} = \frac{1}{\ln (1/\lambda)} ,  \label{eq-f-1-2-Delta-epsilon-v3-oldv2}
\end{align}
which together with~(\ref{eq-f-1-2-Delta-epsilon-v3-old}) means
\begin{align}
f\left(\frac{1}{2\Delta}\right) \geq \frac{1}{\ln (1/\lambda)}. \label{eq-f-1-2-Delta-epsilon-v3}
\end{align}

To analyze Inequality~(\ref{eq-f-1-2-Delta-epsilon-v2}), we now check the monotonicity of $f'(x)$. To this end, the second-order derivatives of $f(x)$ is
\begin{align}
f''(x) & = \frac{ h(x)  }{(1-\lambda^x)^3}  . \label{eq-def-f-second-prime-x}
\end{align}
where we define $h(x)$ as
\begin{align}
h(x) : =  [x\ln\lambda(1+\lambda^x)+2(1-\lambda^x)] \lambda^x \ln\lambda . \label{eq-def-h-x}
\end{align}

To analyze the sign of $f''(x)$ in Eq.~(\ref{eq-def-f-second-prime-x}), we discuss the sign of $h(x)$ in Eq.~(\ref{eq-def-h-x}). Hence, we compute
the derivative of $h(x)$ as $h'(x) = \ln \lambda [1-(1-x\ln \lambda)\lambda^x]  =  \lambda  g(x)$. Given $g(x)>0$ for $0<x\leq 1$, we have $h'(x)>0$ for $0<x\leq 1$. Hence, $h(x)$ strictly increases as $x$ increases for $0<x\leq 1$, implying $h(x)>h(0) = 0$ for $0<x\leq 1$. Using this in Eq.~(\ref{eq-def-f-second-prime-x}), we obtain $f''(x)> 0$ for $0<x\leq 1$, so that $f'(x)$ strictly increases as $x$ increases for $0<x\leq 1$. Thus, we know for any $\epsilon_4  $ satisfying $ 0 < \epsilon_4 < \frac{1}{2\Delta} \leq \frac{1}{2} < 1 $ from  $ \Delta \geq 1 $ that
\begin{align}
\max_{x \in [\epsilon_4,  \frac{1}{2\Delta}]} f'(x) < f'(1) = \frac{1-[1+\ln (1/\lambda)]\lambda}{(1-\lambda)^2}. \label{eq-compare-f-prime-x}
\end{align}
Now we bound the nominator $1-[1+\ln (1/\lambda)]\lambda$ in~(\ref{eq-compare-f-prime-x}). For a lower bound, we use $\ln (1/\lambda) \leq \lambda^{-1} - 1$ given $0 < \lambda < 1 $ to obtain $ 1-[1+\ln (1/\lambda)]\lambda \geq 0$. For an upper bound, we use $\ln (1/\lambda) \geq 1-\lambda$ given $0 < \lambda < 1 $ to obtain $ 1-[1+\ln (1/\lambda)]\lambda \leq 1- (2-\lambda) \lambda = (1-\lambda)^2$. These two bounds imply that $f'(1)$ in~(\ref{eq-compare-f-prime-x}) satisfies $ 0 \leq f'(1) \leq 1 $. Then~(\ref{eq-compare-f-prime-x}) gives
\begin{align}
\max_{x \in [\epsilon_4,  \frac{1}{2\Delta}]} f'(x) < 1. \nonumber
\end{align}
which is used in Inequality~(\ref{eq-f-1-2-Delta-epsilon-v2}) to induce
\begin{align}
f\left(\frac{1}{2\Delta}\right)   \leq  f(\epsilon_4) +  \frac{1}{2\Delta} - \epsilon_4 . \label{eq-f-1-2-Delta-epsilon-v5}
\end{align}
Letting $\epsilon_4 \to 0$ in Inequality~(\ref{eq-f-1-2-Delta-epsilon-v5}), we obtain
\begin{align}
f\left(\frac{1}{2\Delta}\right)  &\leq  \lim_{\epsilon_4 \to 0}   f(\epsilon_4) + \lim_{\epsilon_4 \to 0}  \left(\frac{1}{2\Delta} - \epsilon_4 \right) \nonumber  \\ &  = \frac{1}{\ln (1/\lambda)} + \frac{1}{2\Delta}, \label{eq-f-1-2-Delta-epsilon-v7}
\end{align}
where the last step uses Inequality~(\ref{eq-f-1-2-Delta-epsilon-v3-oldv2}).

Given $\lambda = \frac{\nu}{\mu} $ and $f\left(\frac{1}{2\Delta}\right) = \frac{1}{2\Delta \left[ 1 - \left( \frac{\nu}{\mu} \right)^{1/(2\Delta)}\right]} $ from Eq.~(\ref{eq-def-lambda}) and Eq.~(\ref{eq-def-f-x}), the combination of Inequalities~(\ref{eq-f-1-2-Delta-epsilon-v3})~and~(\ref{eq-f-1-2-Delta-epsilon-v7}) gives the desired result
\begin{align}
 \frac{2}{\ln (\mu / \nu)} \leq  \frac{1}{\Delta \left[ 1 - \left( \frac{\nu}{\mu} \right)^{1/(2\Delta)}\right]}    \leq  \frac{2}{\ln (\mu / \nu)} + \frac{1}{\Delta}  .  \nonumber
\end{align}
\qeda

\subsection{Proof of Lemma~\ref{lem-c-simplified}} \label{subse-lem-c-simplified}

% Lemma~\ref{lem-c-simplified} and Its Proof
% \noindent\textbf{Proof of Lemma~\ref{lem-c-simplified}:} 

For $\delta_5 = \frac{(\epsilon_1 + \epsilon_2) \ln \frac{\mu}{\nu}}{(\epsilon_1 + \epsilon_2) + (1 - \epsilon_1) \cdot (\ln \frac{\mu}{\nu}+1)}$, the term $ \frac{\delta_5 \iffalse_7 \fi }{\ln \frac{\mu}{\nu} - \delta_5 \iffalse_7 \fi }$ equals $\frac{(\epsilon_1 + \epsilon_2) \iffalse_0 \fi   }{(1-\epsilon_1 \iffalse_0 \fi ) \cdot (\ln \frac{\mu}{\nu}+1) }$. \vspace{1.5pt} Given $0< \nu < \mu$, we have $\ln \frac{\mu}{\nu} > 0$, which with $0< \epsilon_1 \iffalse_0 \fi  < 1$ and $\epsilon_2 > 0$ gives
\begin{align}
1+ \frac{\epsilon_1 + \epsilon_2 \iffalse_0 \fi   }{(1-\epsilon_1 \iffalse_0 \fi ) \cdot (\ln \frac{\mu}{\nu}+1) } < 1+ \frac{\epsilon_1 + \epsilon_2 \iffalse_0 \fi   }{1-\epsilon_1 \iffalse_0 \fi } = \frac{1 + \epsilon_2 \iffalse_0 \fi   }{1-\epsilon_1  }  . \nonumber
\end{align}
Thus, Lemma~\ref{lem-c-simplified} is proved. \qeda

\subsection{Proof of Eq.~(\ref{eq-F-t-to-G-t})} \label{Appendix-eq-F-t-to-G-t}

\begin{align}
& \pr{\boldsymbol{F}_{t-\Delta-1} S_{t-\Delta} \ldots S_{t} = \boldsymbol{f}_{t-\Delta-1} s_{t-\Delta} \ldots s_{t}}  \nonumber  \\ & = \sum_{\begin{subarray}{l}
\boldsymbol{f}_{t-\Delta-2}\in \text{Suffix-Set}:\\ \text{suffix}(\boldsymbol{f}_{t-\Delta-2}||s_{t-\Delta-1})=\boldsymbol{f}_{t-\Delta-1}  
\end{subarray}} \nonumber  \\ & \quad \pr{\hspace{-4pt}\begin{array}{l}
\left( \boldsymbol{F}_{t-\Delta-2} S_{t-\Delta-1} \ldots S_{t-1} = \boldsymbol{f}_{t-\Delta-2} s_{t-\Delta-1} \ldots s_{t-1}\right) \\ \land  \left(S_{t} = s_{t} \right)
\end{array}\hspace{-5pt}} \nonumber  \\ &=   \pr{S_{t} = s_{t} } \sum_{\begin{subarray}{l}
\boldsymbol{f}_{t-\Delta-2}\in \text{Suffix-Set}:\\ \text{suffix}(\boldsymbol{f}_{t-\Delta-2}||s_{t-\Delta-1})=\boldsymbol{f}_{t-\Delta-1}  
\end{subarray}} \nonumber  \\ & \quad \pr{ \boldsymbol{F}_{t-\Delta-2} S_{t-\Delta-1} \ldots S_{t-1} = \boldsymbol{f}_{t-\Delta-2} s_{t-\Delta-1} \ldots s_{t-1} }  . \label{eq-G-t-G-t-1}
\end{align}

\begin{align}
& \pi_{\boldsymbol{F}||\boldsymbol{P}}(\boldsymbol{f}_{t-\Delta-1} s_{t-\Delta} \ldots s_{t})   \nonumber  \\ &=   \pr{S_{t} = s_{t} } \times \nonumber  \\ & \sum_{\begin{subarray}{l}
\boldsymbol{f}_{t-\Delta-2}\in \text{Suffix-Set}:\\ \text{suffix}(\boldsymbol{f}_{t-\Delta-2}||s_{t-\Delta-1})=\boldsymbol{f}_{t-\Delta-1}  
\end{subarray}} \pi_{\boldsymbol{F}||\boldsymbol{P}}(\boldsymbol{f}_{t-\Delta-2} s_{t-\Delta-1} \ldots s_{t-1} )  . \label{eq-G-t-G-t-1-stationary1}
\end{align}

\begin{align}
& \frac{\pi_{\boldsymbol{F}||\boldsymbol{P}}(\boldsymbol{f}_{t-\Delta-1} s_{t-\Delta} \ldots s_{t})}{\prod_{i=t-\Delta}^t \pr{S_{i} = s_{i} }}   \nonumber  \\ &=   \sum_{\begin{subarray}{l}
\boldsymbol{f}_{t-\Delta-2}\in \text{Suffix-Set}:\\ \text{suffix}(\boldsymbol{f}_{t-\Delta-2}||s_{t-\Delta-1})=\boldsymbol{f}_{t-\Delta-1}  
\end{subarray}}  \nonumber  \\ & \quad \left( \frac{\pi_{\boldsymbol{F}||\boldsymbol{P}}(\boldsymbol{f}_{t-\Delta-2} s_{t-\Delta-1} \ldots s_{t-1} )}{\prod_{i=t-\Delta-1}^{t-1} \pr{S_{i} = s_{i} }} \cdot \pr{S_{t-\Delta-1} = s_{t-\Delta-1} } \right) . \label{eq-G-t-G-t-1-stationary2}
\end{align}
Using Eq.~(\ref{eq-F-t-F-t-1}) and replacing $t$ therein by $t-\Delta-1$, we have
\begin{align}
& \pr{\boldsymbol{F}_{t-\Delta-1} = \boldsymbol{f}_{t-\Delta-1}}    \nonumber  \\ & = \sum_{\begin{subarray}{l}
\boldsymbol{f}_{t-\Delta-2}\in \text{Suffix-Set}:\\ \text{suffix}(\boldsymbol{f}_{t-\Delta-2}||s_{t-\Delta-1})=\boldsymbol{f}_{t-\Delta-1} 
\end{subarray}} \nonumber  \\ & \quad  \left( \pr{\boldsymbol{F}_{t-\Delta-2} = \boldsymbol{f}_{t-\Delta-2}} \pr{S_{t-\Delta-1} = s_{t-\Delta-1}} \right). \label{eq-F-t-F-t-1-repeat}
\end{align} 
From Eq.~(\ref{eq-G-t-G-t-1-stationary2}) and~(\ref{eq-F-t-F-t-1-repeat}), the transition from $\frac{\pi_{\boldsymbol{F}||\boldsymbol{P}}(\boldsymbol{f}_{t-\Delta-2} s_{t-\Delta-1} \ldots s_{t-1} )}{\prod_{i=t-\Delta-1}^{t-1} \pr{S_{i} = s_{i} }}$ to $\frac{\pi_{\boldsymbol{F}||\boldsymbol{P}}(\boldsymbol{f}_{t-\Delta-1} s_{t-\Delta} \ldots s_{t})}{\prod_{i=t-\Delta}^t \pr{S_{i} = s_{i} }}$ has the same rule as the transition from $\pi_{\boldsymbol{F}}(\boldsymbol{f}_{t-\Delta-2})$ to $\pi_{\boldsymbol{F}}(\boldsymbol{f}_{t-\Delta-1})$, so we can conclude   $\frac{\pi_{\boldsymbol{F}||\boldsymbol{P}}(\boldsymbol{f}_{t-\Delta-1} s_{t-\Delta} \ldots s_{t})}{\prod_{i=t-\Delta}^t \pr{S_{i} = s_{i} }} = \pi_{\boldsymbol{F}}(\boldsymbol{f}_{t-\Delta-1})$, which is exactly the desired result  Eq.~(\ref{eq-F-t-to-G-t}). \qeda

\end{document}